\documentclass[global,numbook,final]{svjour}
\usepackage{times}
\usepackage{amsmath}
\usepackage{amssymb}

\authorrunning{%
  E. Biham, M. Boyer, P. O. Boykin, T. Mor and V. Roychowdhury}
\journalname{Journal of Cryptology}

\spnewtheorem*{nonumlemma}{Lemma}{\bf}{\it}

\let\xor\oplus
\makeatletter
\def\txtline#1{\noalign{\hbox{\strut\hskip\@totalleftmargin {#1}}}} 
\makeatother

\newcommand{\unit}[1]{\mathbf{1}_{\mathrm{#1}}}
\newcommand{\ket}[1]{|{#1} \rangle}
\newcommand{\ketz}[1]{|{#1} \rangle_0}
\newcommand{\zst}[1]{\ket{0}_{\mathrm{#1}}}
\newcommand{\zstx}[1]{\ket{0_x}_{\mathrm{#1}}}
\newcommand{\mst}[1]{\ket{m}_{\mathrm{#1}}}
\newcommand{\keto}[1]{|{#1} \rangle_1}
\newcommand{\ketb}[1]{|{#1} \rangle_b}
\newcommand{\kete}[2]{|{#1} \rangle_{#2}}

\newcommand{\bra}[1]{{\langle {#1}|}}
\newcommand{\brab}[1]{{}_b\langle {#1}|}

\newcommand{\braket}[2]{\langle {#1} | {#2} \rangle}
\newcommand{\pass}[0]{\text{\normalfont pass}}
\newcommand{\swap}[0]{\operatorname{swap}}
\newcommand{\fail}[0]{\text{\normalfont fail}}
\newcommand{\mrv}[1]{\mathbf{#1}}
\newcommand{\info}[0]{\text{\normalfont info}}
\newcommand{\luck}[0]{\text{\normalfont luck}}
\newcommand{\exc}[0]{{\text{\normalfont exc}}}
\newcommand{\sym}[0]{{\text{\normalfont sym}}}
\newcommand{\test}[0]{\text{\normalfont test}}
\newcommand{\allowed}[0]{\text{allowed}}
\newcommand{\secret}[0]{\text{secret}}
\newcommand{\unused}[0]{\text{un-used}}
\newcommand{\rel}[0]{\text{rel}}
\newcommand{\num}[0]{\text{num}}
\def\<{\begin{equation}}
\def\>{\end{equation}}
\allowdisplaybreaks

\makeatletter
\spn@wtheorem{lemm}{Lemma}{\bfseries}{\itshape}
\makeatother

\begin{document}

\title{A Proof of the Security of Quantum Key 
Distribution\thanks{
A shortened version of this paper is published
in \textit{STOC'2000}, and a preliminary full version
appears on the Los-Alamos archive 
http://arxiv.org/abs/quant-ph/9912053~\cite{BBBMR}.}
}

\author{%
   Eli Biham\inst{1} \and
   Michel Boyer\inst{2} \and
   P. Oscar Boykin\inst{3} \and
   Tal Mor\inst{4} \and
   Vwani Roychowdhury\inst{5}}

\institute{%
  Computer Science Department, Technion, \\
  Haifa 32000, Israel. \\ 
  \email{biham@cs.Technion.AC.IL} 
  \and
  DIRO, Universit\'e de Montr\'eal, \\
  CP 6128, Succ. Centre-Ville, \\
  Montr\'eal, H3C 3J7, Canada. \\
  \email{boyer@IRO.UMontreal.CA} 
  \and
  Dept. of Electrical Engineering, UCLA, \\
  Los Angeles, CA 90095-1594, USA. \\
  \email{boykin@ee.ucla.edu} 
  \and
  Computer Science Department, Technion, \\
  Haifa 32000, Israel. \\ 
  \email{talmo@cs.technion.ac.il} \and
  Dept. of Electrical Engineering, UCLA, \\
  Los Angeles, CA 90095-1594, USA. \\
  \email{vwani@ee.ucla.edu}} 
  
\date{\today}

\maketitle

\begin{abstract}
We prove
the security of theoretical quantum key distribution 
against the most general attacks which
can be performed on the channel, by an eavesdropper who has
unlimited computation abilities, and the full power allowed by the
rules of classical and quantum physics.
A key created that way 
can then be used to transmit secure messages such that
their security is also unaffected in the future.
\end{abstract}


\clearpage

\section{Introduction}

Quantum key distribution~\cite{BB84,BBBSS} uses the power of quantum
mechanics to suggest the distribution of a key that is secure
against an adversary with unlimited computation power.
Such a task is
beyond the ability of classical information processing; thus, 
it is the 
main success of the original idea of Wiesner~\cite{Wiesner}
who suggested using quantum mechanics to perform cryptographic tasks.
The extra power gained by the use of quantum bits (quantum two-level
systems, ``qubits'') 
is due to the fact that the state of such a system cannot be
cloned. [Of course, one could use higher level quantum systems as well.]
On the other hand, the security of conventional key distribution
is based on the (unproven) existence of various
one-way functions, and mainly on the difficulty of factoring 
large
numbers, a problem which is assumed to be difficult for a
classical computer, and is proven to be easy for a hypothetical
quantum computer~\cite{Shor}.

The quantum key distribution (QKD) scheme considered in 
our work is the 
protocol of Bennett and Brassard~\cite{BB84}, known as the BB84
protocol. The legitimate users of this (actually, of any)
QKD protocol are conventionally
called Alice (the sender) and Bob (the receiver). Their 
aim is to create and share a secret key.

There are several classes of attacks (see for instance~\cite{BM97a,BBBGM}) 
on quantum key 
distribution that can be performed by an eavesdropper 
having full control of the channel.
The simplest ones are known as individual-particle 
attacks~\cite{FGGNP} in which the transmitted qubits  
are attacked separately, so that the eavesdropper  
can be left 
with some optimal classical information about each transmitted quantum bit.
The eavesdropper can use this classical information in order to 
learn some information about the final secret key.
In contrast, in the most general attack called 
the ``joint attack'', all transmitted quantum particles are 
attacked together, and the eavesdropper's goal is  
to learn as much information as possible about 
the final key, rather than about each transmitted qubit. 
A special class of the joint attack, 
the ``collective attack''~\cite{BM97a} was shown to 
provide more information to the eavesdropper than an 
individual-particle attack~\cite{BMS96}. 
We further explain the differences between the individual-particle attacks,
the collective attacks, and the 
most general attacks (the joint attacks) in 
Subsection~\ref{subsec:eavesdropping}, when we describe 
the two steps of Eve's attack.
Various proofs of security were previously obtained against
collective attacks~\cite{BM97a,BM97b,BBBGM,Mor-thesis} 
(which is a most important subclass of the joint attack), 
and we continue this line of research here to prove the
ultimate security of QKD, against any
attack (under the conventional assumptions of theoretical QKD, 
as explained
below). Note that the eavesdropper is assumed to have unlimited
technology (e.g.,  unlimited computing power, a quantum memory, 
a quantum computer), while 
the legitimate users use practical tools
(or more precisely, 
simplifications of practical tools). 
Such assumptions are required since
the aim of the invention of quantum key distribution
is to obtain a {\em practical} key distribution scheme,
which is proven secure against {\em any} attack, even 
one which is far 
from being practical with current technology.

To prove security against such a super-strong eavesdropper,
conventionally called Eve,   
we develop some important technical tools and we reach some novel
results: We obtain a new {\em information versus disturbance} result,
where the power of quantum information theory is manifested in an
intuitive and clear way.
We show explicitly how the randomness of the choice of bases,
and the randomness of the choice of test-bits provides the desired
security of QKD.
We adopt and generalize sophisticated tools invented in~\cite{BBBGM}:
``Purifications'' which simplify Eve's states,
a bound on accessible information (using Trace-Norm-Difference
of density matrices)
which avoids any complicated optimization of Eve's possible
measurements, and a connection between Eve's 
accessible information and the error-rate she induces.
We add some more simplifications (which were not required in the
analysis of collective attacks in~\cite{BBBGM}):
a reduction to a scheme
in which all qubits are used by Alice and Bob,
and a symmetrization of Eve's attack.

This paper complements the result of Bennett, Mor, 
and Smolin~\cite{BMS96}: That paper shows 
that individual particle attacks are strictly weaker 
(less informative to the eavesdropper) 
than joint attacks\footnote{ 
Many of the leading researchers in experimental quantum
cryptography are unfamiliar with this work of Bennett Mor and
Smolin, and still wrongly state that individual
particle attacks could be as strong as collective/joint attacks.},
and the current paper shows that security can still be obtained
even when the eavesdropper applies the strongest joint attacks.
The current paper also complements the work of Bennett, Brassard,
Cr\'epeau, and Maurer~\cite{BBCM}: That paper shows that 
privacy amplification provides security when the eavesdropper 
is restricted to perform only individual particle attacks, 
and the current paper 
shows that privacy amplification provides security when the eavesdropper
is not restricted, and can apply any joint attack on the particles.

Two other security proofs~\cite{Mayers96,Mayers98},  
and ~\cite{LC98,Lo99}
were reported just prior to ours~\cite{BBBMR}. 
The security result of Lo and Chau~\cite{LC98} 
[note that some of the details were completed or improved in~\cite{Lo99}]
uses novel techniques and is very
important, but it is somewhat limited:
The QKD protocol which is analyzed in~\cite{LC98} requires that 
the legitimate users have quantum memories and fault tolerant
quantum computers, 
technologies which are not yet available to the legitimate
users, and are not expected within the next ten or twenty years,
while the QKD protocol which is analyzed here, the  
BB84 protocol, 
is now demonstrated with some partial success in many labs
(see many references in 
Gisin's reviews~\cite{Gisin-review1,Gisin-review2}). 
Some of the ideas used in~\cite{LC98} appeared earlier,
[e.g., the quantum privacy amplification~\cite{QPA}, and
the quantum repeaters~\cite{Mor96,Mor-thesis}, 
and the use of fault tolerance quantum error correction
for performing quantum privacy 
amplification~\cite{Mor96,Mor-thesis} but Lo and Chau 
succeeded in
using them to yield a novel proof of security from 
classical random sampling
techniques.
The security result of Mayers~\cite{Mayers96,Mayers98} is similar 
to ours in the sense that it proves the 
security of a much more realistic protocol
against an unrestricted eavesdropper, and provides explicit
bounds on the eavesdropper's information. It continues
earlier works such as a 
solution to the error-free case~\cite{Yao95}.

Our proof is different from Mayers,  
was derived independently, 
and may shed more light on the subject.
We analyze the density matrices
which are available to the eavesdropper and 
we prove that it is 
extremely rare that 
these density matrices carry non-negligible
information about the secret key,
and at the same time, Alice and Bob agree to form
a secret key.
In other words, it is extremely rare that Alice and Bob agree to form
a secret key about which these density matrices reveal non-negligible
information.

Two additional proofs were announced
more recently~\cite{SP00,Ben-Or99}.
Shor and Preskill's proof~\cite{SP00} proposes a way to extend Lo and 
Chau's proof so
that it becomes applicable to a more practical protocol, 
hence bypasses the 
main limitation of Lo and Chau's proof.
A written draft of the proof of Ben-Or is 
expected in the near future~\cite{Ben-Or99}.

We base our work on standard assumptions of QKD:
1) We assume the correctness of quantum theory and its
relativistic generalizations, as these were verified with incredible
accuracy in many experiments. 
2) Alice and Bob share an unjammable classical channel. This
assumption is usually replaced by the demand  
that the classical channel is ``unforgeable''; 
an unforgeable channel
can be modified by an eavesdropper but Alice and Bob will notice 
that, with probability exponentially close to 1.
If Alice and Bob share 
a much shorter secret
key to be used for authenticating a standard
classical channel, they can indeed obtain 
an unforgeable channel
(hence the protocol is then a quantum key expansion
protocol, although everyone still call it QKD).
3) Eve cannot attack Alice's and Bob's laboratories.
She can only attack the quantum channel and listen to all
transmissions on the classical channel.
4) Alice sends quantum bits, i.e. two level systems. 
This assumption cannot be fully met in any experimental 
scenario,
but can only be approximated. 

We prove, under those assumptions,  the security of the 
BB84 protocol~\cite{BB84}, against any attack
allowed by the rules of quantum physics.
We prove security for instances in which the error
rate in the transmission from Alice to Bob is up to 7.56\%.

Although experimental QKD is very common
(see for 
instance Gisin's reviews~\cite{Gisin-review1,Gisin-review2}),
at the present time
no experimental system whatsoever is proven 
unconditionally secure.
Some security 
analyses which take into account corrections due to having more than
two levels in the quantum systems 
have been provided (\cite{BMS98,BLMS99}), but research in this area is
still in its early stages.
In fact, many experimental systems are totally insecure due 
to the photon-number-splitting attack~\cite{BLMS99}. 
 
Quantum cryptography~\cite{Wiesner,BB84} is
described in several publications, some of which also
introduce the notations in a more expository way. 
Readers unfamiliar with the basics of 
quantum information processing are referred  to
any recently published textbook on
the subject, e.g., \cite{NielsenChuang2000,Gruska}.   
Here we focus on QKD~\cite{BB84,BBBSS} and specifically 
on the BB84 protocol~\cite{BB84}. 

In BB84 we let 
\begin{description}
\item $\ketz{0}  \equiv \ket{0}$;
\item $\ketz{1} \equiv \ket{1}$;
\item $\keto{0} \equiv \frac{1}{\sqrt{2}}(\ket{0}+\ket{1})$; 
\item $\keto{1} \equiv \frac{1}{\sqrt{2}}(\ket{0}-\ket{1})$,
\end{description}
define four states, such that the first two are orthogonal 
in one basis (known as the computation basis, or the ``z'' basis), 
and the other two are orthogonal in
another basis (the ``x'' basis). 
[Using these ``spin'' notations the bases are 
$\ketz{} \equiv \ket{}_z$, and 
$\keto{} \equiv \ket{}x $.]
Note that the two bases are conjugate, namely,
applying a measurement in one basis on a state belonging to 
the other basis gives a fully random outcome.
In the BB84 protocol Alice and Bob 
use these four possible quantum states. 
Therefore, we shall refer to these states as the BB84 states.

The quantum part of the communication in the BB84 protocol 
contains one step ---
Alice sends Bob a string of qubits, each in one of the 
four BB84 states
(chosen randomly by Alice).
To simplify the analysis, we assume all qubits are sent to Eve,
and then Eve sends all qubits to Bob\footnote{
In case Eve can only hold each qubit for a short time
and must release it before she gets the next, she is
less powerful, so our proof of security covers that 
case as well.}.

The rest of the protocol involves sending classical communication
via the unjammable channel. First Alice sends Bob the basis used for
each photon.
By comparing bases after Alice sends such a state for each qubit
and Bob receives the qubit, 
a common key can be created 
in instances when Alice and Bob used the same
basis. 
Comparing the bases must be performed after Bob receives the qubits, so
that the eavesdropper cannot benefit from having this knowledge
while still holding the qubits.
The common key obtained from the above steps is known as the 
``sifted key''.
A final key is then obtained from the sifted key,
after performing several more steps: 
testing the error rate on some test bits, chosen at random; 
throwing away these test bits, while
Alice and Bob can now have some good estimation of the
error-rate on the remaining shared bits
(called information bits); 
correcting errors on these information bits,
and amplifying the privacy, by creating a shorter final key.

Alternatively, 
if Bob has a memory where he can keep his qubits unchanged
after
receiving them (we call such a memory ``a quantum memory''),
a simpler protocol for obtaining a sifted key is obtained:
Bob waits with the received qubits
till he learns the basis, and then measures in the right bases.
The sifted key is twice as big in this case or the initial string
of qubits can be shortened to half, if the final length of 
the sifted key is to remain
the same.

We prove here  
the security of that simplified protocol in which only
the bits relevant for the sifted key
are discussed; we call it the ``used-bits-BB84''.
We formally describe the used-bits protocol (in detail) in the next
section.
The proof of the security of the 
original BB84 protocol (in which Bob does not have a quantum memory)
easily follows due to a
simple reduction, as we 
show in Appendix~\ref{app-used-bits}. 

In the most general attack on the channel, 
Eve attacks the qubits in two steps. First, 
she lets all qubits pass
through a device that weakly  
probes their state via a quantum unitary transformation. 
Then, after receiving all the classical data, she measures the probe.
Eve's goal is to learn as much information as possible about the final
key without causing Alice and Bob to abort the protocol due to a
failure of the test.
We consider here any attack chosen by Eve, described by
these two steps, and we prove security against any such attack.
We formally explain Eve's most general attack in the next section.

The issue of the security criteria is non-trivial since
one obvious security criterion, namely 
that {\em ``Eve's information given that the test 
passed, is negligible''}, 
does not work; 
this criterion cannot be proven, as a counter 
example exists\footnote{
Namely, there is an attack such that Eve's information is large
even when the test is passed (although in such cases the test is passed very
rarely); Such attacks are studied in Section~\ref{subsec:security}.}
Another natural security
criterion saying 
that {\em ``either Eve's average information is negligible
or the probability that the test is passed is negligible''}, 
also does not work (for a similar reason). 
The criterion that we shall prove in this work says that
{\em ``the event where the test is passed AND Eve's information 
is not negligible, is extremely rare''}.
This security criterion 
is formally presented in the next section.

We will moreover show that
the final key is reliable:
the keys distilled by Alice and Bob (after error correction and
privacy amplification) are
identical except for some exponentially small probability.

Section~\ref{sec:notations} provides a formal description of the
used-bits-BB84 protocol, 
the most general attacks, and the security and reliability criteria.
The rest of the paper contains three main steps leading to the
desired proof of security:
In Section~\ref{Sec:attack} we reduce the problem
of proving security to a simpler problem
of optimizing over all attacks symmetric to the bit values 0 and 1.
In Section~\ref{Sec:infdis} we analyze the information bits
in the bases actually used by Alice and Bob,
and we prove our main {\em information versus disturbance} theorem 
for symmetric attacks; 
the eavesdropper information about the final key
is bounded by the 
probability of errors induced in the {\em other} bases
(namely, errors induced if the other bases were used
by Alice and Bob). 
We then obtain 
in Section~\ref{Sec:compl} an exponentially small 
bound on Eve's information,
proving that the security criterion 
(\ref{sec-crit}) 
described in Section~\ref{sec:notations}
is always satisfied in QKD, provided a good code for error
correction and privacy amplification is used.
Finally, we analyze a specific code, the random linear code,
and 
we prove security for instances in which the error
rate in the transmission from Alice to Bob is up to 7.56\%.
We also analyze the conditions under which this code can
provide data relevant to experimentalists who choose
some parameters (such as the number of photons used for the
communication) and would like to obtain bounds on Eve's
information, on the probability of errors in the final key,
and on the resulting bit-rate of the protocol.
Such explicit bounds are presented here for any error rate
equal to or smaller than 5.50\%.
We summarize these results in Table~\ref{table1:rates}.

We conclude the paper by summarizing the tools used here, and by
suggesting that some of them could be relevant for other proofs
as well. 
Various technical details and proofs of several lemmas
are provided in the appendices.

\clearpage

\section{Notations, the Protocol, Eve's Attack, the Security 
Criteria, and the Main Results}\label{sec:notations} 

\subsection{The used-bits BB84 protocol}\label{subsec:ub-bb84}

Let us describe the used-bits protocol in detail, splitting it into 
{\em creating the sifted key} and {\em creating the final key
from the sifted key}.
This simplified 
protocol assumes that Bob has a quantum memory. 
\begin{enumerate}
\item[I.]Creating the sifted key: 
\item
Alice and Bob choose a large integer $n \gg 1$.
The protocol uses $2n$ bits.

\item
Alice randomly selects two $2n$-bit strings, $b$ and $i$ 
and sends Bob, via a quantum communication channel, 
the string of $2n$ qubits
\[
\ketb{i} = \kete{i_1}{b_1} \kete{i_2}{b_2} \ldots \kete{i_{2n}}{b_{2n}}
\]

\item
Bob tells Alice when he receives the qubits.
[If he received less than $2n$ qubits he adds any missing qubit,
but in an arbitrary state.
If he received more than $2n$ qubits he ignores any extra qubit.
E.g., if qubit number 17 did not arrive Bob will add it (by choosing
its value and basis at random),
and if two qubits arrived instead of one when Bob expects qubit number 17,
then Bob will ignore one of them. Obviously, such cases will
contribute to the error rate, $p_{\rm test}$.]

\item
Alice publishes the bases she used, $b$; this step should be performed
only after Bob received all the qubits.

Bob measures the qubits in Alice's bases to obtain a $2n$-bit 
string $j$.

We shall refer to the resulting $2n$-bit string as the sifted key,
and it would be the same for Alice and Bob, i.e. $j=i$,
if natural errors and
eavesdropping did not exist.

\end{enumerate}

\begin{enumerate} 

\item[II.]
Creating the final key from the sifted key:

\item

Alice chooses at random a $2n$-bit  
string $s$ which has exactly $n$ zeroes and $n$ ones.
There are ${2n \choose n}$ such strings to choose from.

\item 

{}From the $2n$ bits,
Alice selects a subset of $n$ bits, determined by the zeros
in $s$, to be the test bits.
Alice publishes the string $s$, along with 
the values of the test bits (given by an $n$-bit string $i_T$).
The values of Bob's bits on the test bits are given by $j_T$.

The other $n$ bits are the
information bits (given by an $n$-bit string $i_I$). 
They are used for deriving a final key via error
correction codes (ECC) and privacy amplification (PA) techniques.

Later on, Alice will send the ECC and PA information
to Bob, hence Bob needs to correct his errors using the ECC data,
and to obtain a final secret
key equal to Alice's using the PA data.

\item
Bob verifies that the error rate $p_{\test} = |i_T \xor j_T|/n $
in the test bits is lower than
some pre-agreed allowed error-rate $p_{\allowed}$,
and aborts the protocol if the error rate is larger.
The maximal possible allowed error-rate is found in 
Section~\ref{RLC-exist}.

\item\label{it:bobsbits}
Bob also publishes the values of his
test bits ($j_T$). This is not crucial for the
protocol, but it is done to simplify the proof.

\item
Alice selects an $(n,k,d)$ linear
error correcting code $\mathcal{C}$ with $2^k$ code words of $n$ bits
and a minimal Hamming distance $d$ between any two words,
along with the ECC parities on the information bits.
The strategy is that Alice announces an $r \times n$ parity check 
matrix $P_\mathcal{C}$
of $\mathcal{C}$
by announcing its $r=n-k$ rows of $n$ bits 
$v_1, \ldots, v_r$.
This means that the code contains any $i$ 
such that $i \cdot v_q = 0$ for any $q \in \{1 \ldots r\}$.
Formally speaking, 
$\mathcal{C} = \{ i \in \{0,1\}^n \mid i P_\mathcal{C}^\top = 0 \}$,
with $P_\mathcal{C}^\top$ the
transpose of $P_\mathcal{C}$.
Alice then also announces the $r$-bit string 
$\xi = i_I P_\mathcal{C}^\top$ whose bits are 
the parities of her
(random) information string $i_I$ 
with respect to the parity check matrix 
(so the $q$-th bit $\xi_q$
of $\xi$ is 
$\xi_q = i_I \cdot v_q$ for all $1 \leq q \leq r$). 
Bob doesn't announce anything.

We now explain how the code $\mathcal{C}$ is chosen.
The condition on $\mathcal{C}$ is that it corrects
$t \ge (p_{\allowed} + \epsilon_{\rel})  n $ errors,
for some positive (pre-determined)
reliability parameter $\epsilon_{\rel}$.
If an ECC has Hamming weight $d \ge 2t+1$ it will
always correct $t$ errors, and thus the condition 
$d  \ge 2 (p_{\allowed} + \epsilon_{\rel})  n + 1  $ 
is sufficient. Meaning that, any code satisfying this criterion 
is good for
Alice and Bob.

For Random Linear Codes a better bound exists, and
 $d  \ge (p_{\allowed} + \epsilon_{\rel})  n + 1$ 
is also sufficient as noted in~\cite{Mayers98}; 
It is not promised that such a code always corrects $t$ errors,
but it is promised that it corrects $t$ errors with probability
as close to 1 as we want (provided we choose a sufficiently large
$n$).

\item
Bob performs the correction on his information bits $j_I$
as follows:
he finds the $n$-bit string $j^{\rm Bob}$ such that
$j^{\rm Bob} P_\mathcal{C}^\top = \xi$ and such that the Hamming distance
between  $j^{\rm Bob}$ and $j_I$ is minimal. As long as there are at most
$t$ errors in $j_I$ (i.e. $|j_I \oplus i_I| \leq t$)
the obtained string is unique, and Bob finds the right string, namely 
$j^{\rm Bob} = i_I$.
Note that we are not concerned 
here with the efficiency of finding $j^{\rm Bob}$,
but a practical protocol ought to be efficient as well.

\item
Alice selects a privacy amplification function ($\mathcal{PA}$) and
publishes it.  The PA strategy is to publish $m$ strings,
of length $n$ each. These
{\em privacy-amplification parity-check strings} 
$v_{r+1}, \ldots, v_{r+m}$ shall be used as the rows of an $m \times n$
parity matrix $P_\mathcal{PA}$ so that
the final secret key is $a \equiv i_I P_\mathcal{PA}^\top$,
with $a_t = i_I \cdot v_{r+t+1}$ (for $0\le t \le m-1$)).
This strategy is similar to error correction except
that 
the $m$-bit string (namely, the final key) $i_I P_\mathcal{PA}^\top$ 
is kept secret.

The PA strings must be chosen such that the minimal distance
$\hat v$ between any PA parity string $v$ and
any string in the span of their union with the
parity-check-strings of the ECC (the dual to the code)
is at least $\hat v \ge 2 (p_{\allowed} + \epsilon_{\sec}) \ n$.
[This is important for preventing Eve from learning much from 
the error-correcting procedure,
and furthermore from learning something
about the correlations between the bits of the final key.]
Note that, by definition, the minimal 
distance of the space spanned by the ECC
and PA strings $v_1, \ldots, v_{r+m}$, which we
shall denote  $d^\perp$,
is less than the distance $\hat v$; hence if we demand 
$d^\perp \ge 2 (p_{\allowed} + \epsilon_{\sec}) \ n$,
the above desired criterion,  
$\hat v \ge 2 (p_{\allowed} + \epsilon_{\sec}) \ n$,
is automatically satisfied (due to $\hat v \ge d^\perp$).

\item
Bob 
calculates $a= i_I P_\mathcal{PA}^\top$ to finally get the key.

\end{enumerate}

\subsection{Eavesdropping}\label{subsec:eavesdropping}

In the most general attack on the channel, 
Eve attacks the qubits in two steps. First she lets all qubits pass
through a device that weakly 
probes their state via a quantum unitary transformation. 
Then, after receiving all the classical data, she measures the probe.
Note that Eve can gain nothing by measuring the probe earlier, or 
by measuring the qubits
while passing through her. Any such measurement can also be performed
by attaching a probe, applying a unitary transformation, and measuring the 
probe (or part of it) at a later stage. Since there is no gain in performing
a measurement before learning all the classical information that is transmitted
throughout the protocol, the optimal attack (WLoG) is to perform all
measurements after receiving all classical information.
Furthermore,
Eve gains nothing by sending Bob a state that is not a $2n$ qubit state,
so without loss of generality, we assume she sends exactly $2n$ photons:
If Eve sends less than $2n$ qubits, Bob will add the missing qubits
in an arbitrary state (see item I-3 in the protocol),
so Eve could have done it herself. If Eve sends more than $2n$ qubits,
Bob ignores the extra qubits, and again Eve could have done
it herself. 
[An important remark though: the allowed error 
rate in these cases must 
still be limited as described in this work. 
However, in real applications 
the natural losses of qubits become very high due to transmission 
across long distances. 
If one does not wish to limit the distance too much, 
and wishes 
to have security even if losses are much higher than $p_{\allowed}$,
then this is still possible. See a brief explanation in 
Appendix~\ref{app-used-bits}.]

It is important to enable an analysis of Eve's most general attack.
Thus we formally split Eve's attack into her transformation $U$ and her
measurement $\cal{E}$.
\begin{itemize} 

\item[]Eve's transformation, $U$:
Eve attacks the qubits while they are in the channel
between Alice and Bob. Eve can perform any attack allowed by the laws of
physics, the most general one being any unitary
transformation $U$ on Alice's qubits and Eve's probe (an ancilla
initially in a state
$\zst{E}$). 

We are generous to Eve, allowing her to attack all the qubits together
(in practice, she usually needs to release the preceding
qubit towards Bob before
she has access to the next one).

Without loss of generality
we assume that all the noise on the qubits
is caused by Eve's transformation.

A remark:
In individual-particle attacks and in collective attacks Eve's 
transformation is restricted 
so that each transmitted qubit is attacked using a separate,
unentangled probe,
so that the analysis of $U$ is much simplified.
In collective attacks the next step is as general as it is for the joint 
attacks (so that Eve can measure all probes together). 
In contrast, in individual-particle attacks Eve is only allowed to
measure each probe separately from the others. 

\item[] Eve's measurement, $\cal{E}$:
Eve keeps the probe in a quantum memory, meaning that she keeps its state
unchanged.
After Eve receives {\it all} the classical information from
Alice and Bob, including the bases of all bits $b$, 
the choice of test bits $s$, the test bits values, $i_T$ and $j_T$,
the ECC, the
ECC parities $\xi$, and the PA,
she tries to guess the final key using her best
strategy of measurement. The measurement can be done by adding a second
ancilla, and performing a standard projection measurement on Eve's probe
and the ancilla.
This measurement is alternatively described (without the need for this
second ancilla) by the so called
``generalized measurement'' or ``POVM'', $\cal{E}$, which is a set 
of positive operators ${\cal{E}}_e$ such that $\sum_e {\cal{E}}_e =1$.    
When the measurement is
applied onto a density matrix $\rho$ the outcome $e$ is obtained
with probability $p(e) = {\rm Tr}(\rho {\cal{E}}_e)$.
We fix \footnote{
This fixing is allowed due to Davies' theorem~\cite{Davies}.}
the set of possible outcomes $e$, so that it is the same for
all the POVMs used by Eve after she learns
$i_T,j_T,b,s$ and $\xi$.

For more information about POVMs and their connection to standard 
projection measurements 
in an enlarged Hilbert space, 
see~\cite{Peres93,NielsenChuang2000}. 

\end{itemize}
Eve's goal is to learn as much information as possible about the final
key without causing Alice and Bob to abort the protocol due to a
failure of the test.
The task of finding Eve's optimal operation in these two steps is very
difficult. Luckily, to prove security that task {\em need not} be solved,
and it is enough to find bounds on Eve's optimal information (via any
operation she could have done):
In order to analyze her optimal transformation we find bounds for {\em any}
transformation $U$ she could perform, and in order to analyze her optimal
measurement we find bounds for {\em any} measurement $\cal{E}$
she could perform. 

\subsection{What does security mean?}
\label{subsec:security}

We consider here any attack chosen by Eve, described by
$U$ and $\cal{E}$. Let us explain what we mean by saying
that security shall be proven. 

As we already mentioned in the introduction,
the issue of the security criteria is non-trivial.
One obvious security criterion, namely 
that 
{\em ``Eve's information given that the test passed, is negligible''}, 
can be proven wrong (for QKD), and furthermore, 
another natural security
criterion saying that 
{\em ``either Eve's average information is negligible
or the probability that the test is passed is negligible''}, 
also does not work. 

The criterion that we shall prove here says that
{\em ``the event where the test is passed AND Eve's information 
is not negligible, is extremely rare''}.

To be more precise we formally present now these security criteria.
We first provide some relevant information-theoretic notations 
(for some more basic definitions see Appendix~\ref{APP:info-basics}). 
Let $\mrv{A}$ be the random variable whose values are 
Alice's final key,
$a=i_I P^{\top}_{\mathcal{PA}}$, 
and $\mrv{E}$ be a random variable whose 
values $e$
are the outputs
of Eve's measurement $\cal{E}$. Note that $e$ are outcomes of a measurement
that itself is a function of all the classical data provided to Eve,
the ECC and PA (that can be given to Eve in advance), and also 
$i_T, j_T, b, s$, and $\xi$.
However, we usually consider {\em any} attack, therefore for any fixed 
parameters of the attack, $\{U, \cal{E}\}$, the resulting 
$e$ are regular classical values of
a regular classical random variable $\cal{E}$, so all standard rules of
classical information theory (as described in 
Appendix~\ref{APP:info-basics})
apply to them.
Note that our proof never needs to assume that the ECC data 
$P_\mathcal{C}^\top$ and the PA data  
$P_\mathcal{PA}^\top$   
are random, or even that these are initially unknown to Eve.
Therefore these can be chosen in advance and be considered as fixed parameters 
of the protocol.

Let $\mrv{T}$ be the random variable presenting whether the test passed or
failed ($\mrv{T}$ is ``$\pass$'' if $|i_T \oplus j_T| \leq n p_a$ and is
``$\fail$''
otherwise, with $p_a$ denoting the allowed error rate 
$p_a \equiv p_{\allowed}$). Let $c_T = i_T \oplus j_T$ and 
$c_I = i_I \oplus j_I$ be the error syndromes 
on the test and the information
bits.  
Let $I(\mrv{A};\mrv{E})$ be the mutual information
between Alice's final
key and the results of Eve's measurement.
Since some classical data is given to Eve, let 
$\mrv{I}_{Eve}\equiv I(\mrv{A};\mrv{E} \mid i_T,j_T,b,s,\xi)$
be the information Eve has about the key given a particular PA,  ECC
(that remain fixed parameters), $i_T$, $j_T$, $b$, $s$ and $\xi$
(the parity string on the information bits, $\xi = i_I P_\mathcal{C}^\top$).
This information might be large for some specific values (for instance,
if $b$ is fixed, and
Eve has accidently guessed all the bases correctly), but on average
it ought to be negligible in order for the key to be secret.
The average information 
obtained by Eve if a key was always created by Alice
is 
$\langle \mrv{I}_{Eve} \rangle \equiv 
I(\mrv{A};\mrv{E} \mid \mrv{I}_T,\mrv{J}_T,\mrv{B},\mrv{S},\mrv{\Xi})$,
where $\mrv{I}_T$, $\mrv{J}_T$, $\mrv{B}$, $\mrv{S}$ and $\mrv{\Xi}$ 
are the random
variables associated to the random outputs 
$i_T$, $j_T$, $b$, $s$ and $\xi = i_I P_\mathcal{C}^\top$. 
This information cannot be proven to be small, because the fact 
that the test must be passed is not taken into consideration.

We can now formally present our security criteria.
In order to get a better intuition of what security really
means, we also formally present 
in Appendix~\ref{APP:bad-sec-criteria} 
the two security criteria mentioned above,
criteria that are not met by the QKD protocol.
We even prove via counter examples, the {\em SWAP attack}
and the {\em half-SWAP attack}, that these security criteria
indeed don't work\footnote{If Eve is applying the SWAP attack, her information
given that the test is passed will not be small, and the first criterion is not
satisfied;
If Eve is applying the Half-SWAP attack, she gets a lot of
information (half the bits on average), and yet passes the test with high
probability, so the second criterion is not satisfied.
In contrast, the criteria we use in this paper are satisfied by any attack
whatsoever.}.
The SWAP and the half-SWAP examples motivate a more
precise definition of security (first used in~\cite{Mayers98}) that does work
properly, and shall be used in the current work.

\subsubsection{The security criterion:}

We show in this paper that the event where 
the test is passed \emph{and} Eve obtains meaningful 
information about the key is extremely unlikely.  
This is proven here for any attack $\{U,\cal{E}\}$.
Formally, our security criterion is:
\begin{equation}
 P\left[(\mrv{T}=\pass)\wedge (\mrv{I}_{Eve} \ge A_{\info}
 \ e^{-\beta_{\info} n} )\right] \le
  A_{\luck} \  e^{-\beta_{\luck} n}
\ ,
 \label{sec-crit}
\end{equation}
with $A_{\info}$, $\beta_{\info}$, $A_{\luck}$ and $\beta_{\luck}$
positive constants.
Note that this is a criterion for exponential security, 
and a less strict criterion can be defined if one is willing 
to accept polynomial security
(say, with a huge polynomial such as $n^{1000}$).
However, exponential criteria are preferable when possible,
and we succeed to prove here an exponential security criterion.

\subsubsection{An alternative security criterion:}

Let us define $\mrv{I}'_{Eve}$ to be equal to $\mrv{I}_{Eve}$ when 
$\mrv{T}=\pass$ and to be equal to $0$ otherwise. 
Then, the event 
 $\left[(\mrv{T}=\pass)\wedge (\mrv{I}_{Eve} \ge A_{\info}
 \ e^{-\beta_{\info} n} )\right]$ is identical to the event 
$[\mrv{I}'_{Eve} \geq A_\info e^{-\beta_\info n} ]$.
The security criterion can now be
written more concisely as
\begin{equation*} 
P[\mrv{I}'_{Eve} \geq A_\info e^{-\beta_\info n} ] \leq
  A_\luck e^{-\beta_\luck n} \ .
\nonumber
\end{equation*}
The expectancy
of $\mrv{I}'_{Eve}$ which is 
\begin{equation*} \langle \mrv{I}'_{Eve}\rangle =
 \sum_{i_T,j_T,b,s,\xi} \mrv{I}'_{Eve}(i_T,j_T,b,s,\xi)p(i_T,j_T,b,s,\xi) 
\ , 
\nonumber
\end{equation*}
can now be used to define an important security condition:
\begin{equation} 
\label{sec-crit-2}
\langle \mrv{I}'_{Eve}\rangle \le A e^{-\beta n}  \ ,
\end{equation}
with $A$ and $\beta$ positive constants.
As the following lemma shows, the security criterion,
Eq.
(\ref{sec-crit}) is implied by this security condition.
\begin{lemm}\label{lemm-sec-crit-2}
If 
$\langle \mrv{I}'_{Eve}\rangle 
\leq Ae^{-\beta n}$ for $A>0$ then
\begin{equation*}
P[ \mrv{I}'_{Eve} \geq A_\info e^{-\beta_{\info}n} ] \leq
 A_\luck e^{-\beta_\luck n}
\nonumber
\end{equation*}
for all $A_\info$, $A_\luck$, $\beta_\info$, $\beta_\luck$  such
that $A_\info A_\luck = A$, 
 $\beta_\info + \beta_\luck = \beta$ and $A_\luck > 0$.
\end{lemm}
[Note that the security criterion~\ref{sec-crit} is therefore 
implied since the event 
 $\left[(\mrv{T}=\pass)\wedge (\mrv{I}_{Eve} \ge A_{\info}
 \ e^{-\beta_{\info} n} )\right]$ is identical to the event 
$[\mrv{I}'_{Eve} \geq A_\info e^{-\beta_\info n} ]$.]
\begin{proof}
$\mrv{I}'_{Eve}$ is never negative. Therefore,
by Markov's inequality~\cite{Billings86}
(that is $P[X \geq \alpha] \leq \langle X \rangle/\alpha$
for any non-negative random variable $X$),
\begin{align*}
P[\mrv{I}'_{Eve} \geq A_\info e^{-\beta_\info n} ] \leq
  \frac{\langle \mrv{I}'_{Eve} \rangle}{A_\info e^{-\beta_\info n}}
\leq \frac{Ae^{-\beta n}}{A_\info e^{-\beta_\info n}} =
A_\luck e^{-\beta_\luck n} \text{\hspace*{\fill}} & &\qed
\end{align*}
\end{proof}

We gain two things by using this alternative security criteria.
The first is some additional intuition about the security parameter, and
the second is a final form of the criterion which is the one we actually prove
here in the paper.

By definition, 
$ \langle \mrv{I}'_{Eve}\rangle =
 \sum_{i_T,j_T,b,s,\xi} \mrv{I}'_{Eve}(i_T,j_T,b,s,\xi)p(i_T,j_T,b,s,\xi) 
$ 
is equal to
$  \sum_{i_T,j_T: |i_T \oplus j_T|\leq n\,p_a} \sum_{b,s,\xi}
     I(\mrv{A};\mrv{E}\mid i_T,j_T,b,s,\xi)p(i_T,j_T,b,s,\xi) 
$,
thus, it is easy to calculate that   
\begin{equation} 
\label{Average-I-prime-Eve-1}
\langle \mrv{I}'_{Eve}\rangle =  
I(\mrv{A};\mrv{E}\mid \mrv{I}_T,\mrv{J}_T,\mrv{B},\mrv{S},\mrv{\Xi},
  \mrv{T}=\pass)P[\mrv{T}=\pass] 
\ , 
\end{equation}
(see Appendix~\ref{APP:dealing-with-I'} for the details of 
this calculation).
This expression 
provides some intuition regarding the security 
criterion, Eq.(\ref{sec-crit-2}):
It says that if either the probability
to pass the test is negligible or Eve's information given that the
test is passed is negligible, then security is promised.

Using
$c_T$ (the error syndrome on the test bits) 
and using the random variable $\mrv{C}_T \equiv \mrv{I}_T \xor 
\mrv{J}_T$
(the random variable corresponding to the error syndrome),
we can also write  
\begin{equation}
\label{Average-I-prime-Eve-2}
\langle \mrv{I}'_{Eve}\rangle = 
\sum_{c_T|\mrv{T}={\pass}}P\left[\mrv{C}_T=c_T\right]\ 
I(\mrv{A}; \mrv{E}|\mrv{I}_T,\mrv{C}_T=c_T,\mrv{B},\mrv{S},\mrv{\Xi})
\end{equation}
(see Appendix~\ref{APP:dealing-with-I'} for the details of 
this calculation as well).
This is true since the  
random variable $\mrv{C}$ is equivalent
to the random variable $\mrv{J}$ when the 
random variable $\mrv{I}$ is given, and since
summing over all the events  
$\{c_T|\mrv{T}={\pass}\}$ provides exactly the event 
$\{\mrv{T}={\pass}\}$.

This last expression, Eq.(\ref{Average-I-prime-Eve-2}),
tells us that the security criterion 
(\ref{sec-crit-2}) is satisfied if:
\begin{equation} 
\label{sec-crit-3}
\sum_{c_T|\mrv{T}={\pass}}P\left[\mrv{C}_T=c_T\right]\ 
I(\mrv{A}; \mrv{E}|\mrv{I}_T,\mrv{C}_T=c_T,\mrv{B},\mrv{S},\mrv{\Xi})
\le A e^{-\beta n}  \ .
\end{equation}
Thus, this last equation is yet another form of the security criteria.
Indeed,
in Lemmas~\ref{sec5-lemm3} and~\ref{sec5-lemm4}  
in Section~\ref{Sec:compl} we obtain an exponentially small bound on 
$\langle \mrv{I}'_{Eve}\rangle$.
This inequality
then implies that the security criterion
(\ref{sec-crit}) is satisfied, for all attacks 
without any restriction whatsoever,
therefore proving
the security of the used-bits-BB84 and the original BB84 protocols.

To improve the intuition about the different security criteria 
(those that work for QKD 
and also those that do not work) we prove in
Appendix~\ref{APP:Security-Half-SWAP} 
that the Half-SWAP attack can easily be
dealt with, once we use our security criteria; 
meaning that the security
criteria are still satisfied.

\subsection{The main result: a security proof}

In this paper we provide a proof of the 
security of the used-bits BB84 protocol against any attack on the channel.

Formally we prove the following:
\begin{description}
\item
If the allowed error-rate $p_a$, some positive number $\epsilon_{\sec}$, 
and the ECC+PA codes
are chosen such that $p_a + \epsilon_{\sec} \leq \hat{v}/2n$
with $\hat{v} = \min_{r'=r+1}^{r+m} d_H(v_{r'}, V_{r'}^\exc)$ where $d_H$ is 
the Hamming distance, $v_{r'}$ a parity-check string, 
and $V_{r'}^\exc$ the $2^{r+m-1}$ space which is
the span of the ECC and PA
excluding $v_{r'}$ 
(namely, the span of $v_1,\ldots,v_{r'-1},v_{r'+1},\ldots,v_{r+m}$), then
for any $A_{\info} > 0$, $A_{\luck}>0$ such that $A_{\info}A_{\luck} = 2m$
and any $\beta_{\info}$ and $\beta_{\luck}$ such that
$\beta_{\info} + \beta_{\luck} = \epsilon_{\sec}^2/4$,
\begin{equation}
P\left[(\mrv{T}=\pass) \wedge (\mrv{I}_{Eve} \ge A_{\info} 
  \ e^{-\beta_{\info} n}) \right] \le A_{\luck} \  e^{-\beta_{\luck} n} 
\end{equation}
where $\mrv{T} = \pass$ iff $|c_T| \leq n p_a$ and
$\mrv{I}_{Eve} = I(\mrv{A}; \mrv{E} \mid i_T, j_T, b, s, \xi)$.
\end{description}

\subsection{Reliability}\label{subsec:reliability}

It will moreover be shown here that
if the ECC corrects $P_a + \epsilon_{\rel}$ errors then 
the final $m$-bit key is reliable:
The keys distilled by Alice and Bob are
identical except for some exponentially small probability
$A_{\rm rel} \ e^{-\beta_{\rm rel} n}$, with $A_{\rm rel} = 1$ 
and $\beta_{\rm rel} = \epsilon_{\rel}^2/2$.

We shall eventually present here 
an example of a family of ECC+PA codes
such that the final key is secure and reliable, as long as the error
rate $p_a$ is less than 7.56\%, and such that the bit-rate
approaches one when the error-rate approaches zero.
Furthermore, we present a different range of these codes 
such that for large enough\footnote{ 
Namely, not asymptotically large.
For instance, $n$ of the order of $10^4$ or $10^5$.}
but reasonable $n$ the final key is secure and reliable, 
as long as the allowed error
rate $p_a$ is less than 5.50\%; in Table~\ref{table1:rates}
we provide some specific numbers 
that might be interesting to experimentalists who design a QKD protocol.

\clearpage

\section{Eve's Attack}
\label{Sec:attack}

In the used-bits BB84 
protocol Alice encodes a string $i$ in the bases of her
choice $b$ in the state $\ketb{i}$ which she sends to Bob via a quantum channel;
Bob measures a string $j$ using the same set of bases.
In order to perform her attack, 
Eve prepares a probe, E, in a known (ancillary) state, which W.L.G. can
be written as a vector $\zst{E}$ and performs a
unitary transformation $U$ on the state
\[
\zst{E}\ketb{i}
\]
where $\ketb{i}$ is assumed to have been intercepted by Eve. 
The resulting state
$U \zst{E}\ketb{i}$ can be expressed in a unique way as a sum
\begin{equation}\label{eq-Eprimeij}
U \zst{E}\ketb{i} = \sum_j \ketb{E'_{i,j}} \ketb{j}
\end{equation}
where the vectors $\ketb{E'_{i,j}}$ are non normalized vectors in Eve's probe space. 
\begin{equation}\label{eq-Eij}
\ketb{E'_{i,j}}  = \brab{j}U \zst{E}\ketb{i} 
\end{equation}
Eve then sends the
disturbed qubits to Bob, keeping her probe in her hands. 
We call the state above 
\begin{equation}\label{Eve-Bob-state}
\ket{\psi'_{i}} \equiv \sum_{j} \kete{E'_{i,j}}{b} \ketb{j}
\end{equation}
``Eve-Bob's state", 
because it is the 
state in the hands of Eve and Bob together. 

Of course, Eve does not
know the basis $b$ when she performs 
her attack $U$ with initial probe $\zst{E}$.
Actually, Eve-Bob's state is not known to any of the players: 
Alice knows $i$ and $b$, Eve knows $U$ 
(namely, the set of states $\ketb{E'_{i,j}}$) but she knows 
neither $i$, nor $j$ nor $b$,
while Bob knows nothing prior to obtaining $b$ from Alice 
(except his knowledge of the protocol). 
In the next steps Alice sends $b$ to Bob (and Eve), 
and Bob measures and obtains his sifted key $j$. 
Then Alice sends $s$ to Bob (and Eve) and both Alice and 
Bob disclose the test bits $i_T$ and $j_T$. The information
bits are still kept secret.

This section deals with two issues.
1.--- symmetrizing Eve's attack;
2.--- the attack on all bits versus the attack 
induced on the information bits.

Subsection~\ref{Subsec:sec-sym-eves-attack}
presents the symmetrized attack.
Subsection~\ref{Subsec:properties-symm-att}
presents important properties of the symmetric attack.
Subsection~\ref{Subsec:optim-symm-att}
proves that symmetric attacks are at least as good for
Eve as any other attack can be.
Subsection~\ref{Subsec:eves-bobs-state-1}
distills the attack on the information bits,
and finally, Subsection~\ref{Subsec:More-properties-of-symm-att}  
analyzes the symmetrized attacks, when test bits and information bits
are treated separately.

\subsection{Symmetrizing Eve's attack}
\label{Subsec:sec-sym-eves-attack}

For any attack $\{U,\cal{E}\}$, 
we shall now define a different attack $\{U^{\sym},\cal{E}^{\sym}\}$, 
which can be at least as good (for 
Eve) as the attack $\{U,\cal{E}\}$, 
it is symmetric to bit flips, and it is simpler to analyze.
The symmetric attack $\{U^{\sym},\cal{E}^{\sym}\}$ 
is obtained by enlarging Eve's probe,
adding a second probe, M, 
containing $2n$ qubits in a state 
$(1/2^n)\sum_m\mst{M}$,
and transforming it and measuring it as
described below. 
The attack is 
``symmetric'' in a sense that it is unaffected by the choice of $i$
by Alice, and this is true for any basis $b$. 

The symmetrization is done here in a physical way, namely, as a
process that Eve can actually do if she wants to\footnote{
One can also view the symmetrization as a virtual process.
This makes some differences, 
but we do not consider this case here.
}. 
The symmetrization process can be done
in a way that is always beneficial
for Eve, and therefore, 
any attack, no matter how good it is, is 
no better than its optimal symmetrization.
Thus, W.L.G., it is sufficient to prove security against
all symmetric attack.
In order to intuitively understand the design of these 
symmetric attacks (starting from any attack),
we note that for the original attack,
applying the attack ($U$) to a state $i \xor m$ gives 
$U \zst{E}\ketb{i\xor m} = 
\sum_{j'} \ketb{E'_{i\xor m,j'}} \ketb{j'}=
\sum_{j} \ketb{E'_{i\xor m,j \xor m}} \ketb{j \xor m}$
with $j= j' \xor m$.
The symmetrization is achieved by Eve in practice in several steps. 

We first present the symmetrization as if Eve knows the bases $b$:
When the additional ancilla state is $\mst{M}$ 
she applies her original attack after ``shifting'' $i$ by $m$
(namely XORing $i$ with $m$, via bitwise Controlled-NOT gates):
$U \zst{E}\ketb{i\xor m} \mst{M}  = 
\sum_{j} \ketb{E'_{i\xor m,j \xor m}} \ketb{j \xor m} \mst{M}$.
Now we can see that averaging the original attack over
$i$ is equivalent to averaging the shifted attack over all values
$m$.
The averaging over $m$ is easily obtained 
due to starting with a quantum state which is 
an equal superposition of all values of $m$, 
$\zstx{M} \equiv (1/2^n)\sum_m\mst{M}$.
Then Eve could always measure $m$ and continue with the same POVMs
(where each POVM is a function of the values of $i_T,b,\ldots$) 
as in the original attack
obtaining her original asymmetric
attack up to a shift of all values by XORing them with $m$.
Let us refer to this attack
as the ``trivial symmetric attack'' 
$\{U^{\sym},\cal{E}^{\rm trivial}\}$. 
We can also define a slightly stronger and more general  
attack in which Eve measures $m$ on her additional probe, 
but continues with any POVM 
she finds appropriate. 
We call this attack the ``simple symmetrized attack''. 
Obviously, for a given $U$ (and its modified attack, $U^\sym$), 
the optimal {\em simple} symmetrical attack  
is better than the {\em trivial} symmetric attack, 
because potentially more informative POVMs are chosen.
The most general symmetric attack 
$\{U^{\sym},\cal{E}^{\sym}\}$ 
generalizes this 
{\em simple} symmetric attack, as Eve can choose any measurement 
(rather than measuring $m$ first). Clearly, the optimal
symmetric attack (for a given $U$) 
is therefore at least as informative as the 
{\em trivial} and the {\em simple} symmetric attacks.

Note that in the {\em trivial} symmetric attack, 
when Eve's second probe is measured yielding an outcome
$m$, we get back the original attack, up to a shift by $m$.
If the error rate in the original 
attack $U$ is averaged over all $i$ and 
the error rate in the new attack is averaged  
over all $m$, 
the resulting average 
error rate is the same. 
Thus, the {\em trivial} symmetric attack 
induces the same error-rate, and gives
Eve the same information as the original attack.
However, as we just explained,
in the symmetrized attack 
$\{U^{\sym},\cal{E}^{\sym}\}$ 
Eve can also use  
the state $|m\rangle$ in other ways than just measuring $m$.
This modification cannot change the error-rate
due to causality  (Eve's measurement can be done
after Alice and Bob completed their protocol).
On the other hand, the optimal symmetrization 
(optimal POVM, $\cal{E}^\sym$, for each value of
$i_T,b,\ldots$) will  
be at least as good as the
trivial one, meaning that for any value of $i_T,b,\ldots$, it would 
not decrease Eve's information, while it could increase it. 
As a result of these two intuitive observations
dealing with symmetrized attacks is sufficient,
and any other attack cannot be better for Eve.
We render these observations formally 
sound later on in Subsection~\ref{Subsec:optim-symm-att}, 
but we first must deal with the general case in which 
the basis $b$ is not known to Eve by the time she performs
the symmetrization.

The fact that Alice's state is also defined by a basis $b$ 
which is unknown to Eve makes the required 
symmetrization slightly more complex, because we would 
like to obtain 
$i \xor m$ no matter what the basis is.
This is done as follows:
We define the new attack in terms of a previously fixed basis;
we will choose
the computational basis, i.e. the basis $\{ \ketz{i} \}$ 
(for $b = 0$, the zero string).
For each qubit sent by Alice, Eve attaches a new ancillary bit; 
her new
ancilla (Eve's second probe, M)
is thus a $2n$ qubit register, 
whose basis states are called $\mst{M}$. She then
applies independently to each pair 
of qubits (Alice's qubit plus the attached qubit from the probe M)  
the unitary transform satisfying the
equalities 
$S\ketz{0}\ket{0} = \ketz{0}\ket{0}$,
$S\ketz{1}\ket{0} = \ketz{1}\ket{0}$,
$S\ketz{0}\ket{1} = \ketz{1}\ket{1}$ and
$S\ketz{1}\ket{1} = -\ketz{0}\ket{1}$ 
(if the computational basis is $\ket{0_z}, \ket{1_z}$ then 
this corresponds to performing
a controlled $\sigma_x \sigma_z$ transformation on each 
of Alice's qubits using the corresponding
ancillary bit
as control bit). 
If we evaluate $S$ on basis vectors of the alternate basis 
$\keto{0} \equiv \frac{1}{\sqrt{2}}(\ket{0}+\ket{1})$ and
$\keto{1} \equiv \frac{1}{\sqrt{2}}(\ket{0}-\ket{1})$, we get immediately
$S\ket{0}_1\ket{0} = \ket{0}_1 \ket{0}$,
$S\ket{1}_1\ket{0} = \ket{1}_1 \ket{0}$,
$S\ket{0}_1\ket{1} = -\ket{1}_1 \ket{1}$ and
$S\ket{1}_1\ket{1} = \ket{0}_1 \ket{1}$; as a consequence, for each 
such pair of qubits, we can 
summarize the effect of $S$ on basis states by the equality 
(where $i$, $m$ and $b$ are 0 or 1)
\begin{equation*}
S\ketb{i}\ket{m} = (-1)^{(i \oplus b) m} 
\ketb{i \oplus m}\ket{m}
\ .
\end{equation*}
On $2n$ such pairs of qubits, the exponents simply add up and, 
for any 
string $i$, $m$ and $b$ of $2n$ bits we get
\begin{eqnarray}
S_{\mathrm AM} \ketb{i}\mst{M} &=& 
(-1)^{(i \oplus b) \cdot m} \ketb{i \oplus m}\mst{M}\\
S_{\mathrm AM}^\dagger \ketb{i}\mst{M} &=& 
(-1)^{(i \oplus b \oplus m) \cdot m} \ketb{i \oplus m}\mst{M}
\end{eqnarray}
where the subscript for $S$ means it acts on Alice's qubits
(A) and the second probe (M),
where the second equation is deduced from the first by using 
the fact that 
$S^\dagger S = \mathbf{1}$, 
and with $S$ being a $2^{4n} \times 2^{4n}$ matrix.

The symmetrized attack is therefore 
defined by the initial state of the additional
probe
$\zstx{M} \equiv (1/2^n)\sum_m \mst{M}$,
and by the unitary transform
\begin{equation}
\label{usymmdef}
 U^{\sym} \equiv (\unit{E} \otimes S_{\mathrm AM}^\dagger) 
(U_{\mathrm EA} \otimes \unit{M})(\unit{E} \otimes S_{\mathrm AM}) 
\end{equation}
where $U_{\mathrm EA}$ 
is Eve's original attack on Alice's qubits (A) 
and Eve's first probe (E), 
$S$ is applied onto Alice's qubits and Eve's second probe,
and $\unit{E}$ and $\unit{M}$ are the identity on 
Eve's first and second
probe space respectively. 
This completes the definition of the symmetrized attack.

\subsection{Some basic properties of symmetric attacks}
\label{Subsec:properties-symm-att}

\subsubsection{The ``Basic Lemma of Symmetrization'':}
\label{Subsubsec:basic-Lemma-of-symm}

For any attack $U$, and for any basis $b$, 
we write $U^\sym$ slightly differently now by defining 
$\ketb{E^{\sym\, \prime}_{i,j}}$ via
\begin{eqnarray*}
  U^\sym \zst{Eve}\ketb{i} =
U^\sym \zst{E} \zstx{M} \ketb{i} 
\equiv \sum_j \ketb{E^{\sym\, \prime}_{i,j}}\ketb{j}
\end{eqnarray*}
where both probes $\zst{E}$
and $\zstx{M}$ 
have been put together (adjacent to each other)
on the left side, 
to clarify the definition
of these 
$\ketb{E^{\sym\, \prime}_{i,j}}$.

Given any attack $U$, with its $|E'_{i,j}\rangle_b$ the symmetrization
leads to 
these $E^{\sym\, \prime}_{i,j}$s 
that can now be described via the original $E'_{i,j}$s
as follows:
\begin{lemm}
\label{lemma-sym-def}
For any basis string $b$ 
\begin{equation}\label{eq-lemma-sym-def}
\ketb{E^{\sym\, \prime}_{i,j}} = 2^{-n} \sum_m 
   (-1)^{(i \oplus j) \cdot m} \ketb{E'_{i\oplus m, j\oplus m}} \ket{m}
\end{equation} 
\end{lemm}
We refer to this Lemma as {\em the Basic Lemma of Symmetrization}.
\begin{proof}
In order to calculate smoothly, 
we write (again) $\zst{E} \ketb{i}\zstx{M}$ (instead
of $\zst{E}\zstx{M}\ketb{i}$) in the order
the Hilbert spaces 
appear in equation (\ref{usymmdef}) defining $U^\sym$:
\begin{eqnarray*}
\lefteqn{U^\sym \zst{E} \ketb{i}\zst{M} =\ \ }\\
 & & 
2^{-n}(\unit{E} \otimes S)^\dagger 
(U \otimes \unit{M})(\unit{E} 
\otimes S)
\left[\sum_m \zst{E} \ketb{i}\ket{m}\right] \\
&=& 
2^{-n}(\unit{E} \otimes S)^\dagger (U \otimes \unit{M})
\left[\sum_m (-1)^{(i \oplus b) \cdot m} \zst{E} \ketb{i \oplus m}\ket{m}\right]\\
&=& 2^{-n} (\unit{E} \otimes S)^\dagger\left[\sum_{m,j} 
(-1)^{(i \oplus b) \cdot m}
\ketb{E'_{i\oplus m,j \xor m}}\ketb{j \xor m}\ket{m}\right]\\
&=& 2^{-n} \sum_{m, j} (-1)^{(i\oplus b) \cdot m} 
(-1)^{(j \xor m \oplus b \oplus m) 
\cdot m} \ketb{E'_{i \oplus m, j\xor m}}
\ketb{j}\ket{m}\\
&=& 2^{-n} \sum_j\sum_m (-1)^{(i \oplus j) \cdot m} 
\ketb{E'_{i \oplus m,j \oplus m}} \ketb{j}\ket{m}
\end{eqnarray*}
which proves the lemma. \qed
\end{proof}
The Lemma tells us (intuitively)
that Eve gets a similar replacement of $E'_{i,j}$ by 
$E'_{i \xor m, j \xor m}$  whether 
she symmetrizes with respect to the
computational basis or with respect to any other basis. 
This means that symmetrization with respect
to the output bits 0 or 1 results also in 
some form of symmetry 
with respect to the bases.

\subsubsection{Symmetrization and the error-rate:}
\label{Subsubsec:import-symm-att}

For any attack
(symmetric or not) the probability that Bob measures the
string $j$ in basis $b$ if Alice sent $i$ is given by
$p(j \mid i, b) = \braket{E'_{i,j}}{E'_{i,j}}_b$. 
In particular, for symmetric attacks  
$p^{\sym}(j \mid i, b) = 
\braket{E^{\sym\, \prime}_{i,j}}{E^{\sym \, \prime}_{i,j}}_b$. 
As a consequence of the Basic  
Lemma of Symmetrization (Lemma~\ref{lemma-sym-def}) we can now 
establish a link between $p^\sym(j \mid i, b)$,
the probability that (under the symmetrized attack) 
Bob measures $j$ 
in basis $b$ if Alice sent $i$, with $p(j \mid i, b)$, 
the corresponding
probability for the original attack.
For a given $b$ and $i$, the probability of some specific $j=i\xor c$
becomes the probability of $c$.  Thus we can also conclude 
a link between  
$p^\sym(c \mid i, b)$ 
and $p(c \mid i, b)$. 
The two main conclusions of the forthcomming lemma are that
(a) --- the probability (in the symmetrized attack)  
$p^\sym(c \mid i,b)$ for a given $i$, is actually independent 
of $i$, as it is equal to $p(c \mid b)$, and 
(b) ---  
the probability (in the symmetrized attack)  
$p^\sym(c \mid b)$  
is equal to the probability in the original attack,
as it is equal to  $p(c \mid b)$.   
\begin{lemm}
\label{lemma-error-rate-sym}
For any $i$ chosen by Alice and for any $j=i \xor c$
\begin{align}      
 \ p^\sym(j \mid i,b) \equiv    
p^\sym(i \xor c \mid i,b) 
&=  2^{-2n} \sum_{i'} p(i' \xor c \mid i' ,b)
\ ,   \\
 \ p^\sym(c \mid i,b) =    
  p^\sym(i \xor c \mid i,b) =    
p^\sym(j \mid i,b) 
&= p(c \mid b)
\ ,  
\\
p^\sym(c \mid b) 
&= p(c \mid b)
\end{align}
\end{lemm}
\begin{proof}
Using the
fact that the states $\ket{m}$ are orthonormal, we get
\begin{align}
p^\sym(j \mid i,b) &=
\braket{E^{\sym\ \prime}_{i,j}}{E^{\sym\ \prime}_{i,j}}_b \nonumber \\
&=
 2^{-2n} \sum_m 
\braket{E'_{i \oplus m,j\oplus m}}{E'_{i\oplus m ,j\oplus m}}_b
& \text{by Eq. (\ref{eq-lemma-sym-def})} \nonumber \\
&=
2^{-2n} \sum_m p(j\oplus m \mid i \oplus m ,b)\nonumber 
\ ,
\end{align}
By assigning $i' = i \xor m$ this gives 
$p^\sym(j \mid i,b) =  
2^{-2n} \sum_{i'} p(j \xor i' \xor i  \mid i' ,b)
$.
With $c = i \xor j$ we finally get 
$p^\sym(i \xor c \mid i,b) =  
2^{-2n} \sum_{i'} p(i' \xor c  \mid i' ,b)
$. 
This completes the first part of the Lemma.

By definition, the averaging over all $i'$ means that 
$2^{-2n} \sum_{i'} p(i' \xor c \mid i' ,b) \equiv p(c \mid b)$, 
so we get
$p^\sym(i \xor c \mid i,b) = p(c \mid b)$.
We conclude that 
$p^\sym(i \xor c \mid i,b)$ is actually independent of $i$,
namely,  
$p^\sym(j \mid i,b) = p(c \mid b)$. 
For a given $b$ and $i$, $p^\sym(j \mid i, b) 
= p^\sym(i \xor c \mid i,b)
= p^\sym(c \mid i,b)$.
This completes the proof of the second part of the 
lemma.

We now start with
$p^\sym(i \xor c \mid i,b) 
= p(c \mid b)$. Then, 
averaging 
$p^\sym(i \xor c \mid i,b)$ 
over all $i$ means that 
$2^{-2n} \sum_{i} p^\sym(i \xor c \mid i ,b) \equiv p^\sym(c \mid b)$.
However, the summation is over equal terms [$p(c \mid b)$], 
so we finally get
$ p^\sym(c \mid b) =
 p^\sym(i \xor c \mid i ,b) $, proving the last part of the Lemma.
\qed
\end{proof}

\subsection{Symmetric attacks are optimal for the eavesdropper}
\label{Subsec:optim-symm-att}

We now show that for any attack 
$\{U,\cal{E}\}$, 
the attack 
$\{U^{\sym},\cal{E}^{\rm trivial}\}$ 
leaves the same average 
error 
rate and also provides the same 
information to Eve as the original attack. 
The optimal symmetric attack (for a given $U$),
in which the optimization is over all the possible measurements
$\cal{E}^{\sym}$ 
leaves the same average 
error 
rate and provides
information to Eve that is equal or larger than that of
the original attack $U$. 
These results imply (see Lemma~\ref{sec5-lemm2})
that if the security criterion is satisfied for
all symmetric attacks, then it is satisfied for {\em all attacks}.
Let us recall that 
due to causality Bob's outcome will be the
same whatever measurement Eve performs. 
Since symmetrization in one basis yields symmetrization at any
basis, we may assume (W.L.G.) that Eve performed her
symmetrization with respect to the basis used by Alice and Bob. 
In that context,
if Eve uses the {\em trivial} symmetrized attack, 
and measures $\ket{m}$ in the
standard basis, this is simply a replacement of $i$ by 
$i \oplus m$ and $j$ by $j \oplus m$ with
respect to the original attack. 
Continuing by a POVM as in the original attack, 
now yields the same information
as the original attack, while clearly Eve could do better,
as earlier explained.

In the following subsections we make the 
above intuition mathematically solid. [Recall that
the string $s$ 
(where a position equal to 1 corresponds to an information
bit in $i$ 
whilst a 0 indicates a test bit) determines two substrings of $i$,
namely $i_I$ (information bits) and $i_T$ (test bits);
after $s$ is published by Alice we may identify
$\ketb{i}$ with $\ketb{i_T}\ketb{i_I} = \ketb{i_T} \otimes \ketb{i_I}$ 
(this isomorphism
depends on $s$, and is just a permutation of bits); 
note that the same modification
applies to $\ketb{j}$.] 

\subsubsection{Symmetrization does not affect the average error-rate:}
\label{Subsubsec:invariance_prop}

As a corollary 
of Lemma~\ref{lemma-error-rate-sym},
when $s$ is known, we get
\begin{corollary}\label{coroll-Psym-CI-CT}
\begin{align}
P^\sym[ c_I,  c_T  \mid b,s ] &= 
 P[ c_I, c_T \mid b, s] \label{psimcictvspcict} \\
P^\sym[ c_T  \mid b,s ] &= 
 P[ c_T \mid b, s] \ . \label{psimctvspct} 
\end{align}
\end{corollary}
The first equation is a slight modification of the 
third part of Lemma~\ref{lemma-error-rate-sym} 
(due to $s$ being published), 
and the second equation is obtained from the first 
by summing over all $c_I$. 

These results
prove that the average error-rate is not changed
when an attack $U$ is replaced by any symmetric attack $U^\sym$.

\subsubsection{Eve's information is not decreased by symmetrization:}

Let  
$\mrv{E}^\sym$ be the random variable whose values
$e$ are the output of Eve's measurement $\cal{E}^{\sym}$,
and note that the measurement is fixed at the end of the protocol,
hence depends on the value of $\{i_T, c_T, b, s, \xi\}$. 
For any particular attack $U$ and particular 
value $\{i_T, c_T, b, s, \xi\}$, the {\em maximal value} of
$I(\mrv{A}; \mrv{E}^\sym \mid i_T,\mrv{C}_T=c_T, b, s, \xi)$ 
corresponding to Eve's symmetrized attack and 
{\em optimal measurement} is larger than or equal to that obtained 
if she restricts herself to performing the {\em trivial} 
symmetric attack 
(namely, to measuring the $\ket{m}$ probe in the standard 
basis, and repeat the POVM of the original attack).  

Let us denote
$(\mrv{E}', \mrv{M})$ the (multivariate) random variable where for
each particular value of $m$, $\mrv{E}'$ are the random outputs of
the {\em trivial} symmetric attack.
Then, we have by the very definition of the optimal measurement
that
\[
\max_{\{ \cal{E}^{\sym} \}}
I(\mrv{A}; \mrv{E}^\sym \mid i_T,\mrv{C}_T=c_T, b, s, \xi) 
 \geq
I(\mrv{A}; \mrv{E}', \mrv{M} \mid i_T,\mrv{C}_T=c_T, b, s, \xi)
\ , \]
where $\{ \cal{E}^\sym \}$ does not stand for one POVM but for a set
of POVMs, one for each value of $i_T,c_T,b,s,\xi$.
We would like to bound
\begin{equation*}
I(\mrv{A}; \mrv{E} \mid \mrv{I}_T,\mrv{C}_T=c_T,b,s, \mrv{\Xi})
  = 
   \sum_{i_T,\xi} P[i_T, \xi \mid c_T,b,s]
  I(\mrv{A}; \mrv{E} \mid i_T, \mrv{C}_T=c_T,b,s, \xi) 
\ .
\nonumber
\end{equation*}

We must note the important fact that the POVM is only  
fixed at the end of the protocol, hence a different POVM 
$\cal{E}$ 
is chosen for each fixed value of $i_T,\xi$ (as the other parameters 
are fixed here).
The same is true for the {\em trivial} symmetrized attack 
\begin{eqnarray*}
\lefteqn{I(\mrv{A}; \mrv{E}', \mrv{M} \mid \mrv{I}_T,\mrv{C}_T=c_T, b, s,
\mrv{\Xi})}\\
& &  = 
   \sum_{i_T,\xi} P[i_T, \xi \mid c_T,b,s]
  I(\mrv{A}; \mrv{E}', \mrv{M} \mid i_T, \mrv{C}_T=c_T,b,s, \xi) 
\ ,
\end{eqnarray*}
and the same is true for the optimal symmetrized attack 
(for a given $U$)
\begin{eqnarray} \label{MAX-info}
\lefteqn{\max 
I(\mrv{A}; \mrv{E}^\sym \mid \mrv{I}_T,\mrv{C}_T=c_T,b,s,
\mrv{\Xi}) } \nonumber
\\ 
& &\equiv
\sum_{i_T,\xi} P[i_T, \xi \mid c_T,b,s]
\ \max_{\{ \cal{E}^{\sym} \}}
I(\mrv{A}; \mrv{E}^{\sym} \mid i_T, \mrv{C}_T=c_T,b,s, \xi) 
\ .
\end{eqnarray}
With that definition
we are promised that symmetrization is optimal for each 
particular value of $\{i_T,c_T,b,s,\xi\}$ and the resulting information 
is optimal also after summing over $i_T,\xi$:
\[
\max I(\mrv{A}; \mrv{E}^\sym \mid \mrv{I}_T,\mrv{C}_T=c_T,b,s, \mrv{\Xi}) 
 \geq
I(\mrv{A}; \mrv{E}', \mrv{M} \mid \mrv{I}_T,\mrv{C}_T=c_T, b, s, \mrv{\Xi})
\]

Now we are ready to present the main result of this subsection.
An optimal symmetrization of $U$ will not decrease the information 
accessible to Eve in the following sense:
\begin{lemm}
\label{lemma-symm}
For any fixed $U,c_T,b,s$, 
\begin{equation}\label{eq-lemma-symm}
 \max I(\mrv{A}; \mrv{E}^\sym \mid \mrv{I}_T,\mrv{C}_T=c_T,b,s, \mrv{\Xi}) \geq
  I(\mrv{A}; \mrv{E} \mid \mrv{I}_T,\mrv{C}_T=c_T,b,s, \mrv{\Xi})
\end{equation}
\end{lemm}

\begin{proof}
For any given $U$, the optimal symmetric attack is at least as good
as the {\em trivial} symmetric attack for each value
of $c_T,b,s,i_T,\xi$, and therefore also after summing over
$i_T,\xi$. 

Proving formally that the 
{\em trivial} symmetric attack is as good as the original attack
is less trivial\footnote{
Still, it is somewhat similar to the argument given
when we analyzed the case in
which Eve knows the bases.}. Actually, for simplicity,
we only prove the relevant direction,
namely, that the {\em trivial} symmetric attack is at least as good as
the original attack: 
\begin{equation}\label{trivial-better-origin}
I(\mrv{A}; \mrv{E}', \mrv{M} \mid \mrv{I}_T,\mrv{C}_T=c_T, b, s, \mrv{\Xi})
\geq  I(\mrv{A}; \mrv{E} \mid \mrv{I}_T,\mrv{C}_T=c_T,b,s, \mrv{\Xi})
\end{equation}
For the details of that proof, see Appendix~\ref{APP:info-symm-not-decr}.
\quad  \qed
\end{proof}

The above result means that we can use a bound on Eve's average
information in the case of 
a symmetrized attack to apply to the unsymmetrized case.

\subsection{Eve-Bob's state after the basis 
and the test bits are known}
\label{Subsec:eves-bobs-state-1}

When the strings $b$ and $s$ 
are given to Bob (and to Eve) then Eve-Bob's state 
(Eq.\ref{Eve-Bob-state}) ought to be modified. The sifted keys
$\ketb{i}$ and $\ketb{j}$, the resulting error syndrome $c=i \xor j$, 
Eve's attack $U$, and Eve's unnormalized states $E'_{i,j}$
are now expressed differently, so that the 
test bits and information bits are written separately.
Equation (\ref{eq-Eprimeij}) can thus be rewritten as
\begin{equation}
\label{Eve_Full_Attack}
U \zst{E}\ketb{i_T}\ketb{i_I} = 
\sum_j \ketb{E'_{i_T,i_I,j_T,j_I}} \ketb{j_T}\ketb{j_I}
\end{equation}
where the right-hand side corresponds to Eve-Bob's state  
($\ket{\psi'_{i}} $) for a 
given $i=i_T i_I$, and where
\begin{equation}
\label{Eve-E_iTiIjTjI}
\ketb{E'_{i_T,i_I,j_T,j_I}}= 
\brab{j_T}\brab{j_I}U \zst{E}\ketb{i_T}\ketb{i_I} 
\ .
\end{equation}
The probability that Bob measures $\ketb{j_T}\ketb{j_I}$
is 
\begin{equation} \label{P-jTjI-given-ibs}
p(j_T, j_I|i_T,i_I,b,s) = 
\braket{E'_{i_T,i_I,j_T,j_I}}{E'_{i_T,i_I,j_T,j_I}}_b 
\ .
\end{equation}

Once $i_T$ is also given to Eve and Bob,
it is considered as a fixed parameter instead of a variable 
in the equation above.
When $j_T$ is measured, 
the right-hand states 
$ \sum_j \ketb{E'_{i_T,i_I,j_T,j_I}} \ketb{j_T}\ketb{j_I} $ 
are projected onto the particular $j_T$
obtained by the measurement on the test bits, 
and $2^n$ basis states are left in the summation,
corresponding to the $2^n$ possible values of the $n$ 
information qubits 
in Bob's hands.
Formally, the projection is described via
$\langle j_T | \psi'_i\rangle = \sum_{j_I}
\ketb{E'_{i_T,i_I,j_T,j_I}} \ketb{j_I} $. 
The projection should now be 
followed by a normalization of the state,
thus modifying Eve-Bob's state to become
\begin{equation}
\ket{\psi_{i_I}} = \sum_{j_I} 
\frac{1}{\sqrt{p(j_T|i_T,i_I,b,s)}}\ketb{E'_{i_T,i_I,j_T,j_I}} 
\ketb{j_I} \ .
\end{equation}
With $p(j_T|i_T,i_I,b,s) = \sum_{j_I} p(j_T, j_I|i_T,i_I,b,s)$ 
and using Eq.(\ref{P-jTjI-given-ibs}) we get
that the normalization factor (due to the projection on $j_T$) is
the square root of 
\begin{equation}\label{eq-pj-non-symm}
p(j_T|i_T,i_I,b,s) = 
\sum_{j_I} \braket{E'_{i_T,i_I,j_T,j_I}}{E'_{i_T,i_I,j_T,j_I}}_b
\ .
\end{equation}

Let us now define\footnote{
The expression $\kete{E_{i_I,j_I}}{b,s}$ is also a function 
of the parameters $i_T$ and $j_T$ (which are known to Eve by now),
but, writing the expression as 
$\kete{E_{i_I,j_I}}{b,s,i_T,j_T}$ looks cumbersome; 
therefore, 
for convenience, we did not write them in the expression,
while we keep $b,s$ to remind us that the bases and the test 
are known.
} 
\begin{equation}\label{E_projected}
\kete{E_{i_I,j_I}}{b,s} \equiv
\frac{1}{\sqrt{p(j_T|i_T,i_I,b,s)}}\ketb{E'_{i_T,i_I,j_T,j_I}} \ ,
\end{equation}
so that the resulting Eve-Bob's state can be
written more economically in the form
\begin{equation}
\ket{\psi_{i_I}} = \sum_{j_I} \kete{E_{i_I,j_I}}{b,s} \ketb{j_I}
\ .
\end{equation}

{}From Eqs.~(\ref{P-jTjI-given-ibs}, 
~\ref{E_projected}) and 
the conditional probability formula [$p(ab) / p(a) = p(b|a)$] we get
\begin{equation}\label{P-jI-given-jTibs} 
\braket{E_{i_I,j_I}}{ E_{i_I,j_I}}_{b,s}=
p(j_I \ | \ i_I,i_T,j_T,b,s) 
\ ;
\end{equation}
with $c=i\xor j$ this gives 
$\braket{E_{i_I,i_I\xor c_I}}{ E_{i_I,i_I\xor c_I}}_{b,s} 
= p(i_I\xor c_I \ | \ i_I,i_T,j_T,b,s) = 
p(c_I \ | \ i_I,i_T,j_T,b,s)$.

\subsection{Symmetrization --- 
its impact on the test and information bits:}
\label{Subsec:More-properties-of-symm-att}  

We first prove that for symmetrized attacks various expressions become independent of $i_I$:
\begin{lemm}\label{lemm-indep-it}
\begin{equation}\label{eq-indep-it}
p^\sym(j_T \mid i_T, i_I, b, s) = p^\sym(j_T \mid i_T, b, s)
\ .
\end{equation}
\end{lemm}
\proof
As an immediate corollary of Lemma~\ref{lemma-error-rate-sym}
(that says that  $p^\sym(c \mid i,b) = p(c \mid b)$)
when $s$ is known,
we get
\[
p^\sym(c_T,c_I \mid i_T, i_I, b, s) = p[c_T, c_I \mid b,s]
\ .
\]
Recalling that $c_T = i_T \oplus j_T$ and $c_I = i_I \oplus j_I$, 
this implies that
for any $m'_I$
\[
 p^\sym(j_T, j_I \oplus m'_I \mid i_T,i_I \oplus m'_I, b,s) =
 p^\sym(j_T, j_I \mid i_T, i_I, b,s)
\ .
\]
If we sum both sides of this equality over $j_I$ we get
$
p^\sym(j_T \mid i_T, i_I \oplus m_I, b, s) = p^\sym(j_T \mid i_T, i_I, b, s)
$
which means that the probability is independent of $i_I$,
\[ p^\sym(j_T \mid i_T, i_I, b, s) = p^\sym(j_T \mid i_T, b, s)
\ . \]
\qed  

As a corollary
of the above Lemma, 
notice that for symmetric attacks,
\begin{corollary}\label{coroll:i_I}
\begin{equation}
p^\sym(i_I|i_T,j_T,b,s)=1/2^n.
\label{iI-indep}
\end{equation} 
\end{corollary}
Indeed, using the Bayes rule (on $\{j_T; i_I\}$)
\[
p^\sym(i_I \mid i_T,j_T,b,s) = 
\frac{p^\sym(j_T \mid i_I, i_T,b,s)}{p^\sym(j_T \mid i_T, b,s)}
  p^\sym(i_I \mid i_T, b,s) 
 =  \frac{1}{2^n}
\]
where the last equality results from 
Eq.~(\ref{eq-indep-it}) and the fact that all bits of $i$, $b$ and
$s$ are chosen independently  
[so $p^\sym(i_I \mid i_T,b,s) = \frac{1}{2^n}$]. 

Another important consequence of Lemma~\ref{lemm-indep-it} 
is:
\begin{lemm}\label{lemm:indep-of-iI}
For the information bits:
\begin{enumerate}
\item
$\langle E_{i_I,i_I\xor c_I}^{\sym} | 
              E_{i_I\xor k_I,i_I\xor c_I\xor k_I}^{\sym} \rangle_{b,s}$ 
is independent of $i_I$.
\item
$\sum_j\langle E_{i_I,j_I}^{\sym} | 
              E_{i_I\xor k_I,j_I\xor k_I}^{\sym} \rangle_{b,s}$ 
is independent of $i_I$.
\end{enumerate}
\end{lemm}
The proof is given in Appendix~\ref{App:Eij-indep-iI}.

The next step is to show that for symmetrized attacks
various expressions are independent also of $b_I$.
We proved in Lemma~\ref{lemm-indep-it}
that the normalizing factor for fixed 
$i_T$, $j_T$, $b$ and $s$ is the same for all the indices
$i_I$. 
In addition, 
that normalizing factor does not depend on $b_I$ either:
\begin{lemm} \label{lemm:pjtit-indep-sym}
\begin{equation}\label{eq-pjtit-indep-sym}
p^\sym(j_T \mid i_T, b, s) \equiv 
p^\sym(j_T \mid i_T, b_I, b_T,s) = p^\sym(j_T \mid i_T, b_T, s)
\ .
\end{equation}
\end{lemm}
\proof
In fact, Eq.(\ref{eq-pjtit-indep-sym}) is true 
for any attack (symmetrized or not):
\begin{equation}\label{eq-pjtit-indep}
p(j_T \mid i_T, b, s) \equiv 
p(j_T \mid i_T, b_I, b_T,s) = p(j_T \mid i_T, b_T, s)
\ .
\end{equation}
Intuitively, the fact that $i_I$ is {\em not} 
a given parameter 
actually means that we average over it (as 
$p(a) = \sum_b p(a,b) = \sum_b p(b) p(a|b)$).
Once we average over it, the relevant quantum bits 
are traced out, causing independence of $b_I$ as well.  
Thus, in general, $j$ of one subset 
(such as $j_T$) is independent of $b$ of another 
subset (such as $b_I$).
This is formally proven 
in Appendix~\ref{proof-eq-pjtit-indep}.  
Thus follows 
$  p^\sym(j_T \mid i_T, b, s) = p^\sym(j_T \mid i_T, b_T, s) $.
\qed 

As a trivial Corollary
of Lemmas~\ref{lemm-indep-it} and~\ref{lemm:pjtit-indep-sym} 
we get the following:
\begin{corollary} \label{coroll:jT-given-iTbTs}
For symmetrized attacks,
the probability of $j_T$ satisfies
\begin{equation} \label{eq:jT-given-iTbTs}
p^\sym(j_T \mid i_T, i_I, b, s)  
 = p^\sym(j_T \mid i_T, b_T, s)
\ ,
\end{equation}
and therefore, Eq.(\ref{E_projected}) is simplified to
\begin{equation}\label{E_proj_sym}
\kete{E^\sym_{i_I,j_I}}{b,s}  = 
\frac{1}{\sqrt{p^\sym(j_T|i_T,b_T,s)}}
\kete{E^{\sym\, \prime}_{i_T,i_I,j_T,j_I}}{b}
\ .
\end{equation}
\end{corollary}

\clearpage

\section{Information vs. Disturbance}
\label{Sec:infdis}

In this section we analyze the information bits alone (for a given
symmetric attack $U^\sym$, a given input $i_T$ and outcome $j_T$ 
on the test bits, and given bases $b$ and choice of test bits
$s$). When no ambiguity arises, the indices $b$ and $s$ will
be dropped; $\ket{i}$ will denote $\ketb{i}$,
$\ket{i_I}$ will denote $\kete{i_I}{b_I}$ and $\ket{E^\sym_{i_I,j_I}}_{b,s}$ will be denoted $\ket{E_{i_I,j_I}}$.
Our result here applies for any $U^\sym$, hence in
particular {\em for the optimal one}.
The optimization over Eve's measurement is avoided by using the fact 
that trace norm of the difference of two density matrices provides 
an upper bound on the accessible information one could obtain
{\em via any measurement} when having the
two density matrices as the possible inputs.

\subsection{Eve's state}
\label{Subsec:Eve-states}

When Alice sends a state $\ket{i_I} \equiv \kete{i_I}{b_I}$ for the information bits
(where $b_I$ is the string actually used by her and Bob to fix the bases
on information bits),
the state of Eve and Bob together,
$|\psi_{i_I}\rangle = 
        \sum_{j_I}
           \ket{E_{i_I,j_I}}\ket{j_I}
$ 
is fully determined by Eve's
attack and by the data regarding the test bits.
Eve's state in that case  is fully
determined by tracing-out Bob's subsystem $\ket{j_I}$
from 
Eve-Bob's state, and it is
\begin{equation}\label{Eve-states-matrix}
\rho^{i_I}=  \sum_{j_I} \ket{E_{i_I,j_I}}\bra{E_{i_I,j_I}} \ ,
\end{equation}
calculated given $i_T$ and $j_T$.
This state in Eve's hands is a mixed state.  

\subsection{Purification and a related basis}
\label{Subsec:purific}

We can ``purify'' the state while giving more information to Eve by
assuming she keeps the state
\begin{equation}\label{Eve-states-purified}
 \ket{\phi_{i_I}} = \sum_{j_I}\ket{E_{i_I,j_I}}\ket{i_I\xor j_I} 
\end{equation}
where we introduce another subsystem for the 
``purification''. Notice that the indices of $\phi$ and of $E$ are
always information bits ($n$-bit strings). As a consequence, we could
as well have written without ambiguity 
$\ket{\phi_i} = \sum_j \ket{E_{i,j}} \ket{i \oplus j}$
where the sum is taken over all $n$-bit strings $j$ that can serve
as index in $\ket{E_{i,j}}$. We will do this when expressions do not
involve test bits.
The term purification means different things in different papers, thus we
explain it a bit more: A mixed state can also
be obtained from a pure state in an enlarged system (the original system plus
an ancilla), once the ancilla is traced out; the pure state of the
enlarged system (or its density matrix)
is called a purification of the mixed state.
In a more general case, the state in the enlarged system is not 
necessarily
pure, and then we refer to it as a ``lift-up''~\cite{BBBGM} 
of the state of the
original system.

The resulting purified state (i.e., any purification or any lift-up of
Eve's states, 
for instance, the purification $\rho^{i}=\ket{\phi_{i}}\bra{\phi_{i}}$),
is at least as informative to Eve as
$\rho^{i_I}$ (of Eq.~\ref{Eve-states-matrix}) is.  
This is because the density matrix $\rho^{i_I}$ is exactly the
same as Eve's state 
would be if Eve ignored the $i_I\xor j_I$ register of $\phi$.
Thus, any information Eve can obtain from her mixed state is bounded by the
information she could get if the purified state was available to her.

Note that the overlap between these purified states satisfies
\begin{eqnarray}
\braket{\phi_{l}}{\phi_{l\xor k}}&=&\sum_{j}\sum_{j'}
        \braket{E_{l,j}}{E_{l\xor k,j'}}
        \braket{l\xor j}{l\xor k \xor j'} \nonumber \\
        &=&\sum_{j}
        \braket{E_{l,j}}{E_{l\xor k,j \xor k}}
\ ,
\end{eqnarray}
where all the indices are $n$-bit strings.

As a consequence of Lemma \ref{lemm:indep-of-iI} we immediately 
get for the information bits that $\braket{\phi_{l}}{\phi_{l\oplus k}}$ is 
independent of $l$ [meaning, independent of $i_I$, see 
Eq.~(\ref{Eve-states-purified})].
Thus, it is only a function
of $k$ (namely, $k_I$), and we can write this as
\begin{corollary}\label{coroll:Phi}
\begin{equation*}
\Phi_{k} \equiv 
\braket{\phi_{l}}{\phi_{l\oplus k}} \ .  
\nonumber
\end{equation*}
\end{corollary}

For the $2^n$ Hilbert-space spanned by the purified states $\ket{\phi_l}$ (corresponding to information bits),
we define a Fourier basis $\{\ket{\eta}\}$, and show that
it is possible to compute a bound on
Eve's information about the information bits, once the purified states are
expressed in this basis.
\begin{definition} \em
\begin{eqnarray*}
\ket{\eta_{i}}&=& \frac{1}{2^n}\sum_{l} (-1)^{i\cdot l}
\ket{\phi_{l}} \ ; \
d_{i}^2 = \braket{\eta_{i}}{\eta_{i}} \ ; \
\hat\eta_{i} = \eta_{i} / d_{i}
\end{eqnarray*}
\label{eta-def}
\end{definition}
Using the above definitions and $(1/2^n)\sum_l (-1)^{(i \xor j)
\cdot l} = \delta_{ij}$,
Eve's purified state can be rewritten
as:
\begin{equation}\label{eq:phi}
\ket{\phi_i}=\sum_l(-1)^{i\cdot l}\ket{\eta_l}
= \sum_l(-1)^{i\cdot l} d_l \ket{\hat{\eta}_l}
\ .
\end{equation}

Note that
$\braket{\eta_{i}}{\eta_{i}}
=\frac{1}{2^{2n}}\sum_{l}  \sum_k (-1)^{i\cdot k}
\braket{\phi_{l}}{\phi_{l\xor k}}
=\frac{1}{2^{n}}  \sum_k (-1)^{i\cdot k}
\Phi_{k} $.
In terms of Eve's states we can write 
\begin{equation}  
d_i^2 = \braket{\eta_{i}}{\eta_{i}}
 =\frac{1}{2^{2n}}\sum_{l}  \sum_k (-1)^{i\cdot k}
        \sum_{j}
        \braket{E_{l,j}}{E_{l\xor k,j \xor k}}_{b,s}
\ . \label{d-sqr}
\end{equation}
\begin{proposition}
\label{prop-eta-orth}
For symmetrized attacks, $\braket{\eta_j}{\eta_i}=0$ if $i\neq j$. 
\end{proposition}
\begin{proof}
Note that
$\braket{\eta_{j}}{\eta_{i}}
=\frac{1}{2^{2n}}\sum_{l}(-1)^{(i\xor j)\cdot l}\sum_k (-1)^{i\cdot k}
\braket{\phi_{l}}{\phi_{l\xor k}}$.

Since $\braket{\phi_{l}}{\phi_{l\xor k}}\equiv \Phi_k$ 
is independent of $l$,
we see that:
\begin{eqnarray*}
\braket{\eta_{j}}{\eta_{i}}
&=&\frac{1}{2^{2n}}\sum_{l}(-1)^{(i\xor j)\cdot l}\sum_k (-1)^{i\cdot k}
\Phi_k\\
&=&\frac{1}{2^n}\delta_{i,j}\sum_k (-1)^{i\cdot k}\Phi_k\\
&=&\delta_{i,j}\braket{\eta_i}{\eta_i}\quad\qed
\end{eqnarray*}
\end{proof}
The above proposition is used to prove 
Lemma \ref{sd-lemm}.

\subsection{Eve's state and probability of errors
induced on information bits}
\label{Subsec:ESPSIIB}

In this subsection we show that the probability
of any error string Eve would
have induced if the conjugate basis was used for the
information bits, is a simple
function of the $d_{i}$s (of Definition~\ref{eta-def}),
hence a function of the overlap of Eve's purified states.
For any attack ($i_T$ and $j_T$ being fixed once and for all), any $b$ and $s$, we have
\begin{equation}
P[\mrv{C}_I= c_I \ | \ i_I,i_T,j_T,b,s] =
\braket{E_{i_I,i_I\xor c_I}}{ E_{i_I,i_I\xor c_I}}_{b,s}
\label{P-and-E}
\ .
\end{equation}
See Eq.~(\ref{P-jI-given-jTibs}).

For any symmetrized attack
and any $b$ and $s$
the error distribution in the information bits is
\begin{eqnarray}
\lefteqn{P^\sym[\mrv{C}_I = c_I \mid i_T,j_T,b,s]}\nonumber\\
  & &=\sum_{i_I} P^\sym[\mrv{C}_I= c_I\ |  \ i_I,i_T,j_T, b, s]
        p^\sym(i_I \mid i_T,j_T,b,s)\nonumber\\ 
  & &=\frac{1}{2^n}\sum_{i_I}\braket{E^\sym_{i_I,i_I\xor c_I}}{E^\sym_{i_I,i_I\xor c_I}}_{b,s}    
\ ,
\label{use-symmetry}
\end{eqnarray}
namely, 
the average probability of an error syndrome $c_I$ on the information
bits (when the
test bits, basis and sequence are given).
The first equality is derived using standard probability theory 
($p(a) = \sum_b p(a|b) p(b)$)
and the second is 
due to Eq.~(\ref{iI-indep}) and Eq.~(\ref{P-and-E}).

Identity (\ref{use-symmetry}) applies for all strings $b$ and $s$ and, in
particular, for $b^0 = b \oplus s$ we get
\begin{equation}\label{use-symmetry-bis}
P^\sym[\mrv{C}_I = c_I \mid i_T, j_T, b^0, s] =
  \frac{1}{2^n}\sum_{i_I}\braket{E^\sym_{i_I,i_I\xor c_I}}{E^\sym_{i_I,i_I\xor c_I}}_{b^0,s} \ .
\end{equation}
The basis $b^0$ is a basis where the basis for the test bits 
is the same as
$b$, but the basis for each information bit is opposite.
With a little algebra, as shown in Appendix~\ref{app:d-sqrt-is},
we can express $\kete{E^\sym_{i_I,i_I\xor c_I}}{b^0,s}$ in terms
of the  $\kete{E^\sym_{i_I,i_I\xor c_I}}{b,s}$. Then, doing this for the right-hand side
of Eq. (\ref{use-symmetry-bis}) we get the right-hand side of Eq. (\ref{d-sqr}) with
$i = c_I$;
this means that we get the following
\begin{lemm} \label{d-sqr-is-p-err}
\begin{equation} \label{info-vs-dis}
P^\sym\left[\mrv{C}_I = c_I \mid  i_T,j_T,b^0,s\right] = d_{c_I}^2
\ .
\end{equation}
\end{lemm}
The proof is presented in Appendix~\ref{app:d-sqrt-is}. Note
that the $d_i$ used here are those of the symmetrized attack.

Put differently, the term $d_{c_I}^2$ defined in terms of the actual bases
used by Alice and Bob is equal to the probability of the error syndrome $c_I$ on information
bits had Alice and Bob used the conjugate bases on information bits.
As we shall soon see, these $d_i$s actually provide a measure of the information Eve could get from her purified states, therefore leading to 
a novel {\em information versus disturbance} result.

\subsection{Bounds on Eve's information -- the one-bit key case}
\label{Subsec:bounds}

In this subsection 
we much improve upon a result obtained in~\cite{BBBGM} (the result was derived for
the collective attack).
Eve's information about a particular bit of the final key
(even if all other
bits of the final key are given to her) is bounded.
We take into consideration the error-correction data that is given to
Eve, and we do it more efficiently than in~\cite{BBBGM}, hence we
obtain a much better threshold for the allowed error-rate.

Let us first discuss a one-bit final key $a$, 
defined to be the parity of
a substring of the input $i_I$.  The substring is defined using a mask
$v$, meaning that the secret key is $a = v \cdot i_I$.
(In the general case, 
the key is defined as the string 
$a = i_I P_\mathcal{PA}^T$ where $P_\mathcal{PA}$ 
is an $m \times n$
matrix; c.f. subsection \ref{subsec:ub-bb84}, item II.~7).
Bob first corrects his errors using the error correction code data, hence he
learns Alice's string $i_I$.
Eve does not know $i_I$, but she learns
the error correcting code $\mathcal{C}$
used by Alice and Bob as well as $v$ and the
parity bits
$\xi$ sent by Alice to help Bob correct the sequence he received.
All the possible inputs $i_I$ that have the correct parities
$\xi$ for the code $\mathcal{C}$ form a set denoted 
$\mathcal{C}_\xi = \{i_I \mid i_I P_\mathcal{C}^\top = \xi\}$.

When the purification of Eve's state is given by $\ket{\phi_{i}}$
the density matrix is
$\rho^{i}=\ket{\phi_{i}}\bra{\phi_{i}}$.
In order to guess the key
$a = v \cdot i$, Eve must now distinguish between
two {\em ensembles} of states: The ensemble of equally likely
states $\rho^i$ (these states are equally likely  
due to Corollary~\ref{coroll:i_I}), with 
$i_I \in \mathcal{C}_\xi$ (i.e. $i_I P_\mathcal{C}^\top = \xi$)
and key $a = i_I \cdot v =0$,
and the
ensemble of (equally likely)
states $\rho^i$ with $i_I \in \mathcal{C}_\xi$ and
key 
$a= i_I \cdot v=1$.
For $ a\in \{0, 1\}$ these ensembles are represented
by the density matrices $\rho_0 = \rho_0(v,\xi)$ and $\rho_1 = \rho_1(v,\xi)$ defined by:
\begin{eqnarray}
\rho_a(v,\xi)&=&\frac{1}{2^{n-(r+1)}}
\sum_{\stackrel{i_I P_\mathcal{C}^\top = \xi}{i_I\cdot v=a}} \rho^i\label{rho_v_b}
\end{eqnarray}
and Eve's goal is to distinguish between those two.
Note that the two density matrices $\rho_a(v,\xi)$ are the 
lift-ups of the density matrices really known to Eve, namely, matrices 
in which the sum is over the states
of Eq.~(\ref{Eve-states-matrix}) rather than a sum over their purifications.

A good measure for the distinguishability of 
$\rho_0(v,\xi)$ and $\rho_1(v,\xi)$ 
is the optimal mutual
information (known as the accessible information) that one could get
if one needs to guess the bit $a$ by performing an optimal measurement
to distinguish between the two density matrices, 
when the two are given with equal probability (of half).
This information will be called the {\em Shannon Distinguishability}
($SD = SD(\rho_0,\rho_1)$) 
to emphasize that it is a
distinguishability measure. If $v$ is the string used to define
the  one-bit key $\mrv{A}$ sent by Alice, then, due to the optimality of
$SD$, we get (for any symmetric attack) 
\begin{equation}\label{I-lt-SD}
I(\mrv{A};\mrv{E}^\sym|i_T,j_T,b,s,\xi) \leq SD(\rho_0(v,\xi),\rho_1(v,\xi))
\end{equation}
where $\mrv{E}^\sym$ is the random variable corresponding to Eve's actual 
measurement in the symmetrized attack.

Let $v_1, \ldots, v_r$ be the rows of the $r \times n$
parity check matrix $P_\mathcal{C}$ of the $(n,k,d)$ code $\mathcal{C}$
where $r = n-k$. 
The matrix $P_\mathcal{C}$ is assumed of rank $r$ and so, 
the $r$ ``parity-check strings'' $v_1, v_2, \ldots v_r$ 
(that are known to Eve) 
are linearly independent. Let $V_r$ be
the $r$-dimensional linear 
space generated
by $\{v_1, \ldots, v_r\}$. Then, $V_r = 
\{v_\mathbf{s} \mid \mathbf{s} \in \{0,1\}^r\}$
where, by definition\footnote{Note that the vector $\mathbf{s}$ 
is used now to define the possible vectors $v_\mathbf{s}$  
in the span of the parity-check strings [this is in addition to 
$s$ being used as the $2n$-bit string defining the test bits
and the information bits]; The bit $s_l$ is the $l$'th bit 
of $\mathbf{s}$.}
$v_\mathbf{s} = \sum_{l=1}^r s_l v_l$. 
For any
$v_\mathbf{s} \in V_r$, Eve knows $i_I \cdot v_\mathbf{s}$ 
because she
knows all the $\xi_l$ and $i_I \cdot v_\mathbf{s}  = \xi_\mathbf{s}$
where $\xi_\mathbf{s} = \sum_{l=1}^r s_l \xi_l$. 
As a consequence, Eve
has total knowledge of the key if $a = i_I \cdot v_\mathbf{s}$ 
for $v_\mathbf{s} \in V_r$. 
Notice that $V_r$
is
nothing but the dual code $\mathcal{C}^\perp$ of $\mathcal{C}$ which
can be viewed as the set of all the parity strings for $\mathcal{C}$.

For any $v \in \{0,1\}^n$, let
$\hat{v}$ be the minimum Hamming distance $d_H(v,\mathcal{C}^\perp)$
between
$v$ and all the strings in $\mathcal{C}^\perp$. This means that
\[
\hat{v} = 
  \min_{v' \in \mathcal{C}^\perp} d_H(v,v') = 
  \min_{v' \in \mathcal{C}^\perp} |v \oplus v'|
\ .
\]
The value $\hat{v}$ will prove to be a security parameter. 
We use here, as in~\cite{BBBGM}, 
Eve's purified states
$\ket{\phi_{i}} = \sum_l (-1)^{i\cdot l} d_l\ket{\hat \eta_l}$,
and the resulting density matrices 
of Eq.~(\ref{rho_v_b}). 

We now show that
\begin{lemm}\label{sd-lemm} 
For any $\xi \in \{0,1\}^r$, any $(n,k,d)$ code $\mathcal{C}$ with $r \times n$ parity check
matrix $P_\mathcal{C}$ of rank $r = n-k$ and any $v \notin \mathcal{C}^\perp$
the Shannon distinguishability 
between the parity 0 and the parity 1 of the
information bits over the PA string, $v$, 
is bounded above by the following inequality:
\begin{equation}\label{SD-bound}
SD(\rho_0(v,\xi),\rho_1(v,\xi))\le2\sqrt{\sum_{|l|\ge\frac{\hat v}{2}}d_l^2}\ ,
\end{equation}
where $\hat{v} = d_H(v,\mathcal{C}^\perp)$ is the minimum Hamming distance between
$v$ and $\mathcal{C}^\perp$ 
and $\rho_b(v,\xi)$ is defined by Eq. (\ref{rho_v_b}).
\end{lemm}
See proof in Appendix~\ref{APP:tight-bou}.
As that proof was developed from methods 
used in~\cite{BBBGM}
we present in Appendix~\ref{APP:BBBGM-ext}
the preliminary analysis we did for the joint attack,
an analysis that was based on using the tools of~\cite{BBBGM}.
Appendix~\ref{APP:tight-bou} then presents improved tools
leading to 
the result described in Lemma~\ref{sd-lemm}.
Appendix~\ref{APP:tight-bou}
is self contained yet reading 
Appendix~\ref{APP:BBBGM-ext} may help the reader to 
better understand
the motivation and the development  
of the tools used for this proof.

The result of Lemma~\ref{sd-lemm}
gives an upper bound for Eve's 
information about the bit
defined by this privacy amplification string $v$. In order to
get a useful result, namely,
an {\em information versus disturbance}
result, we now prove a proposition in which the bound
on Eve's information is expressed in terms of the probability
of error on the information bits {\em in the conjugate basis}.
\begin{proposition}\label{sd-prop} 
For any $\xi \in \{0,1\}^r$, any $(n,k,d)$ code $\mathcal{C}$ with $r \times n$ parity check
matrix $P_\mathcal{C}$ of rank $r = n-k$ and any $v \notin \mathcal{C}^\perp$
\begin{equation}\label{info-bound}
I(\mrv{A}; \mrv{E}^\sym \mid i_T, j_T, b, s, \xi) \leq
  2 \sqrt{P^\sym\left[|\mrv{C}_I| \geq \frac{\hat{v}}{2} \mid i_T, j_T, b^0, s
\right]}
\end{equation}
where $\hat{v} = d_H(v,\mathcal{C}^\perp)$ is the minimum Hamming distance between
$v$ and $\mathcal{C}^\perp$, $c_I = i_I \oplus j_I$, 
$\xi = i_I P_\mathcal{C}^\top$, the key
is $a = i_I \cdot v$ and $b^0 = b \oplus s$.
\end{proposition}
\begin{proof}
\begin{align*}
I(\mrv{A}; \mrv{E}^\sym \mid i_T, j_T,b,s,\xi) &\leq
  SD(\rho_0(v,\xi),\rho_1(v,\xi)) &\text{by Eq. (\ref{I-lt-SD})} \\
 &\leq 2 \sqrt{\sum_{|l|\ge\frac{\hat v}{2}}d_l^2} 
  &\text{by Lemma~(\ref{sd-lemm})} \\
&= 2 \sqrt{ \sum_{|l| \geq \frac{\hat{v}}{2}} 
  P^\sym[\mrv{C}_I = l \mid i_T,j_T,b^0,s] } 
  &\text{by Lemma~(\ref{d-sqr-is-p-err})} \\
 &= 
  2 \sqrt{
    P^\sym\left[ |\mrv{C}_I| \geq \frac{\hat{v}}{2} \mid i_T, j_T, b^0, s
\right]} 
\ .
 & \qed
\end{align*}
\end{proof}
Notice that the bound obtained in the previous proposition
holds for all $\xi$, that is, it is the same whatever is the syndrome
sent by Alice to Bob to help him correct his information bits. 

Equation (\ref{info-bound}) bounds the information
of Eve (about a one-bit key) using the probability of the
error strings in the other basis, and it
completes the basic {\em information versus disturbance} 
result of our proof.
Previous security proofs (for simpler attacks),
such as~\cite{FGGNP,BM97b,BBBGM} are also based
on various {\em information versus disturbance} arguments, since
the non-classicality of QKD is manifested via such arguments.

The result is expressed using classical terms: Eve's information is
bounded using the
probability of error strings with large Hamming weight.
If only error strings with low weight have non-zero probability,
Eve's information becomes zero.
Such a result is a ``low weight'' property and it resembles
a similar result with this name which
was derived by Yao~\cite{Yao95}
for the security analysis of the error-free
quantum oblivious transfer (and QKD).

\subsection{Bounds on Eve's information -- the $m$-bit key case}
\label{Subsec:bounds-m-bit}

The case of an $m$-bit key $a$ is closely related to the one-bit case.
The only differences are that the upper bound is multiplied by $m$, 
and that
$\hat{v}$ is defined differently in order 
to take into account the privacy amplification
code (in addition to the error-correction code).

In terms of the bound [the R.H.S. of Eq.~(\ref{info-bound})], 
the case of an $m$ bit key $a$ follows from that of a one-bit key if
we use the following lemma:
\begin{lemm}
\label{lemm-info}
Let $\mrv{A} = (\mrv{A}_1, \ldots, \mrv{A}_m)$ be defined by
$m$ random variables.
Let $\mrv{E}$ be any random variable. If
$I(\mrv{A}_1; \mrv{E}) \leq F$
and for all $j$, $1 \leq j \leq m-1$ and all $a_1, \ldots, a_j$, 
$I(\mrv{A}_{j+1} ; \mrv{E} \mid a_1 \ldots a_j) \leq F$
then $I(\mrv{A};\mrv{E}) \leq m F$.
\end{lemm}
\begin{proof}
Note that 
\begin{align*}
I(\mrv{A}_{j+1}; \mrv{E} \mid \mrv{A}_1 \ldots \mrv{A}_j)
&= 
 \sum_{a_1 \ldots a_j}
 P(a_1, \ldots, a_j)
I(\mrv{A}_{j+1} ; \mrv{E} \mid a_1 \ldots a_j)  \\
 &\leq 
  \sum_{a_1 \ldots a_j}  P(a_1, \ldots, a_j) F  \ \  \leq \ \  F
\ .
\end{align*}
The lemma follows from the above and the chain rule for 
information (see Appendix~\ref{APP:info-basics}),
\[
I(\mrv{A}; \mrv{E}) = I(\mrv{A}_1,\mrv{A}_2,\ldots, \mrv{A}_m; \mrv{E}) 
= \sum_{j=1}^m I(\mrv{A}_j; \mrv{E} \mid \mrv{A}_1, \ldots, \mrv{A}_{j-1})
\ .
\]
\hspace*{\fill}\qed 
\end{proof}

Next,
in the particular case at hand, we want to bound Eve's 
information about the
$m$-bit key given the values $i_T$, $j_T$, $b$, $s$ and $\xi$ 
she learned.
This means we want to bound
$I(\mrv{A}; \mrv{E}^\sym \mid i_T, j_T, b, s, \xi)$ where $\mrv{A}$ 
is the $m$-bit
key. 
This is nothing but a mutual information between $\mrv{A}$ and
$\mrv{E}^\sym$ for some fixed (known) values
of random outputs, and the above lemma thus applies.
More precisely, it tells us that if 
some number
$F$ is an upper bound for $I(\mrv{A}_{j+1}; \mrv{E}^\sym \mid i_T,j_T,b,s,\xi,a_1 \ldots a_j)$ then $mF$ will be an upper bound for
$I(\mrv{A}; \mrv{E}^\sym; i_T, j_T, b,s, \xi)$.                         
Announcing $\xi$ and $a_1 \ldots a_j$ is announcing publicly the 
bits $v_1 \cdot i_I$, $\ldots$, $v_{r+j} \cdot i_I$,
which is just the same as using the $r+j$ strings
$v_1, \ldots v_{r+j}$ as parity strings of a code for which
Proposition \ref{sd-prop} applies. 
More formally,
\begin{proposition}\label{info-m-prop1}
Let $v_1, \ldots, v_{r+m}$ be $r+m$ linearly independent $n$-strings
and $V_{r'}$ be the subspace of $\{0,1\}^n$ spanned by
$\{v_1, \ldots, v_{r'}\}$ ($1 \leq r' \leq r+m$). 
Let $P_\mathcal{C}$ be the matrix whose rows
are $v_1, \ldots, v_r$ and $P_\mathcal{PA}$ the one with rows $v_{r+1}, \ldots
,v_{r+m}$. Then for any $\xi \in \{0,1\}^r$
\begin{equation}\label{info-m-bound1}
I(\mrv{A}; \mrv{E}^\sym \mid i_T, j_T, b, s, \xi) \leq
  2 m \sqrt{P^\sym\left[|\mrv{C}_I| \geq \frac{\hat{v}}{2} \mid i_T, j_T, b^0, s
\right]}
\end{equation}
where  $\hat{v} = \min_{r \leq r' <r+m} d_H(v_{r'+1}, V_{r'})$,
$c_I = i_I \oplus j_I$, $\xi = i_I P_\mathcal{C}^\top$, $a = i_I P_\mathcal{PA}^\top$
and $b^0 = b\oplus s$.
\end{proposition}
\begin{proof}
See Appendix~\ref{APP:m-bits}.
\end{proof}

If we modify $\hat{v}$ to any value that is less than or equal to
the minimum over 
all the Hamming distances $d_H(v_{r'+1}, V_{r'})$ 
then 
equation (\ref{info-m-bound1}) is satisfied with the modified $\hat{v}$ as well,
as only the RHS increases.
In particular this is true if 
we follow the definition given in 
Subsection~\ref{subsec:ub-bb84} in item II.~7; 
thus we define
$\hat{v}$ to be (from now on) the minimal distance
between any string $v$ in the set of  
PA parity-check
strings, and any string $v'$ in the span of 
their union with the
parity-check-strings of the ECC (the dual to the code).
This formally means:
\begin{corollary}\label{info-m-coro}
Let $v_1, \ldots, v_{r+m}$ be $r+m$ linearly independent $n$-strings.
Let $P_\mathcal{C}$ be the matrix whose rows
are $v_1, \ldots, v_r$ and $P_\mathcal{PA}$ the one with rows 
$v_{r+1}, \ldots ,v_{r+m}$. 
Let $V_{r'}^\exc$ be the $2^{r+m-1}$-dimensional subspace 
of $\{0,1\}^n$ spanned by a subset
of the $r+m-1$ parity strings which excludes the 
PA string $v_{r'}$
(namely, the span of $v_1,\ldots,v_{r'-1},v_{r'+1},\ldots,v_{r+m}$). 
Then for any $\xi \in \{0,1\}^r$
\begin{equation}\label{info-m-bound}
I(\mrv{A}; \mrv{E}^\sym \mid i_T, j_T, b, s, \xi) \leq
  2 m \sqrt{P^\sym\left[|\mrv{C}_I| \geq \frac{\hat{v}}{2} \mid i_T, j_T, b^0, s
\right]}
\end{equation}
where  $\hat{v} = \min_{r+1 \leq r' \leq r+m} d_H(v_{r'}, V_{r'}^\exc)$,
$c_I = i_I \oplus j_I$, $\xi = i_I P_\mathcal{C}^\top$, $a = i_I P_\mathcal{PA}^\top$
and $b^0 = b\oplus s$.
\end{corollary}

\noindent
[First remark: in fact, for binary linear codes, 
the two $\hat{v}$ defined above,
the one used in Proposition~\ref{info-m-prop1} and the one used
in Proposition\ref{info-m-coro} are equal, 
but this fact is irrelevant for
our paper.

\noindent
Second remark: we could even follow a stricter definition
and replace $\hat{v}$ by $d^\perp$, 
the minimum (non-zero) distance of
the code $V_{r+m}$ of dimension $r+m$ 
(the space spanned by the ECC
and PA strings $v_1, \ldots, v_{r+m}$,
see Subsection~\ref{subsec:ub-bb84}, item II.~7).
Notice that the rows of the
generator matrix of this code are those of $P_\mathcal{C}$ and $P_\mathcal{PA}$.]

\clearpage

\section{Completing the Security Proof}
\label{Sec:compl}

In this section we analyze the attack on the test and information
qubits together (cf Eq.~\ref{Eve_Full_Attack}).
For these states, we bound the weighted average of Eve's 
information 
$\langle \mrv{I}'_{Eve}\rangle$, used in the alternative 
security criteria [see Eq.~(\ref{Average-I-prime-Eve-2})]:
\[
\sum_{c_T|\mrv{T}=\pass}P\left[\mrv{C}_T=c_T\right]\ 
I(\mrv{A};\mrv{E}|\mrv{I}_T,\mrv{C}_T=c_T,\mrv{B},\mrv{S},\mrv{\Xi}) 
\ . \]
We show that the above bound is exponentially small and therefore
Lemma~\ref{lemm-sec-crit-2}
promises us that
security is achieved.
We generalize here previous (and more limited)
proofs~\cite{BMS96,BM97a,BBBGM} that
information about parity bits is exponentially small, 
to be applicable for the most general attack on the channel --- 
the joint attack.
[A remark: 
We freely switch below between $c_T$ and $j_T$ whenever $i_T$ 
is given.] 

\subsection{Applying the bounds to all attacks}

The maximum error rate that still passes the test is 
denoted $p_a$ (or $p_{\allowed}$). This
means that $\mrv{T} = \pass$ if and only if $|c_T| \leq n p_a$. 
For $\hat{v}$ as defined in  
Corollary~\ref{info-m-coro},
and making use of 
that corollary we get, for fixed $b$ and $s$:
\begin{lemm} \em \label{sec5-lemm1}
\begin{eqnarray*}
\lefteqn{\sum_{|c_T|\le n p_a}P^\sym\left[\mrv{C}_T=c_T| b,s\right]
I(\mrv{A};\mrv{E^\sym}\ |\ \mrv{I}_T,\mrv{C}_T=c_T,b,s,\mrv{\Xi}) } \\
&\le& 2m\sqrt{  
P^\sym\left[(|\mrv{C}_I|>\frac{\hat{v}}{2})\wedge 
(\frac{|\mrv{C}_T|}{n}\le p_{a}) \mid b^0,s \right]}
\ .
\end{eqnarray*}
\end{lemm}
The proof is given in Appendix~\ref{APP:sec5-lemm1}. 

Let $U$ (and $\cal{E}$) be some arbitrary attack 
and $\{U^{\sym},\cal{E}^{\sym}\}$ an arbitrary symmetrized attack
resulting from $U$.
As the Lemma above is true for any symmetric attack, it is also 
true for
any $\{U^{\sym},\cal{E}^{\sym}\}$ 
and in particular for 
the optimal one (in which the optimal POVM is performed
for each value of 
$i_T,b,\ldots$)
Thus, we immediately get from Lemma~\ref{sec5-lemm1}
\begin{corollary} \label{sec5-coroll-1}
\begin{eqnarray*}
\lefteqn{\sum_{|c_T|\le n p_a}P^\sym\left[\mrv{C}_T=c_T| b,s\right]
\max  I(\mrv{A};\mrv{E^\sym}\ |\ \mrv{I}_T,\mrv{C}_T=c_T,b,s,\mrv{\Xi}) } \\
&\le& 2m\sqrt{  
P^\sym\left[(|\mrv{C}_I|>\frac{\hat{v}}{2})\wedge 
(\frac{|\mrv{C}_T|}{n}\le p_{a}) \mid b^0,s \right]}
\end{eqnarray*}
\end{corollary}
with the maximum [$\max I(\ )$] defined in Eq.~(\ref{MAX-info}).

We now prove that the above bound, 
with the same definition
of $\hat{v}$, also applies to the original 
unsymmetrized attack ($b$ and $s$ still fixed).
\begin{lemm}\label{sec5-lemm2}
\begin{eqnarray*}
\lefteqn{\sum_{|c_T|\le n p_a}P\left[\mrv{C}_T=c_T| b,s\right]
I(\mrv{A};\mrv{E}\ |\ \mrv{I}_T,\mrv{C}_T=c_T,b,s,\mrv{\Xi}) } \\
&\le& 2m\sqrt{  
P\left[(|\mrv{C}_I|>\frac{\hat{v}}{2})\wedge 
(\frac{|\mrv{C}_T|}{n}\le p_{a}) \mid b^0,s \right]}
\end{eqnarray*}
\end{lemm}
\begin{proof}
This follows from Lemma \ref{lemma-symm}, Corollary
\ref{sec5-coroll-1} and equations (\ref{psimcictvspcict}, \ref{psimctvspct}) from Corollary~\ref{coroll-Psym-CI-CT}:
\begin{align*}
\sum_{|c_T|\le n p_a}& P\left[\mrv{C}_T=c_T| b,s\right]
I(\mrv{A};\mrv{E}\ |\ \mrv{I}_T,\mrv{C}_T=c_T,b,s,\mrv{\Xi})  \\
&=\sum_{|c_T|\le n p_a}P^\sym\left[\mrv{C}_T=c_T| b,s\right]
I(\mrv{A};\mrv{E}\ |\ \mrv{I}_T,\mrv{C}_T=c_T,b,s,\mrv{\Xi})  
 &\text{by Eq. (\ref{psimctvspct})} \\
&\le\sum_{|c_T|\le n p_a}P^\sym\left[\mrv{C}_T=c_T| b,s\right]
\max  I(\mrv{A};\mrv{E^\sym}\ |\ \mrv{I}_T,\mrv{C}_T=c_T,b,s,\mrv{\Xi})  
 &\text{by Lemma \ref{lemma-symm}} \\
&\le 2m \sqrt{
P^\sym\left[ (|\mrv{C}_I|>\frac{\hat{v}}{2})\wedge 
(\frac{|\mrv{C}_T|}{n}\le p_{a}) \mid b^0,s \right]}
 &\text{by Corollary~\ref{sec5-coroll-1}}
\end{align*}
By Eq. (\ref{psimcictvspcict}), 
$P^\sym\left[\mrv{C}_I=c_I,\mrv{C}_T=c_T \mid b,s\right]=
P\left[\mrv{C}_I=c_I, \mrv{C}_T=c_T \mid b,s\right]$
for any basis string, in particular $b^0$;  this concludes the proof.
\qed
\end{proof}
{}From now on, there will be no restriction of symmetry on the attacks.
The results will hold for any attack whatsoever.

\subsection{Exponentially-small bound on Eve's information}
For any $\epsilon_{\sec}$ 
and $p_a$, such that $\hat{v}\ge 
2n(p_{a}+\epsilon_{\sec})$ Lemma \ref{sec5-lemm2} leaves
the following bound:
\begin{corollary}\label{sec5-coroll-2}
\begin{eqnarray*}
\lefteqn{\sum_{|c_T|\le n p_a}P\left[\mrv{C}_T=c_T| b,s\right]
I(\mrv{A};\mrv{E}\ |\ \mrv{I}_T,\mrv{C}_T=c_T,b,s,\mrv{\Xi})} \\
&\hspace*{-3.7mm}\le& 2m \sqrt{
P\left[(\frac{|\mrv{C}_I|}{n}> p_a + \epsilon_{\sec})\wedge 
(\frac{|\mrv{C}_T|}{n}\le p_{a}) \mid b^0,s \right]}
\end{eqnarray*}
\ .
\end{corollary}

Thus far, there is nothing that causes the bound on the right hand
side to be a small number.
The result above is true even if Eve is told in advance the bases of
Alice and Bob (the string $b$),
or if she is told in advance which are the test bits
and which are the used bits (the string $s$),
two cases in which Eve easily obtains full information
about the secret key $a$.

Only Eve's lack of knowledge regarding the random strings $b$ and $s$ provides
an exponentially small number at the right hand side.
Since Eve must fix her attack \emph{before} she knows the basis or
the test-bits choice, 
we compute the average information for a fixed attack
over all bases $b$ and test-bits choice $s$.
Averaging over $b$ means that we sum over all $b$'s and multiply
each term by  the constant
$p(b) = 1/2^{2n}$.
The averaging over $b$ removes the dependence on the particular basis
[due to $\sum_b p(z|b) p(b) = \sum_b p(z,b) = p(z)$].

Averaging over $s$ means that we sum over all $s$'s and multiply
each term by  the constant
$p(s) = 1/{2n \choose n}$.
The averaging over $s$ removes the dependence on the particular choice
of which bits are the test bits
[due to $\sum_s p(z|s) p(s) = \sum_s P(z,s) = p(z)$].

\begin{lemm} \em 
\label{sec5-lemm3}
Let $\mrv{T} = \pass$ iff $|c_T| \leq np_a$, and let 
$\mrv{I}'_{Eve}$ be the random variable
equal to $\mrv{I}_{Eve} = I(\mrv{A};\mrv{E}\mid i_T,j_T,b,s,\xi)$ 
when $\mrv{T} = \pass$ and $\mrv{I}'_{Eve} = 0$ otherwise. 
Then for any
$\epsilon_{\sec}$ and $p_a$ such that
$p_a + \epsilon_{\sec} \leq \hat{v}/2n$ 
we get
\[
\langle \mrv{I}'_{Eve}\rangle \leq
 2m\sqrt{
         P\left[(\frac{|\mrv{C}_I|}{n}>p_{a}+\epsilon_{\sec}) \wedge
        (\frac{|\mrv{C}_T|}{n}\le p_{a})\right]}
\ .
\]
\end{lemm}

\begin{proof} 
We already proved 
(Eq.~\ref{Average-I-prime-Eve-2}) that
 \[\langle \mrv{I}'_{Eve}\rangle = 
\sum_{c_T|\mrv{T}={\pass}}P\left[\mrv{C}_T=c_T\right]\ 
I(\mrv{A}; \mrv{E}|\mrv{I}_T,\mrv{C}_T=c_T,\mrv{B},\mrv{S},\mrv{\Xi})\]
where $\mrv{T} =\pass$ iff
$|c_T| \leq np_a$. Expanding the right-hand side, we get
\[
 \langle \mrv{I}'_{Eve}\rangle = 
\sum_{b,s} p(b,s)\sum_{|c_T|\le n p_a}P\left[\mrv{C}_T=c_T|b,s\right]
 I(\mrv{A};\mrv{E}\ |\ \mrv{I}_T,\mrv{C}_T=c_T,b,s,\mrv{\Xi}).
\]
Using Corollary~\ref{sec5-coroll-2} we obtain the first bound below; 
then
using the fact that $\sum_i p_i \sqrt{x_i}\le \sqrt{\sum_i p_i x_i}$,
and that $p(b,s) = p(b^0,s) = 2^{-2n}p(s)$ ($b$ and $s$ being
chosen independently) we get the second bound; 
finally noting that summing over $b$ is the same
as summing over $b^0$,  we get the third bound:
\begin{align*}
\langle \mrv{I}'_{Eve}\rangle
&\le \sum_{b,s}p(b,s)2m \sqrt{
P\left[(\frac{|\mrv{C}_I|}{n}>p_a + \epsilon_{\sec})\wedge 
(\frac{|\mrv{C}_T|}{n}\le p_{a}) \mid b^0,s \right]}\\
&\le 2m \sqrt{ \sum_{b,s}2^{-2n}p(s) P\left[(\frac{|\mrv{C}_I|}{n}>p_a +
\epsilon_{\sec})\wedge 
(\frac{|\mrv{C}_T|}{n}\le p_{a}) \mid b^0,s \right]}\\
&= 2m \sqrt{ P\left[(\frac{|\mrv{C}_I|}{n}>p_a + \epsilon_{\sec})\wedge 
(\frac{|\mrv{C}_T|}{n}\le p_{a}) \right]} \ . &\qed
\end{align*} 
\end{proof}

For a long string, the test bits and the
information bits should have a similar number of errors
if the test is picked
at random.
The probability that they have
different numbers of errors should go to
zero exponentially fast as shown in the following lemma.
\begin{lemm} \em
For any $\epsilon >0$,\ \  
$P\left[(\frac{|\mrv{C}_I|}{n}>p_{a}+\epsilon) \wedge
(\frac{|\mrv{C}_T|}{n}\le p_{a})\right] \le e^{-\frac{1}{2}n\epsilon^2}$.
\label{sec5-lemm4}
\end{lemm}
\begin{proof}
This follows directly from Hoeffding's law of large numbers \cite{Hoeffding}. The details are 
given in Appendix~\ref{law-large-num}.
\end{proof}

\subsection{The main results}
We are now in a position to state and prove our main results.

\begin{proposition}\label{mainprop}
If $p_a$ and $\epsilon_{\sec}$ and the ECC+PA codes
are such that $p_a + \epsilon_{\sec} \leq \hat{v}/2n$
with $\hat{v} = \min_{r'=r+1}^{r+m} d_H(v_{r'}, V_{r'}^\exc)$ where $d_H$ is 
the Hamming distance, $v_{r'}$ is a parity-check string, 
and $V_{r'}^\exc$ is the $2^{r+m-1}$ space which is
the span of $v_1,\ldots,v_{r'-1},v_{r'+1},\ldots,v_{r+m}$,  then
\[
\langle \mrv{I}'_{Eve} \rangle \leq
2m\sqrt{ e^{-\frac{1}{2}n\epsilon_{\sec}^2}} 
\]
where $\mrv{I}'_{Eve} = \mrv{I}_{Eve}$ if
$|c_T| = |i_T \oplus j_T| \leq n p_a$ (test passed) and 
$\mrv{I}'_{Eve} = 0$ otherwise.
\end{proposition}
\begin{proof}
This follows immediately from Lemma \ref{sec5-lemm3} and Lemma \ref{sec5-lemm4}.
\end{proof}

\begin{theorem}\label{security-theorem}
If $p_a$ and $\epsilon_{\sec}$ and the ECC+PA codes
are such that $p_a + \epsilon_{\sec} \leq \hat{v}/2n$
with $\hat{v} = \min_{r'=r+1}^{r+m} d_H(v_{r'}, V_{r'}^\exc)$ where $d_H$ is 
the Hamming distance, $v_{r'}$ is a parity-check string, 
and $V_{r'}^\exc$ is the $2^{r+m-1}$ space which is
the span of $v_1,\ldots,v_{r'-1},v_{r'+1},\ldots,v_{r+m}$,  then
for any $A_{\info} > 0$, $A_{\luck}>0$ such that $A_{\info}A_{\luck} = 2m$
and any $\beta_{\info}$ and $\beta_{\luck}$ such that
$\beta_{\info} + \beta_{\luck} = \epsilon_{\sec}^2/4$,
\begin{equation}
P\left[(\mrv{T}=\pass) \wedge (\mrv{I}_{Eve} \ge A_{\info} 
  \ e^{-\beta_{\info} n}) \right] \le A_{\luck} \  e^{-\beta_{\luck} n} 
\end{equation}
where $\mrv{T} = \pass$ iff $|c_T| \leq n p_a$ and
$\mrv{I}_{Eve} = I(\mrv{A}; \mrv{E} \mid i_T, j_T, b, s, \xi)$.
\end{theorem}
\begin{proof}
This follows from Proposition \ref{mainprop} if we
let $A = 2m$ and $\beta = \epsilon_{\sec}^2/4$ in  Lemma \ref{lemm-sec-crit-2}.
\end{proof}

Let us recall that, in addition to the 
security, one must also guarantee
the reliability of the final key. Namely we need
to make sure
that Alice's final key and 
Bob's final key are (almost always) identical.
Note that Lemma~\ref{sec5-lemm4} can be rewritten:
\[
P\left[(\mrv{T} =\pass) \wedge  (|\mrv{C}_I| > 
(p_a + \epsilon_{\rel})n\right)] \leq e^{-\frac{1}{2}\epsilon_{\rel}^2}
\]
This also means that
\begin{corollary}
\label{reliability-corollary}
The probability that the test is passed and that
there are more than 
$(p_a + \epsilon_{\rel})n$ errors in the information string 
is exponentially small;
it is bounded by
\[ h = e^{-\frac{1}{2}n\epsilon_{\rel}^2}\ . \]
\end{corollary}
Once the ECC is chosen such that 
$(p_a + \epsilon_{\rel})n$ errors in the information string 
are corrected, Alice's and Bob's final keys identical
except for an exponentially small probability bounded by $h$.
This result means that $A_{\rel} = 1$ and 
$\beta_{\rel} = \epsilon_{\rel}^2 / 2$,  
in the reliability criterion of 
Subsection~\ref{subsec:reliability}.

\subsection{The existence of codes that provide 
security and reliability}\label{RLC-exist}

The above bound on Eve's information is exponentially small,
provided there is a family of good linear ECCs satisfying also
the requirement
that $\hat{v} \geq 2n(p_a + \epsilon_{\sec})$ when PA strings are added. 
What we formally need is a family 
of (linear) ECC+PA codes satisfying the following two conditions:
\begin{enumerate}
\item[(1)] The ECC can correct up to
$t = p_{\rm allowed}+\epsilon_{\rel}$ errors.
For this to happen, we demand that the 
minimum distance $d$ between the code words of the ECC
satisfy $d \geq 2t+1$.
Hence, a
  $d \geq 2t+1 = 2n(p_{\allowed} + \epsilon_{\rel}) +1$
is sufficient.
This code can correct all the
errors in the information string,
except for an exponentially small probability bounded by
$ h $ (of Corollary~\ref{reliability-corollary})
of having more errors
in the information string than expected.
\item[(2)] The minimum
distance
$d^\perp$,
of the code words in the span of the dual code and the
PA strings (hence, the augmented dual code
is of dimension  $r+m$) should have a minimum distance
$d^\perp \geq 2n(p_{\allowed} + \epsilon_{\sec})$.
\end{enumerate}

We discuss below the class of linear codes called 
random linear codes. 
Such codes cannot be easily decoded
hence their practical usefullness is limited.
It may well be that such codes can be replaced 
by the much more practical codes ---
the Reed-Solomon codes, 
without losing the security and reliability proven below.
However, analyzing Reed-Solomon codes is beyond the
scope of this work. 

For random linear codes (RLC's) the two requirements
mentioned above can easily be satisfied.
We can generate an $m$-bit secret key 
if we pick an $(n,n-r)$ RLC, where $r$ and $m$ satisfy 
\begin{eqnarray*}
H_2(2p_{a}+2\epsilon_{\rel}+1/n) &<&r/n \\
H_2(2p_{a}+2\epsilon_{\sec})+
H_2(2p_{a}+2\epsilon_{\rel}+1/n)
&<&1-R_{\secret}\ ,
\end{eqnarray*}
with  $H_2$ the entropy, and $R_{\secret}\equiv m/n$ the bit-rate
(namely, the efficiency of the QKD scheme). 
If these conditions are not met then the random linear
code provides neither reliability nor security; 
see Appendix~\ref{app:codes-exist}.
At the limit of large $n$ and $\epsilon$'s close to zero
we get as a bound $2 H_2(2 p_a) < 1$.
Then, 
$p_{\allowed}<5.50\%$
satisfies the bound and hence this is our first threshold 
[see Appendix~\ref{app:codes-exist} for the detailed calculation].
It is the threshold in the case in which we
want to have an exact bound on Eve's information and on the
reliability of the final key, as a function of parameters chosen
by the designer of the QKD protocol. This is important 
for a designer who needs to choose 
a sufficiently large $n$ (that is not assumed to go to infinity);
then Eve's information is bounded as in Proposition~\ref{mainprop}
and the reliability is bounded as 
in~Corollary~\ref{reliability-corollary}.

Note that if we let $p_{\allowed}$  be sufficiently  close to zero 
then (for sufficiently large $n$ and small $\epsilon$'s)
a bit-rate $R_{\secret}$ close
to one can be obtained. Specific values of Eve's information,
the probability of error in the final key, and the resulting
bit-rate are provided in Table~\ref{table1:rates};
this is done by choosing 
$\epsilon_{\sec} = \epsilon_{\rel} = \epsilon$ 
(for the sake of simplicity).
As the parameters $n$, $\epsilon$, and $p_\allowed$ can be chosen by
the designer of the protocol, we present here 3 values of the 
reliability/security parameter, and we then calculate\footnote{
The term $1/n$ that appears in the parameter 
$[2p_\allowed + 2\epsilon + 1/n]$
is negligible except in the two cases where the entire 
term approaches 11.0\%.} 
the reliability as a function
of $n$, and we calculate\footnote{
We choose a maximal bit rate by solving 
$H_2(2p_{a}+2\epsilon)+
H_2(2p_{a}+2\epsilon+1/n)
=0.99-R_{\secret}$.  
} the maximal bit-rate as a function of $P_\allowed$.

\begin{table}
$$\begin{tabular}{lrccc}
\hline
&&$\epsilon=0.5\%$&$\epsilon=1\%$&$\epsilon=2\%$\\
\hline
Reliability&$n=12500 $&&0.54&1/12\\
Bound ($h$) &$n=50000 $&0.54&1/12&1/22026\\
&$n=200000 $&1/12&1/22026&$4\cdot 10^{-18}$\\
&$n=800000 $&1/22026&$4\cdot 10^{-18}$&$\approx 10^{-70}$\\
&$n=3200000$&$4\cdot 10^{-18}$&$\approx 10^{-70}$&\\
\hline  \\
\hline
&&$\epsilon=0.5\%$&$\epsilon=1\%$&$\epsilon=2\%$\\
\hline
Rate ($R_{\secret} =m/n$)&$P_{\allowed}=2.0\%$&41.7\%&33.5\%&18.5\%\\
&$P_{\allowed}=3.5\%$&18.5\%&11.7\%&0.007\%$^*$\\
&$P_{\allowed}=5.0\%$&0.007\%$^*$&$^\dagger$&$^\dagger$\\
\hline
\multicolumn5l{$^\dagger$ Out of the allowed range (negative rate)}\\
\multicolumn5l{$^*$ For the case of $2P_\allowed + 2\epsilon = 11.0\%$ we
calculate $R_{\secret}$ by solving}\\ 
\multicolumn5l{$H_2(2p_{a}+2\epsilon)+
H_2(2p_{a}+2\epsilon+1/n)
=0.9999-R_{\secret}$.  
Here, security and reliability}\\ 
\multicolumn5l{can be obtained only with $n > 10^6$ or so}
\end{tabular}$$
\caption{Summary of the characteristics of a QKD protocol that uses RLC:
The ``Reliability Bound'', $h$, is calculated according 
to Corollary~\ref{reliability-corollary},
and the maximal bit rate $R_{\secret}$ is calculated by solving 
$H_2(2p_{a}+2\epsilon)+
H_2(2p_{a}+2\epsilon+1/n)
=0.99-R_{\secret}$ (with two exceptions, denoted with $^*$ in the table).  
The parameters in this table are 
closely related to the parameters used in experiments:
$n$ is related to the number of photons obtained by Bob; 
$2n$ photons are used according to the used-bits-BB84 protocol 
and slightly more than $4n$ in the conventional BB84. 
The error rate considered here is achieved in many experimental setups,
but might limit the distance of transmission.
A photon rate of 1000 photons per second 
(if we count the photons obtained by Bob) was also reported in various
experiments, 
so the resulting secret-key bit-rate $R_{\secret}$ can be sufficient for 
many practical usages.}
\label{table1:rates}
\end{table}

The ``Reliability Bound'' $h$ is calculated according 
to Corollary~\ref{reliability-corollary},
and (due to the equal $\epsilon$'s) we can then get the bound on 
Eve's information (according to Proposition~\ref{mainprop}), which is exactly
$2m \sqrt{h}$.
We consider the numbers we got for the ``Reliability Bound'' in the table
to be  ``Good''  
when the probability
of error is $1/22026$ or below. 
However, with $h = 1/22026$, 
Eve's information is $2m$ times $1/148$ which means that the users
cannot really enjoy the allowed bit-rate, 
and must use a much smaller value for
$m$, as Eve could then learn too much.
When the ``Reliability Bound'' is 
$4\cdot 10^{-18}$ or $\approx 10^{-70}$ 
there is clearly no problem at all with Eve's
information, and $m$ can be as large as the allowed bit-rate enables.

For RLC one can actually obtain a better threshold for the allowed
error rate (as first noticed by Mayers~\cite{Mayers98}), 
by modifying requirement (1) so that:
\begin{enumerate}
\item[(1')] The ECC can correct up to
$p_{\rm allowed}+\epsilon_{\rel}$ errors, with probability
as close to 1 as we wish.
\end{enumerate}
Namely, for any $\hat\delta$, the ECC can correct up to 
$p_{\rm allowed}+\epsilon_{\rel}$ errors, with probability
smaller than $\hat\delta$. 
For RLC this is true (due to Shannon's bound, 
see for instance~\cite{MS-book}) for any code having a 
minimum distance 
$d \geq t+1 = n(p_{\allowed} + \epsilon_{\rel}) +1$
(rather than $d \geq 2t+1 $, that 
promises the success of correcting all errors),
provided that $r/n > H_2(p_{\allowed} + \epsilon_{\rel})$,
and that a sufficiently large $n$ is chosen.

We show in Appendix~\ref{app:codes-exist} that   
requirements (1')  and (2) 
can be satisfied and one can generate an $m$-bit secret key, 
if one picks an $(n,n-r)$ RLC, where $r$ and $m$ satisfy the following:
\begin{eqnarray*}
H_2(p_{a}+\epsilon_{\rel}+1/n) &<&r/n \\
H_2(2p_{a}+2\epsilon_{\sec})+
H_2(p_{a}+\epsilon_{\rel}+1/n)
&<&1-R_{\secret}\ ,
\end{eqnarray*}
where $R_{\secret}\equiv m/n$.
In the limit of large $n$ and $\epsilon$'s close to zero
we get as a bound $H_2(2 p_a) + H_2(p_a) < 1$.
Then, 
$p_{\allowed}<7.56\%$
satisfies the bound and hence this is our improved
threshold (which is identical to the threshold calculated by 
Mayers~\cite{Mayers98}).
Note that 
Eve's information is still 
bounded to be exponentially small
due to~Theorem~\ref{security-theorem}, but the reliability 
is now bounded only asymptotically as we did not find an explicit
formula for the probability $\hat\delta$ 
of having an error (as a function of $n$)
when the distance
is $d > t+1$.

Asymptotically, with a rate $R_{\secret}<1-H_2(p_a)-H_2(2p_a)$ the
final key is secure
and reliable for the given ECC+PA.  Note, as $p_a$ goes to zero,
$R_{\secret}$ goes to $1$, which means that
(asymptotically) almost all the information bits are
secret.

This threshold is based on the properties of the code, and other codes
might give worse thresholds, but might have other desired properties. 
Random linear codes are not so useful as their decoding cannot be done
efficiently.
It is possible to make use of methods for approximate decoding (in which we are not
always promised that the closest code word is chosen after the error correction),
but the bound on reliability then need some adjustments.
It might be better to replace the RLC by a
code that can be decoded efficiently (e.g., Reed-Solomon
concatenated code, with a random seed), and add random PA strings. The Hamming distance
between the PA check-strings and the ECC check-strings is still
bounded below in the same way as for the RLC 
(see~\cite{Mayers98}).

Finally, it is interesting to note that the bound
$H_2(p_a) + H_2(p_a) < 1$ (which was neither reported by us nor by Mayers) 
leads to the threshold
of 11\%, and such threshold was 
reported and proven by Shor and Preskill~\cite{SP00}. 
This probably means that 
the alternative proof presented there can, in some sense,
modify requirement (2) in a way similar to the modification 
done here to change from (1) to (1') above.
However, we could not see how the same modification could apply to our
proof.

A well-known 
way to improve the threshold further is to allow two-way communication 
as part of the ECC+PA process. This technique is known as key distillation, 
see the basic idea described 
in~\cite{Brassard-Salvail}. The analysis of Eve's density matrices
becomes much more complicated in such a case, and we do not yet know if our
proof can easily be adjusted to allow that\footnote{
After the submission of our paper, Gottesman and Lo proved  
that the Shor-Preskill proof of security can be adjusted to deal with 
such a key distillation, yielding an improved threshold for $p_\allowed$;  
see quant-ph/0105121).}.

\section{Summary}
\label{Sec:summary}

We proved the security of the Bennett-Brassard (BB84) protocol for
quantum key distribution. Our proof is based on analyzing Eve's
reduced density matrices, on a novel 
{\em information versus disturbance} 
 result, on the optimality of symmetric
attacks, on laws of large numbers, and on various techniques that
simplify the analysis of the problem.  

Many of the ideas and the tools developed here can be found relevant
when proving the security of other QKD schemes:
the analysis of Eve's reduced density matrices, the purifications
of her states, the usage of that purification for finding
a relevant information versus disturbance bound, 
the use of Hoeffding's law of large numbers,
the trace-norm-difference bound, etc.
Other tools, such as the reduction to the used-bits-BB84
protocol, and the extensive usage of symmetry could also provide
some important insight, but are somewhat more specific to the BB84
scheme.

\section{Acknowledgement}

The work of T.M.~is supported 
in part by the Israel MOD Research and Technology Unit. 
The work of M.B. is supported in part by
the Natural Sciences and Engineering Research Council ({\sc
NSERC}) of Canada.
The work of E.B.~is supported in
part by the European Commission through the IST Programme under contract
IST-1999-11234.
The work of P.O.B.,~T.M., and V.R.,~is supported in 
part by the Defense 
Advanced Research Projects
Agency (DARPA) project MDA972--99--1--0017,
by the U.S. Army Research Office/DARPA DAAD19--00--1--0172,
by Grant No.~530-1415-01
from the DARPA Ultra program, and 
by Grant No.~961360 from the Jet Propulsion Lab.

\clearpage


\clearpage

\appendix

\section{Security of BB84} \label{app-used-bits}

In the paper we prove that used-bits-BB84 is secure.
Let us now present the original BB84 protocol and prove,
by reduction, that its
security follows immediately from the security of the used-bits-BB84
protocol.

The differences between the protocols are only in the first part.
The first part of the BB84 protocol is as follows:

\begin{enumerate}

\item[I.]Creating the sifted key: 

\item

Alice and Bob choose a large integer $n \gg 1$, and a number
$\delta_{\num}$, such that
$ 1 \gg \delta_{\num} \gg  1/\sqrt{(2 n)}$.
The protocol uses $n'' = (4+\delta_{\num})n$ bits.

\item

Alice randomly selects two $n''$-bit strings, $b''$ and $i''$, 
which are then
used to create qubits:
The string $b''$ determines the basis $0\equiv z$, and $1 \equiv x$ of the
qubits.
The string $i''$ determines the value (0 or
1) of each of the $n''$ qubits (in the appropriate bases).

\item

Bob randomly selects an $n''$-bit string, $b^{''\rm Bob}$, which determines Bob's
later choice of bases for measuring each of the $n''$ qubits.

\item

Alice generates $n''$ qubits according to her selection of
$b''$ and $i''$, and sends
them to Bob via a quantum communication channel.

\item
After receiving the qubits, Bob measures in the basis $b^{''\rm Bob}$.

\item
Alice and Bob publish the bases they used; this step should be performed
only after Bob received all the qubits.

\item
All qubits with different bases are discarded by Alice and Bob.
Thus, Alice and Bob finally have $n' \approx n''/2$ bits
for which they used the same bases $b'$.
The $n'$-bit string 
would be identical for Alice and Bob
if Eve and natural
noise do not interfere.

\item\label{it:abort}
Alice selects the first $2n$
bits from the $n'$-bit string,  
and the rest of the $n'$ bits are discarded.
If $n'<2n$ the protocol is aborted 
(a fake random key can be chosen in this case via the unjammable classical
channel,
so that the key is not secret; however the probability for this to happen is
exponentially small).

We shall refer to the resulting $2n$-bit string as the sifted key.

\end{enumerate}

The second part of the protocol is identical to the second part of the
used-bits-BB84 protocol.
To prove that BB84 is secure let us modify BB84 by a few steps in a way
that each step can only be helpful to Eve, and the final protocol is the
used-bits-BB84.
Each item below describes a different protocol, obtained by 
modifying the previous protocol.

Recall that Alice and Bob choose their strings of basis $b''$ and 
$b^{''\rm Bob}$ in
advance. Recall that the two strings are random.
Thus, the first modification below has no influence at all on the
security or the analysis of the BB84 protocol.
Note that after the first modification Alice knows
the un-used bits in advance. 
The second and the third modifications are
done in a way that Eve can only gain, hence
security of the resulting protocol provides the security of BB84.
The last modification is only ``cosmetic'', 
in order to derive precisely
the used-bits-BB84 protocol. This modification changes nothing in
terms of Eve's ability.

\begin{itemize}

\item 
Let Bob have a quantum memory. 
 Let Alice choose $b^{''\rm Bob}$ instead of Bob at step 3. 
 When Bob receives the qubits at step 5, let
 him keep the qubits in a memory, and tell Alice he received them.
 In step 6, 
let Alice announce $b^{''\rm Bob}$ to Bob, and Bob measures 
in bases $b^{''\rm Bob}$.

{}From the announcements of $b''$ and $b^{''\rm Bob}$
Bob knows which are the used and the un-used bits,
as determined in steps 7 and 8.
Now, at the end of step 8, Alice and Bob know
all the un-used bits, so they ignore them,
to be left with $2n$ bits.

Note that in this modified protocol,
Alice can calculate which are the un-used bits already at step 3
(if she wishes to know this).

\item
Let Alice calculate the un-used bits and 
announce them already at the end of
step 3. Let her also announce their bases ($b_{\unused}^{\rm Alice}$ and 
$b_{\unused}^{\rm Bob}$) and 
bits-values $i_{\unused}$.
Obviously, such announcements can only help Eve to gain more
information (and maybe even to chose a better attack).
Thus this step only reduces the security, so if the protocol defined
here is secure, so is the original BB84 protocol.

\item 
Let Alice generate and send to Bob
only the used bits in step 4, 
and let her ask Eve to send the un-used bits (by telling her
which these are, and also the preparation data for
the relevant subsets, that is---$b_{\unused}^{\rm Alice}$ and 
$i_{\unused}$).
Knowing which are the used bits, and knowing their bases 
and values 
can only help Eve in designing her attack, 
thus security can only be
reduced by this step.

Since Bob never uses the values of the unused bits in the  
protocol (he only ignores them),
he doesn't care if Eve doesn't provide him these bits or provide them to
him without following Alice's preparation request. 

After Bob receives the used and unused bits, let him give Eve the
unused qubits (without measuring them),
and ask her to measure them in bases $b_{\unused}^{\rm Bob}$.
Having these qubits can only help Eve in designing her optimal final 
measurement,
thus security can only be reduced by this step.

Since Bob never use the values of the unused bits 
in the rest of the protocol,
he doesn't care if Eve doesn't provide him these values correctly or at
all.

\item

Since Alice and Bob never made any use of the unused bits, 
Eve could have
them as part of her ancilla to start with, 
and Alice could just create $2n$
bits, send them to Bob, and then tell him the bases.

The protocol obtained after this reduction, is a protocol in which 
Eve has
full control on her qubits and on the unused qubits.
Alice and Bob have control on the preparation and measurement of 
the used
bits only.  This is the used-bits BB84, for which we prove 
security in the
text.

\end{itemize}

One important remark is that the exponentially small 
probability that $n' <  2n$ 
in Step~\ref{it:abort} 
(so that the protocol is aborted due to
insufficient number of bits in the sifted key) now becomes a
probability that Eve learns the key.

Another important remark is that the issue of high loss rate of qubits
(e.g., due to losses in transmission or detection) can also be handled
via the same reduction.
Thus, our proof could apply also to a more practical BB84 protocol 
where high losses are allowed.  
The required modification to the protocol then is that Bob now will not add
missing qubits, in step I.3 of the used-bits BB84 protocol, and in an additional
step (prior to step I.4.) he will inform Alice of the bits he did not obtain.

By the way, another practical aspect
is imperfect sources (in which the created states are not
described by a two-level system). This subject
is the issue of recent subtlety
regarding the security of practical schemes~\cite{BMS98,BLMS99},
and it is not discussed in
this current work.

\section{Information Theoretic Basics and Results}\label{app:Info-theory}

\subsection{Basics of information theory~\cite{CoverThomas}}
\label{APP:info-basics}

Let $\mrv{X}$ and $\mrv{Y}$ be random variables whose values are indexed
by $x$ and $y$ respectively, appearing with probabilities
$p(x)$ and $p(y)$.
The entropy of a random variable is 
$H(\mrv{X}) = - \sum_x p(x) \log_2 p(x)$. For two variables
$H(\mrv{X}|y) = - \sum_x p(x|y) \log_2 p(x|y)$ and
$H(\mrv{X}|\mrv{Y}) \equiv  \sum_y p(y) H(\mrv{X}|y)$.
For any two random variables $\mrv{X}$ and $\mrv{Y}$, the mutual information
$I(\mrv{X};\mrv{Y}) = H(\mrv{X}) - H(\mrv{X} \ |\ \mrv{Y})$ 
describes the decrease in the entropy of $\mrv{X}$
due to learning $\mrv{Y}$; 
This function $I$ is symmetric to swapping $\mrv{X}$ and $\mrv{Y}$.

For three random variables $\mrv{A}$, $\mrv{E}$, and $\mrv{X}$ given to be $x$, 
the conditional 
mutual information is
$I(\mrv{A};\mrv{E} \mid x) =  
 H(\mrv{A} \mid x) - H(\mrv{A} \mid  \mrv{E}, x)$ 
Then, the conditional 
mutual information for the three random variables is
$I(\mrv{A};\mrv{E} \mid \mrv{X}) \equiv \sum_x p(x)
I(\mrv{A};\mrv{E} \mid x)$. Another case which is relevant is
with four random variables $\mrv{A}$, $\mrv{E}$, $\mrv{X}$ and $\mrv{Y}$ 
given to be equal to $y$, 
$I(\mrv{A};\mrv{E} \mid \mrv{X},y) = \sum_x p(x|y)
I(\mrv{A};\mrv{E} \mid x,y)$.

An important tool is the chain rule
$I(\mrv{A},\mrv{B} ; \mrv{C}) = 
I(\mrv{A} ; \mrv{C}) + 
I(\mrv{B};\mrv{C} \mid \mrv{A}) $. 
As a corollary from the chain rule and the positivity of mutual information,
one gets
$I(\mrv{A},\mrv{B} ; \mrv{C}) \geq
I(\mrv{B};\mrv{C} \mid \mrv{A}) $. 

\subsection{Bad Security Criteria}\label{APP:bad-sec-criteria}

\subsubsection{A first bad security criterion and the SWAP attack:}

What one might like to obtain as a security criterion is
that Eve's information given that the test is passed, is negligible.
Formally, this puts a restriction on the values of $j_T$:
for any $i_T$, 
only $j_T$ such that
$|j_T \oplus i_T| \leq n p_a$ are allowed. Then, the criterion is
\begin{equation}
I(\mrv{A};\mrv{E}\mid \mrv{I}_T,\mrv{J}_T,\mrv{B},\mrv{S},\mrv{\Xi},
  \mrv{T}=\pass)
\le A \ e^{-\beta n} 
\end{equation}
with $A$ and $\beta$ positive constants,
and $I(\mrv{A};\mrv{E} \mid \mrv{I}_T,\mrv{J}_T,\mrv{B},\mrv{S},\mrv{\Xi},
  \mrv{T}=\pass)
= \sum_{i_T,j_T,b,s,\xi} p(i_T,j_T,b,s,\xi \mid \mrv{T}={\pass})
I(\mrv{A};\mrv{E} \mid i_T,j_T,b,s,\xi, \mrv{T}={\pass})$,
with $c_T = i_T \oplus j_T$, and 
$  \mrv{T}=\pass$ meaning that $c_T \le n p_a$.

Unfortunately,
the above bound is too demanding and is
\emph{not} satisfied in quantum cryptography.
Given that the test is passed, Eve can still have
full information.  Consider the {\em SWAP
attack}: Eve takes 
Alice's qubits and puts them 
into a quantum memory.  She sends random
BB84 states to Bob. Eve measures the qubits she kept after learning
their bases, hence gets full information about Alice's final key.
In this case, 
Bob will almost always abort the protocol because
it is very unlikely that his bits will pass the test.  However,
in the rare event when the test is passed,
Eve has full information about Alice's key.
So, given
the test is passed (a rare event), information is still $m$ bits,
and the above criterion cannot be satisfied.  

\subsubsection{A second bad security criteria and the half-SWAP attack:}

Another potential security criterion says the following:
``For any attack, either Eve's average information is negligible
or the probability that the test is passed is negligible''.
Namely, if Eve tries an
attack that would give her non-negligible 
information about a final key,
she has to be extremely lucky in order to pass the test.
This security criterion can be formally written as
$\langle \mrv{I}_{Eve} \rangle P(\mrv{T} = \pass)  
\le A \  e^{-\beta n}$ 
with $A$ and $\beta$ positive constants.
This suggested security criterion is different from 
the previously suggested one, 
and it is satisfied by the SWAP attack mentioned above.

Unfortunately, as observed in an earlier
(archive) version of~\cite{Mayers98},
this criterion is also inappropriate. Consider the {\em half-SWAP attack} in
which Eve does nothing with probability half, and performs the SWAP attack
with probability half.  This half-SWAP attack gives an average information of
exactly m/2, and it passes the test with probability larger than half.
Obviously these two cases, getting a non-negligible information,
and passing the
test with high probability, 
will not happen in the same event, hence security can still be achieved,
but it must be defined
via less demanding criteria, such as those two used in the paper.

\subsection{Alternative Security Criteria}\label{APP:good-sec-condition}

\subsubsection{Finding different expressions for 
$\langle \mrv{I}'_{Eve}\rangle $:}
\label{APP:dealing-with-I'}

First, we prove Eq.(\ref{Average-I-prime-Eve-1})
namely, that 
$\langle \mrv{I}'_{Eve}\rangle =
I(\mrv{A};\mrv{E}\mid \mrv{I}_T,\mrv{J}_T,\mrv{B},\mrv{S},\mrv{\Xi},
  \mrv{T}=\pass)P[\mrv{T}=\pass]$.

By expanding of 
$\langle \mrv{I}'_{Eve}\rangle $
we get:
\begin{align*}
\langle \mrv{I}'_{Eve}\rangle &=
  \sum_{i_T,j_T: |i_T \oplus j_T|\leq n\,p_a} \sum_{b,s,\xi}
     I(\mrv{A};\mrv{E}\mid i_T,j_T,b,s,\xi)p(i_T,j_T,b,s,\xi) \\
 &= \sum_{i_T,j_T,b,s,\xi}
     I(\mrv{A};\mrv{E}\mid i_T,j_T,b,s,\xi)
        p(i_T,j_T,b,s,\xi\mid \mrv{T}=\pass)P(\mrv{T}=\pass) \\
 &= \sum_{i_T,j_T,b,s,\xi}
     I(\mrv{A};\mrv{E}\mid i_T,j_T,b,s,\xi,\mrv{T}=\pass)
        p(i_T,j_T,b,s,\xi\mid \mrv{T}=\pass)P(\mrv{T}=\pass) \\
 &= [\sum_{i_T,j_T,b,s,\xi}
     I(\mrv{A};\mrv{E}\mid i_T,j_T,b,s,\xi,\mrv{T}=\pass)
        p(i_T,j_T,b,s,\xi\mid \mrv{T}=\pass)]P(\mrv{T}=\pass) \\
  &= 
    I(\mrv{A};\mrv{E}\mid \mrv{I}_T,\mrv{J}_T,\mrv{B},\mrv{S},\mrv{\Xi},
   \mrv{T}=\pass)
        P[\mrv{T}=\pass]
\end{align*}
Indeed, $p(i_T,j_T,b,s,\xi \mid \pass)p(\pass) = p(i_T,j_T,b,s,\xi,\pass)$
and this value is equal to $p(i_T,j_T,b,s,\xi)$ if $|i_T\oplus j_T|\leq np_a$
and is 0 otherwise. When the value is not 0, then the condition $\pass$
is automatically satisfied and can be put in the right-hand side of the
mutual information.

Second, we prove
in full details Eq.(\ref{Average-I-prime-Eve-2})
namely, that 
$\langle \mrv{I}'_{Eve}\rangle =
 \sum_{|c_T| \leq n p_a} P[\mrv{C}_T = c_T] 
             I(\mrv{A}; \mrv{E} \mid \mrv{I}_T, \mrv{C}_T = c_T 
              ,\mrv{B},\mrv{S},\mrv{\Xi})$.

Note that 
$\mrv{I}'_{Eve}$ is the random variable equal to
 $I(\mrv{A};\mrv{E} \mid i_T,j_T,b,s,\xi)$ when $|i_T \oplus j_T| \leq n p_a$ 
(i.e. when $\mrv{T} = \pass$) and 
to 0 otherwise.
As a consequence,
\begin{eqnarray*}
\langle \mrv{I}'_{Eve}\rangle &=&
 \sum_{i_T,j_T,b,s,\xi} \mrv{I}'_{Eve}(i_T,j_T,b,s,\xi)p(i_T,j_T,b,s,\xi) \\
&=& \sum_{|c_T| \leq n p_a} \sum_{i_T,b,s,\xi}
                I(\mrv{A}; \mrv{E} \mid i_T, \mrv{C}_T = c_T ,b,s,\xi)
                P\left[i_T,\mrv{C}_T = c_T,b,s,\xi\right]\\ 
&=& \sum_{|c_T| \leq n p_a} \sum_{i_T,b,s,\xi}
                I(\mrv{A}; \mrv{E} \mid i_T, \mrv{C}_T = c_T ,b,s,\xi)
                P\left[i_T,b,s,\xi \mid c_T\right]
                  P\left[\mrv{C}_T=c_T\right] \\ 
&=& \sum_{|c_T| \leq n p_a} P[\mrv{C}_T = c_T] 
             I(\mrv{A}; \mrv{E} \mid \mrv{I}_T, \mrv{C}_T = c_T 
              ,\mrv{B},\mrv{S},\mrv{\Xi})
\end{eqnarray*}

\subsubsection{Security against the Half-SWAP Attack:}
\label{APP:Security-Half-SWAP}

In the half-SWAP attack 
Eve has a probe $\ket{p}$ where $p$ is a $2n$-bit
string. With probability half she applies the unitary transform $U_0\ket{p}\ket{i}_b =
\ket{p}\ket{i}_b$ (she does nothing and then sends $\ket{i}_b$ to Bob) and
with probability half she applies the unitary transform $U_1\ket{p}\ket{i}_b =
\ket{i}_b\ket{p}$ (swap) and sends $\ket{p}$ to Bob, keeping the probe in
the state $\ket{i}_b$. We can present a fully-quantum attack, and let Eve use
an additional single-qubit
probe $\ket{e_0}$ initially in the state $H\ket{0}$,
so that her full probe contains $2n+1$ qubits. 
Her attack is defined by the unitary transform
\begin{align*}
 U\ket{0}\ket{p}\ket{i}_b &= \ket{0}\ket{p}\ket{i}_b \\
 U\ket{1}\ket{p}\ket{i}_b &= \ket{1}\ket{i}_b\ket{p} 
\end{align*}
which means that she uses her additional qubit $|e_0\rangle$ to decide whether
she swaps or not (using $2n$ Controlled-SWAP gates). 
Let us describe 
Eve's
measurement: she measures her new bit $e_0$ in the standard basis and then,
if she gets $e_0 = 1$, she
measures the ``probe'' $\ket{e_1} = \ket{i}_b$
in the basis $b$ and gets $i$, else she measures
her original probe  $\ket{p}$ in the standard basis and gets $p$. Her two outputs 
$(e_0,e_1)$, equal to either $(0,p)$ or $(1,i)$,
define the random variable $\mrv{E} = (\mrv{E}_0, \mrv{E}_1)$ (respectively).
Formulated that way, the half-SWAP
attack fits better our framework. Notice that $p$ and $a$ (Alice's final key) are
completely uncorrelated and that $i$ determines completely $a$ after the ECC and
PA steps are completed.

Now let us look at our security criteria, and observe 
$I(\mrv{A};\mrv{E}\mid \mrv{I}_T,\mrv{J}_T,\mrv{B},\mrv{S},\mrv{\Xi},\pass)
        P[\mrv{T}=\pass]$. Of course $p(\pass) = 1/2$. It is however a big
mistake to believe that 
$I(\mrv{A};\mrv{E}\mid \mrv{I}_T,\mrv{J}_T,\mrv{B},\mrv{S},\mrv{\Xi},\pass)$
is equal to $m$ or $m/2$. Eve's information is equal to $m$ if the
following \textbf{two conditions} are satisfied: 
\begin{itemize}
\item the test is passed 
\item  she applied the SWAP attack
\end{itemize}
otherwise, she gets 0 information. So Eve's information is $m$ times the
probability that both the test is passed and she applied the SWAP attack,
which is equal to 1/2 times the probability of passing the test when she
swaps. This is exponentially small.

In order to make this intuitive reasoning formal,
let us use (a particular case of) the chain rule for 
mutual information (see Appendix~\ref{APP:info-basics}):
\[
  I(\mrv{E} ; \mrv{A}) \equiv 
  I(\mrv{E}_0,\mrv{E}_1 ; \mrv{A}) = I(\mrv{E}_0; \mrv{A}) +
      I(\mrv{E}_1; \mrv{A} \mid \mrv{E}_0)
\]
Now, $\mrv{E}_0$ corresponds to a random bit generated by Eve, independently
of $i$ and thus independently of $a$. 
As a consequence $I(\mrv{E}_0; \mrv{A} \mid i_T,j_T,b,s,\xi) = 0$ and 
thus $I(\mrv{E}_0;\mrv{A}\mid \mrv{I}_T,\mrv{J}_T,\mrv{B},\mrv{S},\mrv{\Xi},\pass) = 0$. 
This implies that
\[
I(\mrv{E};\mrv{A}\mid \mrv{I}_T,\mrv{J}_T,\mrv{B},\mrv{S},\mrv{\Xi},\pass)p(\pass) =
I(\mrv{E}_1;\mrv{A}\mid \mrv{E}_0, \mrv{I}_T,\mrv{J}_T,\mrv{B},\mrv{S},\mrv{\Xi},\pass)p(\pass)
\]
Now
\[
 I(\mrv{E}_1; \mrv{A} \mid \mrv{E}_0,i_T,j_T,b,s,\xi) 
 = \sum_{e_0} 
      I(\mrv{E}_1; \mrv{A} \mid e_0,i_T,j_T,b,s,\xi)p(e_0\mid i_T,j_T,b,s,\xi) 
\]
If $e_0 = 0$ then $\mrv{E}_1$ is just the
dummy output that is independent of $a$ and
and as a consequence
$I(\mrv{E}_1; \mrv{A} \mid e_0,i_T,j_T,b,s,\xi) = 0$.
On the other hand, if $e_0 = 1$ (written ``swap'' hereunder) then, Eve gets full information, i.e.
the $m$ bits of the key.
We are thus left with the equality
\[
I(\mrv{E}; \mrv{A} \mid i_T,j_T,b,s,\xi) = 
       m\, p(\swap \mid i_T,j_T,b,s,\xi)
\]
where, of course, Bob's outputs $j_T$ will depend heavily on the swap!
We can now expand
\begin{align*}
I(\mrv{E}; \mrv{A} \mid \mrv{I}_T,\mrv{J}_T,\mrv{B},\mrv{S},\mrv{\Xi},\pass) p(\pass)
 &= m \sum_{|i_T\oplus j_T|\leq n p_a} \sum_{b,s,\xi}  p(\swap \mid i_T,j_T,b,s,\xi)p(i_T,j_T,b,s,\xi) \\
&= m\, p(\swap \wedge \pass) \\
&= m\, p(\pass \mid \swap) p(\swap) \\
&= \frac{m}{2} p(\pass \mid \swap)
\end{align*}
which is exponentially small.

In fact, the half-SWAP attack
does not even make 
$I(\mrv{E}; \mrv{A} \mid \mrv{I}_T,\mrv{J}_T,\mrv{B},\mrv{S},\mrv{\Xi},\pass)$
large since this is equal to
\[
\frac{m}{2} p(\pass \mid \swap) \frac{1}{p(\pass)} = m\, p(\pass \mid \swap)
\]
meaning that the first inappropriate security criteria is actually satisfied
correctly if the Half-SWAP attack is used.

\section{A Few Technical Lemmas}\label{app:lemmas}

\subsection{A Proof of Lemma~\ref{lemma-symm}}
\label{APP:info-symm-not-decr}

We prove here~Eq.\ref{trivial-better-origin}.
It is actually possible to prove equality\footnote{
This is done by proving that 
$I(\mrv{A}; \mrv{M} \mid \mrv{I}_T,\mrv{C}_T=c_T, b, 
s, \mrv{\Xi}) = 0$. See the chain rule used in the first inequality
below.
}, 
but for our purpose inequality is as
good, so we do not bother with proving equality. 
\begin{proof}

Using the chain rule described 
in Appendix~\ref{APP:info-basics}, we get
\begin{eqnarray*}
\lefteqn{
I(\mrv{A}; \mrv{E}' ,\mrv{M} \mid \mrv{I}_T,\mrv{C}_T=c_T, b, s, \mrv{\Xi})} \\ 
 &\geq&
I(\mrv{A}; \mrv{E}' \mid \mrv{M}, \mrv{I}_T,\mrv{C}_T=c_T, b, s, \mrv{\Xi}) \\
&=&
 \sum_{i_T,\xi,m} P'\left[i_T,\xi,m \mid c_T,b,s\right]
I(\mrv{A}; \mrv{E}' \mid m, i_T, c_T, b, s, \xi) \\
&=&
 \sum_{i_T,\xi,m} P'\left[i_T,\xi \mid c_T,b,s,m\right]
I(\mrv{A}; \mrv{E}' \mid m, i_T, c_T, b, s, \xi)p(m)
\end{eqnarray*}
For any fixed $m$, the effect of the symmetrizing transformation $S$ 
is to replace $i$ by $i \oplus m$,
($c_T$ remaining fixed).
In particular $i_T$ becomes $i_T \oplus m_T$ and 
$i_I$ becomes $i_I \oplus m_I$ and so
$\xi$ becomes $(i_I \oplus m_I)P_\mathcal{C}^\top = \xi \oplus 
m_I P_\mathcal{C}^\top$
and so
\begin{eqnarray*}
P'(i_T,\xi \mid c_T,b,s,m) &=& P(i_T \oplus m_T, \xi \oplus m_I P_\mathcal{C}^\top \mid 
  c_T,b,s) \\
I(\mrv{A}; \mrv{E}' \mid m, i_T,\mrv{C}_T=c_T,b,s,\xi) &=&
I(\mrv{A}; \mrv{E} \mid i_T \oplus m_T ,\mrv{C}_T=c_T,b,s,\xi \oplus m_I P_\mathcal{C}^\top)
\end{eqnarray*}
If we let $i'_T = i_T \oplus m_T$, $\xi' = \xi \oplus m_I P_\mathcal{C}^\top$  and use
the fact that the same value of $\xi'$ is obtained $2^{n-r}$ times, we get
\begin{eqnarray*}
\lefteqn{
I(\mrv{A}; \mrv{E}' ,\mrv{M} \mid \mrv{I}_T,\mrv{C}_T=c_T, b, s, \mrv{\Xi})} \\*
 &\geq&
 \sum_{i_T,\xi,m} P'[i_T,\xi \mid c_T,b,s,m]
I(\mrv{A}; \mrv{E}' \mid m, i_T, \mrv{C}_T=c_T, b, s, \xi)p(m) \\*
&=& 
 2^{n-r} \sum_{i_T,\xi,i'_T,\xi'} P[i'_T, \xi' \mid c_T,b,s]
I(\mrv{A}; \mrv{E} \mid i'_T , \mrv{C}_T=c_T, b, s, \xi')p(m)  \\*
&=& 
 2^{n-r} 2^{n+r} \sum_{i'_T,\xi'} P[i'_T, \xi' \mid c_T,b,s]
I(\mrv{A}; \mrv{E} \mid i'_T , \mrv{C}_T=c_T, b, s, \xi')2^{-2n}  \\*
&=& 
 \sum_{i'_T,\xi'} P[i'_T, \xi' \mid c_T,b,s]
I(\mrv{A}; \mrv{E} \mid i'_T , \mrv{C}_T=c_T, b, s, \xi')  \\*
&=&
I(\mrv{A}; \mrv{E} \mid \mrv{I}_T , \mrv{C}_T=c_T, b, s, \mrv{\Xi}) \quad\qed
\end{eqnarray*}
\end{proof}

\subsection{A Proof of Lemma \ref{lemm:indep-of-iI}}
\label{App:Eij-indep-iI}

Using 
the Basic Lemma of Symmetrization 
(Eq.~\ref{eq-lemma-sym-def})
and the fact that the $\ket{m}$ form
an orthonormal basis, 
\begin{equation}\label{Eij-braket}
\braket{E^{\sym\,\prime}_{i,j}}{E^{\sym\,\prime}_{i',j'}}_b 
= 2^{-2n}
 \sum_m (-1)^{(i \oplus j \oplus i'\oplus j') \cdot m}
 \braket{E'_{i\oplus m, j\oplus m}}{E'_{i'\oplus m, j'\oplus m}}_b
\ .
\end{equation}
By replacing $i,j, i'$ and $j'$ 
by $i\oplus u$, $j\oplus u$, $i' \oplus u$ and $j' \oplus u$ in this
formula, we get
$\braket{E^{\sym\,\prime}_{i\xor u 
,j\xor u}}{E^{\sym\,\prime}_{i'\xor u,j'\xor u}}_b 
= 2^{-2n}
 \sum_m (-1)^{(i \oplus j \oplus i'\oplus j') \cdot m}
 \braket{E'_{i\xor u\oplus m, j \xor u\oplus m}}{E'_{i'\xor u\oplus m, j'\xor u\oplus m}}_b$.
Defining $w = u\xor m$ we get
$\braket{E^{\sym\,\prime}_{i\xor u 
,j\xor u}}{E^{\sym\,\prime}_{i'\xor u,j'\xor u}}_b 
= 2^{-2n}
 \sum_w (-1)^{(i \oplus j \oplus i'\oplus j') \cdot w\xor u}
 \braket{E'_{i\xor w, j \xor w}}{E'_{i'\xor w, j'\xor w}}_b$, 
and using (\ref{Eij-braket}) we finally get
\begin{equation}
\label{basic-symm-eq}
\braket{E^{\sym\,\prime}_{i \oplus u,j\oplus u}}{E^{\sym\,\prime}_{i'\oplus u, j'\oplus u}}_b =
  (-1)^{(i \oplus j \oplus i' \oplus j')\cdot u}
  \braket{E^{\sym\,\prime}_{i,j}}{E^{\sym\,\prime}_{i',j'}}_b
\ .
\end{equation}

Considering information and test bits,
if we let $u = u_I u_T$ with $u_T = 0$ and 
use the fact (Lemma~\ref{lemm-indep-it})
that the normalizing
factor for a symmetrized attack depends only on 
$i_T$, $j_T$, $b$ and $s$ (so we can divide 
both sides by the same
normalization factor),  we deduce from (\ref{basic-symm-eq})
the identity
\begin{equation}
\label{basic-symm-eq-info}
\braket{E^\sym_{i_I \oplus u_I,j_I\oplus u_I}}{E^\sym_{i_I'\oplus u_I, j_I'\oplus
u_I}}_{b,s} =
  (-1)^{(i_I \oplus j_I \oplus i_I' \oplus j_I')\cdot u_I}
  \braket{E^\sym_{i_I,j_I}}{E^\sym_{i'_I,j'_I}}_{b,s}
\ .
\end{equation}

For any $n$-bit string $u_I$, we get by Eq. (\ref{basic-symm-eq-info}),
by letting $i'_I = i_I\oplus k_I$, $j'_I= j_I\oplus k_I$ that $(i_I\oplus j_I\oplus i'_I\oplus j'_I)\cdot u_I = 0$ and so
\[
 \braket{E^\sym_{i_I \oplus u_I,j_I \oplus u_I}}{E^\sym_{i_I \oplus k_I \oplus u_I,j_I \oplus k_I \oplus u_I}}_{b,s} = \braket{E^\sym_{i_I,j_I}}{E^\sym_{i_I\oplus k_I,j_I\oplus k_I}}_{b,s}
\ . \]

By writing $j= i\xor c$ we get 
\[
 \braket{E^\sym_{i_I \oplus u_I,j_I \xor c_I \oplus u_I}}{E^\sym_{i_I \oplus k_I \oplus u_I,i_I \xor c_I \oplus k_I \oplus u_I}}_{b,s} = \braket{E^\sym_{i_I,i_I \xor c_I}}{E^\sym_{i_I\oplus k_I,i_I\oplus 
c_I \oplus k_I}}_{b,s}
\ , \]
so that the first part of the Lemma is proven
[$\langle E_{i_I,i_I\xor c_I}^{\sym} | 
              E_{i_I\xor k_I,i_I\xor c_I\xor k_I}^{\sym} \rangle$ 
is independent of $i_I$.]

Summing over $c_I$ 
and changing back to $j_I$ we get that 
$\sum_j\langle E_{i_I,j_I}^{\sym} | 
              E_{i_I\xor k_I,j_I\xor k_I}^{\sym} \rangle$ 
is also independent of $i_I$.

\subsection{A Proof of 
Eq. (\ref{eq-pjtit-indep})}\label{proof-eq-pjtit-indep}

We show that 
$p(j_T \mid i_T,b_I,b_T,s) = p(j_T \mid i_T, b'_I, b_T,s)$
for any choice of basis $b'_I$ on information bits.
For any 
basis $b'_I$, the change of basis between $b'_I$ and 
$b_I$ is expressed by a unitary matrix $U = (u_{i'_I,i_I})$ such that
$\kete{i'_I}{b'_I} = \sum_{i_I} u_{i'_I,i_I} \kete{i_I}{b_I}$,
$\kete{i_I}{b_I} = \sum_{i'_I} {u^\dagger_{i_I,i'_I}} \kete{i'_I}{b'_I}$ and, of course, $UU^\dagger = U^\dagger U = 1$.
{}From the defining equation 
$ \ketb{E'_{i_T,i_I,j_T,j_I}} =
\brab{j_T}\brab{j_I}U \zst{E}\ketb{i_T}\ketb{i_I} $
(Eq.~\ref{Eve-E_iTiIjTjI}) and the above, we get
\begin{equation}\label{e-prime-b-prime}
\kete{E'_{i_T,i'_I,j_T,j'_I}}{b_T,b'_I} =
 \sum_{i_I,j_I} u_{i'_I,i_I} u^\dagger_{j_I,j'_I}
   \kete{E'_{i_T,i_I,j_T,j_I}}{b_T,b_I}
\end{equation}

For any $b$, we have $p(j_T \mid i_T,b,s) =
\sum_{i_I} p(j_T \mid i_T,i_I,b,s)p(i_I \mid i_T,b,s)$.  
As $p(i_I \mid i_T,b,s) = 1/2^n$ (since these values are chosen at random by Alice) we can deduce, using Eq.~(\ref{eq-pj-non-symm}) 
$p(j_T|i_T,i_I,b,s) = 
\sum_{j_I} \braket{E'_{i_T,i_I,j_T,j_I}}{E'_{i_T,i_I,j_T,j_I}}_b$,
that 
\begin{equation}\label{p-jt-mid-it}
p(j_T \mid i_T, b,s) = 
\frac{1}{2^n} \sum_{j_I,i_I}
\braket{E'_{i_T,i_I,j_T,j_I}}{E'_{i_T,i_I,j_T,j_I}}_b
\end{equation}

If we apply Eq. (\ref{p-jt-mid-it}) in the particular case where
the basis is $b'_I,b_T$, 
and we expand its right-hand side using
Eq.~(\ref{e-prime-b-prime}), then, because of the unitarity of $U$, 
the six sums reduce to two, yielding a term that is exactly equal to 
the right-hand side of Eq.~(\ref{p-jt-mid-it}) 
with basis $b = b_I,b_T$. That is:
\begin{equation}\label{p-jt-mid-it-b'}
p(j_T \mid i_T, b'_I,b_T,s) = 
\frac{1}{2^n} \sum_{j_I,i_I}
\braket{E'_{i_T,i_I,j_T,j_I}}{E'_{i_T,i_I,j_T,j_I}}_b
\end{equation}
\qed

\subsection{A Proof of Lemma \ref{d-sqr-is-p-err}}
\label{app:d-sqrt-is}

We start from Eq. (\ref{use-symmetry-bis}), namely
\begin{equation} \label{app-eq:use-symmetry-bis}
P^\sym\left[\mrv{C}_I = c_I \mid i_T, j_T, b^0, s\right] =
  \frac{1}{2^n}\sum_{i'_I}\braket{E^\sym_{i'_I,i'_I\xor c_I}}{E_{i'_I,i'_I\xor c_I}}_{b^0,s}.
\end{equation}
with $b^0 = b \oplus s$. From Hadamard, we know that the unitary
matrix $U = (u_{i'_I,i_I})$ used to express $\kete{i'_I}{\bar{b}_I}$ in terms
of the $\kete{i_I}{b_I}$ is defined by $u_{i'_I,i_I} =
 2^{-n/2}(-1)^{i'_I \cdot i_I}$ and, for that particular choice of $b'_I$, 
Eq. (\ref{e-prime-b-prime}) reduces to 
\[
\kete{E^{\sym\, \prime}_{i_T,i'_I,j_T,j'_I}}{b_T,\bar{b}_I} = \frac{1}{2^n}
 \sum_{i_I,j_I} (-1)^{i'_I \cdot i_I} (-1)^{j_I \cdot j'_I}
   \kete{E^{\sym\, \prime}_{i_T,i_I,j_T,j_I}}{b_T,b_I}
\]
Due to Corollary~\ref{coroll:jT-given-iTbTs} 
$p^\sym(j_T|i_T,b_T,s)$ is independent of $b_I$,  
so both sides 
can be divided by the same normalization 
factor, and this
implies that 
\[
\kete{E^\sym_{i'_I,j'_I}}{b^0,s} = \frac{1}{2^n}
 \sum_{i_I,j_I} (-1)^{({i'_I \cdot i_I} + {j_I \cdot j'_I})}
   \kete{E^\sym_{i_I,j_I}}{b,s} \ .
\]
Then, going back to Eq. (\ref{app-eq:use-symmetry-bis}) and replacing $\kete{E^\sym_{i'_I,i'_I \oplus c_I}}{b^0,s}$ 
by those values, leaves 
\begin{eqnarray*}
& & P^\sym[\mrv{C}_I= c_I \mid i_T,j_T,b^0,s] \\
 &=& \frac{1}{2^n}\sum_{k_I} \sum_{i_I,j_I} \sum_{i'_I,j'_I}\frac{1}{2^{2n}}
(-1)^{(i_I\xor i'_I)\cdot k_I \oplus (j_I\xor j'_I)\cdot (k_I\xor c_I)}
\braket{E^\sym_{i_I,j_I}}{E^\sym_{i'_I,j'_I}}_{b,s}\\
 &=& \frac{1}{2^{3n}}\sum_{i_I,i'_I,j_I,j'_I}
        \left(\sum_{k_I} (-1)^{k_I\cdot (i_I\xor i'_I \xor j_I\xor j'_I)}
        \right)
        (-1)^{c_I\cdot (j_I\xor j'_I)}
        \braket{E^\sym_{i_I,j_I}}{E^\sym_{i'_I,j'_I}}_{b,s}\\
\txtline{The sum over $k_I$ is non zero only when
        $i_I\xor i'_I = j_I \xor j'_I \stackrel{\Delta}{=} h_I$, }
\txtline{and then it is
        $2^n$, so}
 &=& \frac{1}{2^{2n}}\sum_{i_I,j_I,h_I} (-1)^{c_I\cdot h_I}
        \braket{E^\sym_{i_I,j_I}}{E^\sym_{i_I\xor h_I,j_I\xor h_I}}_{b,s}\\
&=&\braket{\eta_{c_I}}{\eta_{c_I}} = d_{c_I}^2
\end{eqnarray*}
where the last equalities are due to the calculation
of the norm of $\eta$
in Eq.~(\ref{d-sqr}). 

\subsection{A Proof of Proposition~\ref{info-m-prop1}}
\label{APP:m-bits}

We prove here Proposition~\ref{info-m-prop1}
that claims a bound on the $m$-bit key given a bound on 1-bit key.
\begin{proof}
Let $F(x) = 2\sqrt{P^\sym[|\mrv{C}_I| \geq \frac{x}{2} \mid i_T, j_T, b^0,s]}$.
For
any $r'$ such that $r \leq r' < r+m$, let $\mathcal{C}'$  be the 
code whose parity check
matrix $P_{\mathcal{C}'}$ has the rows $v_1, \ldots, v_{r'}$. Then
$P_{\mathcal{C}'}$ has rank $r'$ and $\mathcal{C}'$ is an $(n,k',d')$
code with $k' = n -r'$. Moreover $v_{r'+1} \notin \mathcal{C'}^\perp =
V_{r'}$.
As a consequence, Proposition \ref{sd-prop} applies and gives that
\[
I(\mrv{A}' ; \mrv{E}^\sym \mid i_T, j_T, b,s,\xi') \leq
  F(\hat{v}_{r'+1})
\]
for $a' = v_{r'+1} \cdot i_I = a_{j+1}$ with $j = r'-r$, 
$\xi' = i_I P_{\mathcal{C}'}^\top = 
  \xi_1\ldots \xi_r a_1\ldots a_j$,
$\hat{v}_{r'+1} = d_H(v_{r'+1}, V_{r'})$ and $b^0 = b\oplus s$.
This can be rewritten
\[
I(\mrv{A}_{j+1} ; \mrv{E}^\sym \mid i_T, j_T, b,s,\xi, a_1 \ldots a_j) \leq
  F(\hat{v}_{r'+1})
\]
and the result follows from Lemma \ref{lemm-info} by taking 
$F = \max_{r \leq r' < r+m} F(\hat{v}_{r'+1}) = F(\hat{v})$ for
$\hat{v} = \min_{r \leq r' < r+m} \hat{v}_{r'+1}$.
\hspace*{\fill}\qed
\end{proof}

\subsection{A Proof of Lemma \ref{sec5-lemm1}}
\label{APP:sec5-lemm1} 

The Lemma says:
\begin{eqnarray*}
\lefteqn{\sum_{|c_T|\le n p_a}P^\sym\left[\mrv{C}_T=c_T| b,s\right]
I(\mrv{A};\mrv{E^\sym}\ |\ \mrv{I}_T,\mrv{C}_T=c_T,b,s,\mrv{\Xi}) } \\
&\le& 2m\sqrt{  
P^\sym\left[(|\mrv{C}_I|>\frac{\hat{v}}{2})\wedge 
(\frac{|\mrv{C}_T|}{n}\le p_{a}) \mid b^0,s \right]}
\end{eqnarray*}

\begin{proof} 
If we expand $\mrv{I}_T$ and $\mrv{\Xi}$ in the expression
$I(\mrv{A};\mrv{E^\sym}\ |\ \mrv{I}_T,\mrv{C}_T=c_T,b,s,\mrv{\Xi})$ 
then we get
\begin{align*}
\sum_{|c_T|\le n p_a}& P^\sym\left[\mrv{C}_T=c_T| b,s\right]
I(\mrv{A};\mrv{E^\sym}\ |\ \mrv{I}_T,\mrv{C}_T=c_T,b,s,\mrv{\Xi})  \\
&=\sum_{|c_T|\le n p_a,i_T,\xi} p^\sym(i_T,C_T=c_T,\xi\mid b,s) 
  I(\mrv{A}; \mrv{E}^\sym \mid i_T,c_T,b,s,\xi) \\
&=\sum_{|i_T\oplus j_T|\le n p_a,i_T,\xi} p^\sym(i_T,j_T,\xi\mid b,s) 
  I(\mrv{A}; \mrv{E}^\sym \mid i_T,j_T,b,s,\xi) \\
&=\sum_{|i_T\oplus j_T|\le n p_a,i_T,\xi} p^\sym(i_T,j_T \mid b^0,s) 2^{-r}
  I(\mrv{A}; \mrv{E}^\sym \mid i_T,j_T,b,s,\xi) 
 \ .
\end{align*}
The last equality requires a detailed explanation:
First, notice that $p^\sym(j_T \mid i_T, b, s, \xi) =
p^\sym(j_T \mid i_T,b,s)$ because the probability 
$p^\sym(j_T \mid i_T, b,s, i_I)$ is
independent of $i_I$ by Lemma~(\ref{lemm-indep-it})  
and the condition
$\mrv{\Xi} = \xi$ means $i_I P_\mathcal{C}^\top = \xi$, which is a condition
on $i_I$. As a consequence, using the fact that 
(for any attack) $p(\xi \mid b,s) =
2^{-r}$, $p(i_T \mid b, s) = 2^{-n}$ and 
$p(i_T, \xi  \mid b,s) =  p(i_T \mid b, s) p(\xi \mid b,s) $ 
[so that 
$p(i_T, \xi  \mid b,s) =  2^{-(n+r)}$],
we get
\begin{align*}
p^\sym(i_T,j_T,\xi \mid b, s) &=  
p^\sym(j_T \mid i_T, b, s, \xi) p^\sym(i_T, \xi \mid b ,s) & \\
&= p^\sym(j_T \mid i_T, b, s, \xi)2^{-(n+r)} & \\
&= p^\sym(j_T \mid i_T, b, s)2^{-(n+r)}   &\text{by the above}\\
&= p^\sym(j_T \mid i_T,b^0, s)2^{-(n+r)} &\text{by Lemma~(\ref{lemm:pjtit-indep-sym})} \\
&= 2^{-(n+r)} [p^\sym(i_T,j_T \mid b^0, s)/
p^\sym(i_T \mid b^0, s)]  &\text{by definition of $p(A\mid B)$} \\
&= p^\sym(i_T, j_T \mid b^0, s)2^{-r}  
&\text{due to  $p(i_T \mid b^0,s)= 2^{-n}$} 
\end{align*}
The result 
\begin{align*}
\sum_{|c_T|\le n p_a}& P^\sym\left[\mrv{C}_T=c_T| b,s\right]
I(\mrv{A};\mrv{E^\sym}\ |\ \mrv{I}_T,\mrv{C}_T=c_T,b,s,\mrv{\Xi})  \\
&=\sum_{|i_T\oplus j_T|\le n p_a,i_T,\xi} p^\sym(i_T,j_T \mid b^0,s) 2^{-r}
  I(\mrv{A}; \mrv{E}^\sym \mid i_T,j_T,b,s,\xi) \\
&\le\sum_{|i_T\oplus j_T|\le n p_a,i_T,\xi} 2^{-r} p^\sym(i_T,j_T \mid b^0,s) 
2m \sqrt{P^\sym\left[ |\mrv{C}_I| \geq \frac{\hat{v}}{2} \mid i_T,j_T,b^0,
     s\right]}
 \ 
\end{align*}
now follows immediately from corollary~(\ref{info-m-coro}). 
Using the fact
that square-root is a convex function
$\sum p_i \sqrt{x_i} \leq \sqrt{\sum p_i x_i}$ 
so we get    
\begin{align*}
\sum_{|c_T|\le n p_a}& P^\sym\left[\mrv{C}_T=c_T| b,s\right]
I(\mrv{A};\mrv{E^\sym}\ |\ \mrv{I}_T,\mrv{C}_T=c_T,b,s,\mrv{\Xi})  \\
&\le  2m \sqrt{\sum_{|i_T\oplus j_T|\le n p_a,i_T,\xi} 2^{-r} p^\sym(i_T,j_T \mid b^0,s) 
P^\sym\left[ |\mrv{C}_I| \geq \frac{\hat{v}}{2} \mid i_T,j_T,b^0,
     s\right]}
 \ 
\end{align*}
Finally, 
we get rid of the $2^{-r}$ factor by summing over $\xi$ 
(each equally likely), and 
we complete the proof using 
\begin{align*}
  \sum_{|i_T \oplus j_T| \leq np_a ,i_T,\xi}& 
P^\sym\left[|\mrv{C}_I| \geq \frac{\hat v}{2}
\mid i_T,j_T,b^0,s\right] p^\sym(i_T,j_T \mid b^0, s) 2^{-r}
   \\
&=
 P^\sym\left[|\mrv{C}_I| \geq \frac{\hat{v}}{2}, |\mrv{C}_T|
  \leq np_a \mid b^0, s\right] & \qed
\end{align*}

\end{proof}

\subsection{A Proof of Lemma \ref{sec5-lemm4}}
\label{law-large-num}
Let
\[
P\left[\left(\frac{|\mrv{C}_I|}{n} > p_a + \epsilon \right) \wedge
         \left(\frac{|\mrv{C}_T|}{n} \leq p_a\right) \right]
=\sum_b p(b)h_b(p_a,\epsilon) 
\]
with
\[
   h_b(p_a,\epsilon)
    = 
   P\left[\left(\frac{|\mrv{C}_I|}{n} > p_a + \epsilon \right) \wedge
         \left(\frac{|\mrv{C}_T|}{n} \leq p_a\right) \mid b \right]
\]
This $h_b(p_a,\epsilon)$
is the probability that the information bits have $\epsilon$
more than the allowed error rate, when the test bits have less than the
allowed error rate averaged over all choices of test and 
information bits, for a particular basis $b$, and is given by  
\[  \sum_{c} P\left[\left(\frac{|\mrv{C}_I|}{n} > p_a + \epsilon\right) \wedge
         \left(\frac{|\mrv{C}_T|}{n} \leq p_a\right) \mid \mrv{C}=c,b \right]
         P\left[\mrv{C}=c \mid b\right]
\]
where $c$ is over all possible error strings on all bits, test and
information. Note that in principle $P\left[\mrv{C}=c \mid b\right]$,
can be calculated but we shall soon see that there is no need for it.

Now we must note that 
$$P\left[\left(\frac{|\mrv{C}_I|}{n} > p_a + \epsilon\right) \wedge
         \left(\frac{|\mrv{C}_T|}{n} \leq p_a\right) \mid \mrv{C}=c,b \right]$$
does not depend on the attack. And in fact, in the aforementioned expression,
the basis $b$ is superfluous. Once the error string $c$ is fixed, the values
$\frac{|c_I|}{n}$ and $\frac{|c_T|}{n}$ depend uniquely on the 
random string $s$. In fact $\frac{|c_I|}{n}$ is the average of a random
sampling without replacement of $n$ bits taken from the $2n$ bits $c$
whose mean $\mu$ is $\frac{|c|}{2n}$. From 
Hoeffding \cite{Hoeffding} we know
that
\begin{equation}
\label{first:hoeffding}
P\left[\frac{|\mrv{C}_I|}{n}- \mu \ge \frac{\epsilon}{2} \mid c,b \right]\le
e^{-\frac{1}{2} n\epsilon^2}
\end{equation}
By definition $|c| = |c_I| + |c_T|$ and so 
\[
\mu = \frac{|c|}{2n} = \frac{|c_I|}{2n} + \frac{|c_T|}{2n}
\]
Replacing $\mu$ by its value in (\ref{first:hoeffding}) and simplifying, 
equation (\ref{first:hoeffding}) becomes
\begin{equation}
\label{good:hoeffding}
P\left[\frac{|\mrv{C}_I|}{n} \ge
       \frac{|\mrv{C}_T|}{n} + \epsilon \mid \mrv{C}=c,b \right] \le
e^{-\frac{1}{2} n\epsilon^2}
\end{equation}
Now, since 
\[
   (\frac{|c_I|}{n} > p_a + \epsilon ) \wedge
         (\frac{|c_T|}{n} \leq p_a) \Longrightarrow
    \frac{|c_I|}{n} \ge \frac{|c_T|}{n} + \epsilon
\]
we deduce from (\ref{good:hoeffding}) that
\[
   P\left[\left(\frac{|\mrv{C}_I|}{n} > p_a + \epsilon\right) \wedge
         \left(\frac{|\mrv{C}_T|}{n} \leq p_a\right) \mid \mrv{C}=c,b \right]
         \leq 
e^{-\frac{1}{2} n\epsilon^2}
\]
and consequently,
\[
 h_b(p_a,\epsilon) = 
   P\left[\left(\frac{|\mrv{C}_I|}{n} > p_a + \epsilon \right) \wedge
         \left(\frac{|\mrv{C}_T|}{n} \leq p_a\right) \mid b \right] \leq
e^{-\frac{1}{2} n\epsilon^2}
\] and
\[
P\left[\left(\frac{|\mrv{C}_I|}{n} > p_a + \epsilon \right) \wedge
         \left(\frac{|\mrv{C}_T|}{n} \leq p_a\right) \right]
 \leq 
e^{-\frac{1}{2} n\epsilon^2}
\]

\section{Eve's Information Versus the disturbance}
\label{app:BBBGM}  

In this appendix we do not prove Lemma~\ref{sd-lemm} immediately.
We prove it later on, in the second
subsection (the tight bound). 
For simplicity of the presentation, we first prove
another Lemma which leads to a loose bound (with an additional 
factor of $2^r$), for which the derivation is simpler.
The bulk of the loose bound was derived in~\cite{BBBGM}, and is adapted here
to the analysis of the joint attack. The tight
bound is an improvement over that derivation yielding a much better
threshold for $p_{\allowed}$:
The loose bound leads to a threshold of  
less than 1\%, while the threshold for the tight bound
is 7.56\%.
One can skip directly to
the second subsection if desired.

Both the loose and the tight bound are derived using   
the fact that
the Shannon distinguishability between the parity 0 density matrix,
$\rho_0$, and the parity 1 density matrix, $\rho_1$,  
is bounded 
(\cite{BBBGM,FG})
by the trace norm of $\rho_0-\rho_1$ 
and using the fact 
that  
we can easily calculate this trace norm when the purified states
are given by Eq.~(\ref{eq:phi}). 

\subsection{The Loose Bound}
\label{APP:BBBGM-ext}

Exploiting the techniques developed in~\cite{BBBGM}
(to prove security against any collective attack) we now present a bound
which is applicable to the joint attack.

We have already defined a purification of Eve's state:
$\ket{\phi_{i_I}} = \sum_l (-1)^{i_I\cdot l} \ket{\eta_l}$
The density matrix for such a $\ket{\phi_{i_I}}$ is
\begin{equation}
\rho^{i_I}=\ket{\phi_{i_I}}\bra{\phi_{i_I}}
        =\sum_{l,l'}
                (-1)^{i_I \cdot (l\xor l')} d_l d_{l'}
                \ket{\hat\eta_l}\bra{\hat\eta_{l'}}
\end{equation}
Recall that the final key is computed as $b=v\cdot i_I$. 
Eve does not know
$i_I$, but she knows $v$, and she knows
(from the 
announced ECC parity string $\xi$)
that $i_I$ is in the coset 
$\mathcal{C}_\xi$. 
Hence, in order to know
the key, Eve must distinguish between the states 
$i_I = i_\xi \oplus c$ in 
$\mathcal{C}_\xi$ that 
give parity $b=0$ and the states 
$i_I = i_\xi \oplus c$ in 
$\mathcal{C}_\xi$ that 
give parity $b=1$. 
For $b\in \{0,1\}$ the reduced density
matrix is
\begin{eqnarray*}
\rho_b(v,\xi)&=&\frac{1}{2^{n-(r+1)}}
\sum_{\stackrel{c \in \mathcal{C}}{v\cdot (i_\xi \oplus c)=b}} 
\rho^{i_\xi \oplus c}\\
        &=&\frac{1}{2^{n-(r+1)}}
\sum_{\stackrel{c \in \mathcal{C}}{v \cdot (i_\xi \oplus c)=b}}
                \sum_{l,l'}
                (-1)^{(i_\xi \oplus c) \cdot (l\xor l')} d_l d_{l'}
                \ket{\hat\eta_l}\bra{\hat\eta_{l'}}
\end{eqnarray*}
where the sum 
is over values $c$ that satisfy both the condition of being a
code word, and the condition
of leading to the particular parity $b$ for
the PA.

\begin{lemm} 
Let $\mathcal{C}$ be any linear code in $\{0, 1\}^n$ and $a \in \{0,1\}^n$ be 
such that $a \notin \mathcal{C}^{\perp}$ then
\begin{equation} \label{zerosum}
\sum_{c \in \mathcal{C}} (-1)^{c \cdot a} = 0
\end{equation}
\end{lemm}
\begin{proof} 
Let $\{w_1, \ldots, w_k\}$ be a basis of $\mathcal{C}$. Define $t \in \{0,1\}^k$
by $t_\alpha = w_\alpha \cdot a, \ 1 \leq \alpha \leq k$; 
$a \notin \mathcal{C}^{\perp}$ means that $t$ is not the zero string.
Let now 
$h: \{0,1\}^k \rightarrow \mathcal{C}$ be defined by 
$h(s) = \sum_{1 \leq \alpha \leq k} s_{\alpha}w_{\alpha}$; then
$h(s) \cdot a = \sum s_{\alpha} w_{\alpha} \cdot a = \sum s_{\alpha} t_\alpha 
= s \cdot t$ and so
$$\sum_{c \in \mathcal{C}} (-1)^{c \cdot a} =
  \sum_{s} (-1)^{h(s) \cdot a} =
  \sum_{s} (-1)^{s \cdot t} = 0$$ \qed
\end{proof}
\begin{lemm}\label{sd-lemm-bbbgm}
For any $(n,k,d)$ code $\mathcal{C}$ with $r \times n$ parity check matrix
$P_\mathcal{C}$
of rank $r = n - k$, any $\xi \in \{0,1\}^r$ and any 
$v \in \{0,1\}^n$
the Shannon distinguishability $SD(\rho_0(v,\xi),\rho_1(v,\xi))$
where 
\[
\rho_b(v,\xi)=\frac{1}{2^{n-(r+1)}}
\sum_{\stackrel{i_I P_\mathcal{C}^\top = \xi}{i_I\cdot v=b}} \rho^i
\]
between the parity 0 and the parity 1 of the
information bits over any PA string, $v$, 
is bounded above by the following inequality:
\begin{equation}
SD(\rho_0(v,\xi),\rho_1(v,\xi))\le 2^{r+1}\sqrt{
\sum_{|l|\ge\frac{\hat v}{2}}d_l^2}
\ ,
\end{equation}
where $\hat v$ is the  minimum 
distance between $v$ and the code $\mathcal{C}^\perp$, i.e.
the  minimum
weight of $v\xor v'$ for any $v' \in \mathcal{C}^\perp$.
\end{lemm}

\begin{proof}  
The Shannon distinguishability between the parity 0 and the parity 1
is bounded by the trace norm of $\rho_0(v,\xi)-\rho_1(v,\xi)$, see~\cite{BBBGM,FG}.
Let us calculate the required bound:
\begin{eqnarray*}
\lefteqn{\rho_0(v,\xi)-\rho_1(v,\xi)} \\*
     &=&\frac{1}{2^{n-(r+1)}}
         \sum_{c \in \mathcal{C}} (-1)^{(i_\xi \oplus c)\cdot v}
                \sum_{l,l' }
                (-1)^{(i_\xi \oplus c)\cdot (l\xor l')} d_l d_{l'}
                \ket{\hat\eta_l}\bra{\hat\eta_{l'}}\\
        &=&\frac{1}{2^{n-(r+1)}} \sum_{l,l'}
                \left(\sum_{c \in \mathcal{C}}
                        (-1)^{(i_\xi \oplus c) \cdot (l\xor l'\xor v)}\right)
                d_l d_{l'} \ket{\hat\eta_l}\bra{\hat\eta_{l'}}\\
        &=&\frac{1}{2^{n-(r+1)}} \sum_{l,l'}
               (-1)^{i_\xi \cdot (l \oplus l' \oplus v)}
                \left(\sum_{c \in \mathcal{C}}
                        (-1)^{c \cdot (l\xor l'\xor v)}\right)
                d_l d_{l'} \ket{\hat\eta_l}\bra{\hat\eta_{l'}}\\
\end{eqnarray*}
{}From equation (\ref{zerosum}) we know the sum over $\mathcal{C}$ is zero
except when $l \oplus l' \oplus v \in \mathcal{C}^{\perp} = V_r$,
i.e. when $l' = l \oplus v \oplus v_{\mathbf s}$ for 
some $v_{\mathbf s} \in V_r$. As a consequence:
\begin{eqnarray*}
\rho_0(v,\xi)-\rho_1(v,\xi)&=& 2 \sum_{v_s \in V_r } (-1)^{i_\xi \cdot v_s}
  \sum_l d_l d_{l\xor v \xor v_s}
                \ket{\hat\eta_l}\bra{\hat\eta_{l\xor v \xor v_s}}
\end{eqnarray*}
As already said, the trace norm of this matrix serves as a bound on 
the information Eve receives~\cite{BBBGM,FG}.
\begin{eqnarray*}
SD(\rho_0(v,\xi),\rho_1(v,\xi))&\le& \frac{1}{2}Tr|\rho_0(v,\xi)-\rho_1(v,\xi)| 
\end{eqnarray*}

Using the above and making use of the triangle inequality for
the trace norm, the following is obtained (where $SD(\rho_0(v,\xi),\rho_1(v,\xi))$ is
denoted $SD_v$ for short): 
\begin{eqnarray*}
SD_v&\le& 
Tr \left| \sum_{v_s \in V_r} (-1)^{i_\xi \cdot v_s}
  \sum_l d_l d_{l\xor v \xor v_s}
                \ket{\hat\eta_m}\bra{\hat\eta_{m\xor v \xor v_s}} \ \right| \\
&=&\frac{1}{2}Tr\left| \sum_{v_s \in V_r} (-1)^{i_\xi \cdot v_s}
 \sum_{l}
                d_l d_{l\xor v \xor v_s}
                \left(\ket{\hat\eta_l}\bra{\hat\eta_{l\xor v \xor v_s}}
+ \ket{\hat\eta_{l\xor v \xor v_s}}\bra{\hat\eta_l}\right) \right| \\ 
&\le& \sum_{v_s \in V_r}
\sum_l d_l d_{l\xor v \xor v_s}(\ \frac{1}{2}Tr\Big|\ 
\ket{\hat\eta_l}\bra{\hat\eta_{l\xor v \xor v_s}}
+ \ket{\hat\eta_{l\xor v \xor v_s}}\bra{\hat\eta_l}\ \Big|\ ) \\
&=&\sum_{v_s \in V_r }
\sum_l d_l d_{l\xor v \xor v_s}\sqrt{1 -
[\Im(\braket{\hat\eta_l}{\hat\eta_{l\xor v \xor v_s}})]^2}\\
&\le&\sum_{v_s \in V_r }
\sum_l d_l d_{l\xor v \xor v_s}
\end{eqnarray*}
where the sign $\Im$ means the imaginary part.
In the above, we made use of the fact that the trace norm is exactly
computable for needed matrix.
Now we will concern ourselves with bounding each of the terms
$\sum_l d_l d_{l\xor w_s}$,
where $w_s = v \xor v_s$. 
\begin{eqnarray*}
\sum_l d_l d_{l\xor w_s}&=&
        \sum_{|l|>\frac{|w_s|}{2}} d_l d_{l\xor w_s}
        + \sum_{|l|\le\frac{|w_s|}{2}} d_l d_{l\xor w_s} \\
&=&\sum_{|l|>\frac{|w_s|}{2}} d_l d_{l\xor w_s}
        + \sum_{|l'\xor w_s|\le\frac{|w_s|}{2}} d_{l'\xor w_s} d_{l'} \\
\end{eqnarray*}
If $|l' \oplus w_s| \leq \frac{|w_s|}{2}$ then 
$  |w_s| = 
             |l' \oplus w_s \oplus l'| 
        \leq |l' \oplus w_s| + |l'| 
        \leq \frac{|w_s|}{2} + |l'|$ 
and so $|l'| \geq \frac{|w_s|}{2}$.
Therefore, 
\begin{eqnarray*}
\sum_{|l|>\frac{|w_s|}{2}} d_l d_{l\xor w_s}
        + \sum_{|l'\xor w_s|\le\frac{|w_s|}{2}} d_{l'\xor w_s} d_{l'}
&\le&\sum_{|l|\ge\frac{|w_s|}{2}} d_l d_{l\xor w_s}
        + \sum_{|l'|\ge\frac{|w_s|}{2}} d_{l'\xor w_s} d_{l'} \\
&=&2\sum_{|l|\ge\frac{|w_s|}{2}} d_l d_{l\xor w_s} \\
&=&\frac{1}{\alpha}\sum_{|l|\ge\frac{|w_s|}{2}}2 d_l 
(\alpha d_{l\xor w_s}) \\
&\le&\frac{1}{\alpha} \sum_{|l|\ge\frac{|w_s|}{2}} [d_l^2
 + \alpha^2 d_{l\xor w_s}^2] \\
&=&\alpha\sum_{|l|\ge\frac{|w_s|}{2}} d_{l\xor w_s}^2 
+ \frac{1}{\alpha}\sum_{|l|\ge\frac{|w_s|}{2}}d_l^2 \\
\end{eqnarray*}
where the last three steps are true 
for any real $\alpha$, and real $d_l,d_{l\xor w_s}$.

Due to the fact that the $d_l^2$ form a probability distribution,
any sum of them is less than or equal to unity.
\begin{eqnarray*}
\sum_l d_l d_{l\xor w_s}
&\le&\alpha+\frac{1}{\alpha}\sum_{|l|\ge\frac{|w_s|}{2}}d_l^2
\\
&\le&\alpha+\frac{1}{\alpha}\sum_{|l|\ge\frac{\hat{v}}{2}}d_l^2
\end{eqnarray*}
where $\hat{v} = \min_{v_s} | v \xor v_s |$ 
(remember that $w_s = v \oplus v_s$). Summing over all $v_s \in V_r$
and setting $\alpha=\sqrt{\sum_{|l|\ge\frac{\hat{v}}{2}}d_l^2}$ now leaves:
\begin{equation}
SD_v\le 2^{r+1}
\sqrt{\sum_{|l|\ge\frac{\hat{v}}{2}}d_l^2}
\end{equation}\qed
\end{proof}  

Following the proof of the above Lemma, one can guess that it is not a
tight bound since we sum over $2^r$ terms while most
of them do not contribute to the sum (or contribute negligible values).
This understanding led us to 
reach a tighter bound.

\subsection{Eve's Information about one bit -- Tight Bound}
\label{APP:tight-bou}

We will now make a finer analysis of Eve's state after she learns
the parity matrix and 
parity string $\xi$.
We start again from the equality:
\begin{equation}
\ket{\phi_{i_I}} = \sum_l (-1)^{i_I\cdot l} \ket{\eta_l}
\end{equation}
Let $v_1, \ldots, v_r$ be the rows of $P_\mathcal{C}$,  and $v_{r+1} = v$.
It is assumed that the sequence $v_1, \ldots, v_{r+1}$ is linearly 
independent; it can thus be extended to a basis $v_1, \ldots, v_n$ of $\{0,1\}^n$.
For any $r'$ let $V_{r'}$ be the span of $\{v_1, \ldots, v_{r'}\}$ and
$V_{r'}^c$ be the span of $\{v_{r'+1}, \ldots, v_n\}$. For all $r'$, the spaces
$V_{r'}$ and $V_{r'}^c$ are complementary; this means that any element 
$l \in \{0,1\}^n$ has a {\em unique representation}  $l = m \oplus n$
with $m \in V_{r'}^c$ and $n \in V_{r'}$. 

For $\xi \in \{0,1\}^r$, let $i_\xi$ denote some fixed
$n$-bit string such that
$i_\xi P_\mathcal{C}^\top = \xi$ (existence is guaranteed by the fact that
$P_\mathcal{C}$ has maximal rank).  For any $i_I \in \mathcal{C}_\xi$ we
have $(i_I - i_\xi)P_\mathcal{C}^\top = \xi - \xi =0$ and so
$i_I - i_\xi \in \mathcal{C}$ and thus, for any $n \in V_r = \mathcal{C}^\perp$,$(i_I - i_\xi) \cdot n = 0$ i.e. $i_I \cdot n = i_\xi \cdot n$.

Putting those 
remarks together we get:
\begin{eqnarray*}
\ket{\phi_{i_I}}&=& \sum_{m\in V_r^c}\sum_{n\in V_r}
 (-1)^{i_I\cdot (m\xor n)} \ket{\eta_{m\xor n}}\\
&=&\sum_{m\in V_r^c}(-1)^{i_I\cdot m}\sum_{n\in V_r}
 (-1)^{i_I\cdot n} \ket{\eta_{m\xor n}} \\
&=&\sum_{m\in V_r^c}(-1)^{i_I\cdot m}\sum_{n\in V_r}
 (-1)^{i_\xi \cdot n} \ket{\eta_{m\xor n}} \\
&=&\sum_{m\in V_r^c}
                (-1)^{i_I \cdot m} \ket{\eta'_m}
\end{eqnarray*}
where $\eta'_m$ is defined, for each $m \in V_r$, by
\begin{equation}\label{eta-prime-define}
\ket{\eta'_m} = \sum_{n\in V_r}
(-1)^{i_\xi \cdot n} \ket{\eta_{m\xor n}}
\end{equation}

Let us write 
\[ \eta'_m = d'_m \hat\eta'_m \]
with the $\hat{\eta'}_m$s normalized 
so that 
$ d'{}^2_m = \braket{\eta'_m}{\eta'_m}$,
and the density matrix 
for $\ket{\phi_{i_I}}$ reduces to:
\begin{eqnarray*}
\rho^{i_I}&=&\ket{\phi_{i_I}}\bra{\phi_{i_I}}\\
        &=&\sum_{m,m'\in V^c_r}
                (-1)^{i_I \cdot (m\xor m')} d'_m d'_{m'}
                \ket{\hat\eta'_m}\bra{\hat\eta'_{m'}}
\end{eqnarray*}
Due to 
Proposition \ref{prop-eta-orth} (the orthogonality of the 
$\eta_m$s), 
we get that $\braket{\eta_{m \oplus n_1}}{\eta_{m \oplus n_2}}=0$ except
when $n_1 \oplus n_2 = 0$. Together with
Eq.~(\ref{eta-prime-define})   
  this implies  
\begin{equation}\label{eq-d-prime}
 d'{}^2_m = \sum_{n\in V_r} d_{m\xor n}^2 
\ .
\end{equation}

Recall that the final key is computed as $b = v\cdot i_I$. 
Of course, Eve does not
know
$i_I$, but she knows $v$ and she knows (from the announced 
ECC parity string $\xi$) that 
$i_I \in \mathcal{C}_\xi = \{ i_\xi \oplus c \ | \ c \in \mathcal{C} \}$.
Eve wants to determine $b$.
For $b\in \{0,1\}$ the reduced density
matrix is
\begin{eqnarray*}
\rho_b(v,\xi)&=&\frac{1}{2^{n-(r+1)}}
\sum_{\stackrel{c \in \mathcal{C}}{(i_\xi \oplus c) \cdot v=b}} 
 \rho^{i_\xi \oplus c}\\
        &=&\frac{1}{2^{n-(r+1)}}
\sum_{\stackrel{c \in \mathcal{C}}{(i_\xi \oplus c) \cdot v=b}}
                \sum_{m,m'\in V^c_r}
                (-1)^{(i_\xi \oplus c) \cdot (m\xor m')} d'_m d'_{m'}
                \ket{\hat\eta'_m}\bra{\hat\eta'_{m'}}
\end{eqnarray*}
We can now prove
\begin{theopargself}
\begin{nonumlemma}[\ref{sd-lemm}]
The Shannon distinguishability 
between the parity 0 and the parity 1 of the
information bits over any PA string, $v$, 
is bounded above by the following inequality:
\begin{equation}
SD(\rho_0(v,\xi),\rho_1(v,\xi))\le
2\sqrt{
\sum_{|l|\ge\frac{\hat v}{2}}d_l^2}\ ,
\end{equation}
where $\hat{v} = d_H(v, V_r)$ is the  minimum weight of $v\xor v_s$ for any $v_s \in V_r$.
\end{nonumlemma}
\end{theopargself}
\begin{proof}  
The Shannon distinguishability between the parity 0 and the parity 1
is bounded by the trace norm of $\rho_0(v,\xi)-\rho_1(v,\xi)$:
\begin{eqnarray*}
\lefteqn{\rho_0(v,\xi)-\rho_1(v,\xi) = } \\
 & &\frac{1}{2^{n-(r+1)}}
\sum_{c \in \mathcal{C}} (-1)^{(i_\xi \oplus c)\cdot v}
                \sum_{m,m'\in V^c_r}
                (-1)^{(i_\xi \oplus c)\cdot (m\xor m')} d'_m d'_{m'}
                \ket{\hat\eta'_m}\bra{\hat\eta'_{m'}} = \\
& &\frac{1}{2^{n-(r+1)}} \sum_{m,m'\in V^c_r}
                \left(\sum_{c \in \mathcal{C}}
                        (-1)^{(i_\xi \oplus c)\cdot (m\xor m'\xor v)}\right)
                d'_m d'_{m'} \ket{\hat\eta'_m}\bra{\hat\eta'_{m'}} = \\
& &\frac{1}{2^{n-(r+1)}} \sum_{m,m'\in V^c_r}
                (-1)^{i_\xi \cdot (m\xor m'\xor v)}
                \left(\sum_{c \in \mathcal{C}}
                        (-1)^{c \cdot (m\xor m'\xor v)}\right)
                d'_m d'_{m'} \ket{\hat\eta'_m}\bra{\hat\eta'_{m'}}\\
\end{eqnarray*}
Applying equality (\ref{zerosum}) the sum indexed by $c$ is zero except
when $m \oplus m' \oplus v \in \mathcal{C}^\perp = V_r$. But  
$m \oplus m' \oplus v \in V_r^c$ because $m, m' \ {\rm and} \ 
v \in V_r^c$.
This implies $m \oplus m' \oplus v \in V_r \cap V_r^c =
\{0\}$ and thus $m' = m \oplus v$.
Of course, with $m \oplus m' \oplus v = 0$, the sum indexed by $c$ is 
$2^k = 2^{n-r}$ and the coefficient $(-1)^{i_\xi \cdot (m \oplus m' \oplus v)}$
is 1. Therefore $\rho_0(v,\xi) - \rho_1(v,\xi)$ takes the very simple form:
\begin{equation}\label{eq-simple-form}
\rho_0(v,\xi)-\rho_1(v,\xi)=  2 \sum_{m\in V^c_r}
                d'_m d'_{m\xor v}
                \ket{\hat\eta'_m}\bra{\hat\eta'_{m\xor v}}
\end{equation}
We now claim that
\begin{align}
&V_r^c = V_{r+1}^c \cup \{ m\oplus v \mid m \in V_{r+1}^c\} &
  \text{(disjoint union)} \label{eq-disjoint}\\
&\text{if } d_H(m,V_r) < \frac{\hat{v}}{2} \text{ then } d_H(m\oplus v, V_r) \geq
 \frac{\hat{v}}{2} &\text{for any $m \in \{0,1\}^n$} \label{eq-distance}
\end{align}
Claim (\ref{eq-disjoint}) follows from the fact that $v_{r+1} = v$, 
$V^c_{r}$ is the span of
$\{v_{r+1}, \ldots, v_n\}$ and $V^c_{r+1}$ is the span of 
$\{v_{r+2}, \ldots, v_n\}$, and that those elements are all linearly
independent. 
\noindent
As for claim (\ref{eq-distance}) if $d_H(m, V_r) < \hat{v}/2$ and
$d_H(m\oplus v, V_r)  < \hat{v}/2$, then there is $n$ and $n'$ in $V_r$
such that $|m \oplus n| < \hat{v}/2$ and $|m \oplus v \oplus n'| < \hat{v}/2$.
This implies that $|m \oplus n \oplus m \oplus v \oplus n'| < \hat{v}$.
However $m \oplus n \oplus m \oplus v \oplus n' = n \oplus n' \oplus v$
and $n \oplus n' \in V_r$ and this contradicts the fact that 
$\hat{v}= d_H(v,V_r)$

Now, using claim (\ref{eq-disjoint}), we can rewrite Eq (\ref{eq-simple-form}):
\begin{eqnarray*}
\rho_0(v,\xi)-\rho_1(v,\xi)&=& 2 \sum_{m\in V^c_{r+1}}
                d'_m d'_{m\xor v} \left\{
                \ket{\hat\eta'_m}\bra{\hat\eta'_{m\xor v}} +
                \ket{\hat\eta'_{m\xor v}}\bra{\hat\eta'_m}\right\}
\end{eqnarray*}
As usual,
the trace norm of this matrix serves as a bound on the information Eve
receives.
It is
\begin{eqnarray*}
SD(\rho_0(v,\xi),\rho_1(v,\xi))&\le& \frac{1}{2}Tr|\rho_0(v,\xi)-\rho_1(v,\xi)| 
\end{eqnarray*}
Writing $SD_v$ instead of $SD(\rho_0(v,\xi),\rho_1(v,\xi))$
for short:
\begin{eqnarray*}
SD_v
    &\le& Tr \left| \sum_{m\in V^c_{r+1}}
                d'_m d'_{m\xor v} \left\{
                \ket{\hat\eta'_m}\bra{\hat\eta'_{m\xor v}}
               +\ket{\hat\eta'_{m\xor v}}\bra{\hat\eta'_m} \right\}\right| \\
    &\leq& \sum_{m\in V^c_{r+1}}
                d'_m d'_{m\xor v} Tr \left| \,
                \ket{\hat\eta'_m}\bra{\hat\eta'_{m\xor v}}
               +\ket{\hat\eta'_{m\xor v}}\bra{\hat\eta'_m} \, \right| \\
    &=&\sum_{m\in V^c_{r+1}}
         2 d'_m d'_{m\xor v}
  \sqrt{1 - [\Im(\braket{\hat\eta'_m}{\hat\eta'_{m\xor v}})]^2}\\
    &\le& \sum_{m\in V^c_{r+1}}
                2 d'_m d'_{m\xor v}  
\end{eqnarray*}
where the sign $\Im$ means the imaginary part.
Now we wish to give a bound in terms of the original values $d_l$.  
Using the fact that for any $\alpha > 0$ and any $x$, $y$ 
(which are real numbers),
$0 \leq (\alpha^\frac{1}{2}x - \alpha^{-\frac{1}{2}}y)^2 = \alpha x^2 + y^2/\alpha - 2xy$,
we get the general inequality $2xy \leq \alpha x^2 + \frac{1}{\alpha}y^2$ and so
\begin{align*}
SD_v &\leq \sum_{m\in V^c_{r+1}} 2 d'_m d'_{m\xor v} &\text{ }  \\
  &\leq \sum_{\stackrel{m\in V^c_{r+1}}{d_H(m, V_r) \geq \hat{v}/2}}
     2 d'_m d'_{m\xor v} +
  \sum_{\stackrel{m\in V^c_{r+1}}{d_H(m, V_r) < \hat{v}/2}}
     2 d'_m d'_{m\xor v} \\
  &\leq \sum_{\stackrel{m\in V^c_{r+1}}{d_H(m, V_r) \geq \hat{v}/2}}
      \left[\alpha d^{\prime 2}_{m\xor v} +
      \frac{1}{\alpha} d^{\prime 2}_m \right] +
  \sum_{\stackrel{m\in V^c_{r+1}}{d_H(m, V_r) < \hat{v}/2}}
      \left[\alpha d^{\prime 2}_{m} +
      \frac{1}{\alpha} d^{\prime 2}_{m\xor v} \right] \\
  &\leq \alpha \sum_{m \in V_r^c}  d^{\prime 2}_{m}  +
   \frac{1}{\alpha} 
   \sum_{\stackrel{m\in V^c_r}{d_H(m, V_r) \geq \hat{v}/2}}
        d^{\prime 2}_{m} 
   \text{\hspace*{2cm}by Eqs. (\ref{eq-disjoint},\ref{eq-distance})}\\
  &\leq \alpha
    \sum_{l \in V_r^c \oplus V_r} d^2_l
 + \frac{1}{\alpha} 
   \sum_{\stackrel{m\in V^c_r, n\in V_r}{d_H(m, V_r) \geq \hat{v}/2}}
        d^2_{m\oplus n} 
   \text{\hspace*{2cm}by Eq. (\ref{eq-d-prime})}\\
  &\leq \alpha + \frac{1}{\alpha} \sum_{|l| \geq \frac{\hat{v}}{2}} d^2_l
\end{align*}
Now we fix 
$\alpha=\sqrt{\sum_{|l|\ge\frac{\hat v}{2}}d_l^2}$ and obtain:
\begin{equation}
SD_v\le 2\sqrt{
\sum_{|l|\ge\frac{\hat v}{2}}d_l^2}
\end{equation}
\hfill \qed 
\end{proof}  

Note that 
$\hat{v} = d_H(v, V_r)$ 
where $r$ is the number of parity check strings.

\section{Existence of Codes for Both Reliability and Security}
\label{app:codes-exist}

Choosing a code which is good
when $n$ is large (for constant error rate) is
not a trivial problem in ECC. A Random Linear Code (RLC)
is one such code, however, it does not
promise us that the distances are as required,
but only gives the desired
distances with probability as close to one as we want.
With RLC, we find that the threshold below which a secure key
can be obtained is $p_{\allowed} \le 7.56 \%$.

In order to correct $t$ errors with certainty,
a code must have a minimal
Hamming distance between the code words $d \ge 2t +1$ so that all
original code
words, even when distorted by $t$ errors, can still be identified
correctly.
For any $c_T$ which passes the test,
we are promised (due to Lemma~\ref{sec5-lemm4})
that the probability of having
$t = |c_I| > n(p_{\allowed} + \epsilon_{\rel}) $
errors is smaller than
$ h = e^{-\frac{1}{2} n\epsilon_{\rel}^2}$.

Thus, we need to choose a RLC that promises a Hamming distance at
least $d$ such that
$p_{\allowed} + \epsilon_{\rel} <   t/n = \frac{d-1}{2n}$,
and then the
$t$ errors are corrected except for a probability
smaller than $ h =  e^{-\frac{1}{2} n\epsilon_{\rel}^2}$.
However, RLC can never promise a specific minimal distance with certainty, 
but can only promise it
with probability exponentially close to one:
For any $n,r=n-k$, and for $\delta$ such that $H_2(\delta)< r/n$,
an arbitrary {\it random linear code} $(n,k,d)$
satisfies $d/n\ge \delta$,
except for a probability (see~\cite{Gallager},
Theorem 2.2)
\begin{equation}
P[d/n<\delta] \le
        \frac{c(\delta)}{\sqrt{n}} 2^{n(H_2(\delta)-r/n)} 
\stackrel{\Delta}{=} g_1 
\end{equation}
where $c(\delta) = \frac{1}{1-2\delta}
\sqrt{\frac{1-\delta}{2\pi\delta}}$.

If we choose $\delta = 2(p_{\allowed} + \epsilon_{\rel}) + 1/n$
then we are promised that the errors are corrected,
except for some probability (bounded by $h$)
that the error rate is larger than expected, and 
some probability (bounded by $g_1$) that a bad random code was chosen.

Using such a code,
$\epsilon_{\rel}$ is now a function of $\delta$ so that
$ \epsilon_{\rel} = \delta/2 - 1/(2n) - p_{\allowed}$ and therefore,
\begin{eqnarray}
h =  e^{-(n/8)(\delta-\frac{1}{n}-2p_{\allowed})^2}
\end{eqnarray}
and almost all such codes correct all the errors.
One could conclude that the code is reliable except for a probability
$g_1 + h$, but this is not the case here; 
although the code is randomly produced, it can still be checked
in advance,
and used only if it satisfies the condition on $d$.
Thus the term $g_1$ does not need to be added\footnote{
We can still add the term $g_1$ and this saves us the need to find the
minimal distance of the code. 
} to the reliability bound,
and the bound is then given by $h$ alone.

Recall that we choose $\epsilon_{\sec}$ such that
$|v|\ge 2n(p_{\allowed}+\epsilon_{\sec})$.
Let $|v|$ be the minimal distance between one PA string and any other
parity check string (or linear combination) taken from ECC and PA.
Clearly, the Hamming weight of the dual code of the ECC, once the PA is
also added, provides a lower bound on $|v|$.
Thus, it is sufficient to demand
$d^\perp \ge 2n(p_{\allowed}+\epsilon_{\sec})$ in order to prove
security.
Choosing a RLC for the ECC and PA,
one cannot be completely sure that the distance
indeed satisfies the constraint, but this shall be true~\cite{Gallager} 
with probability
exponentially close to one (and can be checked in advance).
We use the dual code $(n,r^\perp,d^\perp)$, where $r^\perp=n-r-m$.
Such codes satisfy $d^\perp/n\ge \delta^\perp$, except for a fraction of
\begin{equation}
P[d^\perp/n<\delta^\perp] \le
\frac{c(\delta^\perp)}{\sqrt{n}} 2^{n(H_2(\delta^\perp)-(n-r-m)/n)}
= g_2 
\end{equation}
with $\delta^\perp = 2(p_{\allowed}+\epsilon_{\sec})$.

Assuming that Eve gets full information (namely, $m$ bits) when the code
fails we get due to the above and Proposition~\ref{mainprop}
\begin{equation}
\langle \mrv{I}'_{Eve} \rangle 
\leq 
m\left(2 e^{-\frac{1}{4}n\epsilon_{\rm sec}^2}+g_2\right)
\end{equation}
but we can get rid of $g_2$ by checking the code in advance\footnote{
Or we can add that term to Eve's information and this saves
us the need to find the
minimal distance of the dual code.}.
If we demand that 
\begin{eqnarray*}
H_2(\delta)-r/n&<&0 \\
H_2(\delta^{\perp})+r/n+m/n-1&<&0 
\ ,
\end{eqnarray*}
then both $g_1$ and $g_2$ are exponentially small.
Written another way:
\begin{eqnarray*}
H_2(2p_{\allowed}+2\epsilon_{\rel}+1/n) &<&r/n \\
H_2(2p_{\allowed}+2\epsilon_{\sec})+ r/n
&<&1-R_{\secret}
\end{eqnarray*}
where $R_{\secret}\equiv m/n$.

In order to find the threshold on
$p_{\allowed}$ we combine 
these two equations together 
\begin{equation}
H_2(2p_{\allowed}+2\epsilon_{\sec})+
H_2(2p_{\allowed}+2\epsilon_{\rel}+1/n)
<1-R_{\secret}
\ .
\end{equation}
In the limit of large $n$ and the two $\epsilon$'s close to zero,
we get that 
$p_{\allowed}<5.50\%$
satisfies the bound and hence this is our threshold.
[We can then chose the appropriate $r/n$ so that both $g_1$ and $g_2$ functions 
are exponentially small.]

Asymptotically, a final key with a bit-rate 
$R_{\secret}<1-H_2(2p_a)-H_2(2p_a)$ is secure
and reliable for the given ECC+PA chosen at random.  
Note, as $p_a$ goes to zero,
$R_{\secret}$ goes to $1$, which means all the information bits are
secret (asymptotically).

The above result can be improved (as noticed first 
by Mayers~\cite{Mayers98}) 
by taking RLC with distance
$d=t+1$ instead of $d=2t+1$.
Namely, $d-1 \ge n(p_{\allowed} +
\epsilon_{\rel})$ (without the factor of 2).
Due to Shannon's bound~\cite{MS-book} 
such a code can also correct
$ t = n(p_{\allowed} + \epsilon_{\rel})$ errors
with probability of failure smaller than $\hat\delta$
(for any $\hat\delta$).
This is true provided that $r/n > H_2(p_{\allowed} + \epsilon_{\rel})$,
and that a sufficiently large $n$ is chosen, but we did not find an
explicit connection between $n$ and $\hat\delta$, 
as we did with the other
probabilities $g_1$, $g_2$, and $h$.

The above is true except for an
exponentially small probability $g_1'$ that the code got the wrong
distance~\cite{Gallager}, 
and an exponentially small probability $h'$ that the code
is fine yet there
are more errors in the information bits than expected.

Choosing now $\delta = p_{\rm allowed} + \epsilon_{\rm rel} + 1/n$,
the term $g_1'$ is still the same as before, but with a
different $\delta$ then before. The condition for $g_1'$ to be 
exponentially small
becomes now
\[ H_2(p_{\allowed}+\epsilon_{\rel}+1/n) <r/n \ . \]
The term $h'$ (telling us the probability of having more errors on the information 
bits than expected from the test results) is 
\[ 
h' =  e^{-(n/2)(\delta-\frac{1}{n}-p_{\allowed})^2}
\ .
\]
One could conclude that the code is reliable except for a probability
$g_1' + h' + \hat\delta$, but (again) 
the term $g_1'$ can be removed if we check
the code in advance to make sure it has the right distance.
The bound is thus given by $h' + \hat\delta$.
However, we do not have an exponentially small
expression for $\hat\delta$ (as a function of $n$) and it is only
known that we can render the error as small as we want
by taking a sufficiently large $n$.

For the security proof 
we choose $\epsilon_{\sec}$ such that
$|v|\ge 2n(p_{\allowed}+\epsilon_{\sec})$, and we demand
$d^\perp \ge 2n(p_{\allowed}+\epsilon_{\sec})$.
Choosing a RLC for the ECC and PA,
one cannot be completely sure that the distance
indeed satisfies the constraint, but this shall be true with probability
exponentially close to one (and can be checked in advance).
As before, we use the dual code $(n,r^\perp,d^\perp)$, where $r^\perp=n-r-m$.
Such codes satisfy $d^\perp/n\ge \delta^\perp$, except for a fraction of
\begin{equation}
P[d^\perp/n<\delta^\perp] \le
\frac{c(\delta^\perp)}{\sqrt{n}} 2^{n(H_2(\delta^\perp)-(n-r-m)/n)}
= g_2' 
\end{equation}
with $\delta^\perp = 2(p_{\allowed}+\epsilon_{\sec})$.
As before, we can get rid of $g_2'$ 
by checking the code in advance.

In order for $g_2'$ to be exponentially small 
we demand 
\[
H_2(\delta^{\perp})+r/n+m/n-1 < 0 \ , \]
so finally:
\begin{eqnarray*}
H_2(p_{\allowed}+\epsilon_{\rel}+1/n) &<&r/n \\
H_2(2p_{\allowed}+2\epsilon_{\sec})+
H_2(p_{\allowed}+\epsilon_{\rel}+1/n)
&<&1-R_{\secret}
\end{eqnarray*}
where $R_{\secret}\equiv m/n$.

In the limit of large $n$ and $\epsilon$'s close to zero,
$p_{\allowed}<7.56\%$
satisfies the bound and hence this is our imporved threshold.
With this threshold we have an explicit bound on Eve's information,
but only an asymptotic bound for the probability of failing in terms
of reliability.

\end{document}